\documentclass[ecta,draft]{econsocart}
\RequirePackage[colorlinks,citecolor=blue,linkcolor=blue,urlcolor=blue]{hyperref}
\usepackage{bbm}
\usepackage{booktabs}
\usepackage{multirow}
\usepackage{amsmath}
\usepackage{amssymb}
\usepackage{graphicx}
\usepackage{subcaption}
\usepackage{placeins}

\startlocaldefs

\theoremstyle{plain}
\newtheorem{assumption}{Assumption}
\newtheorem{prop}{Proposition}

\theoremstyle{remark}

\endlocaldefs

\makeatletter\AtBeginDocument{\renewcommand\NAT@sep{;}}\makeatother
\begin{document}
\begin{frontmatter}

\title{Does the Gender Wage Gap Originate at Labor Market Entry? Evidence from South Korea}
\runtitle{Gender wage gap at labor market entry}

\begin{aug}
\author[id=au1,addressref={add1}]{\fnms{Dongwoo}~\snm{Kim}\ead[label=e1]{dongwook@sfu.ca}}
\address[id=add1]{%
\orgname{Simon Fraser University and Korea University}}
\end{aug}

\support{I thank participants at the Korean Economic Review International Conference 2026 for helpful comments and discussions. I gratefully acknowledge support from the Social Sciences and Humanities Research Council of Canada under Insight Grant 435-2024-0322. I have no competing interests to declare. All errors are my own.}

\begin{abstract}
When in the lifecycle does a large gender wage gap emerge? South Korea has the largest aggregate gender pay gap (29\%) in the OECD. Among recent college graduates the conditional gap is only 3.3\% over 2008--2019, falling from 3.4\% to 2.5\%. Correcting for differential selection into employment with semiparametric and machine learning methods leaves it unchanged. Nonparametric and semiparametric bounds contain the point estimates. Selection-corrected quantile regression shows that the remaining entry-level gap is concentrated at the top of the distribution. In Korean panel data the corrected gap widens severalfold across the prime-age workforce, where the selection correction becomes first order. Korea's gender wage disparity is mostly generated after labor market entry.
\vspace{1em}
\end{abstract}

\end{frontmatter}

\newpage

\section{Introduction}\label{sec:intro}

When in the lifecycle does the gender wage gap emerge? Answering this question is important for understanding the sources of gender inequality and for designing effective policy interventions. The aggregate national gender wage gap, a widely cited statistic, does not answer this question because it conflates initial differences at labor market entry with the cumulative effects of career interruptions, occupational re-sorting, child penalties, and generational differences. This paper investigates whether large gender gaps originate at labor market entry in South Korea. \cite{oecd2024} reports a gender pay gap of 29.3\% for full-time Korean workers in 2023, about 2.6 times the OECD average of 11.3\%, and \cite{oecd2026wellbeing} identifies it as the widest among its member countries. Korea has institutional conditions, such as strong gender norms, a substantial gender employment gap, and sharp occupational segregation, that could have generated a large entry-level gap. Therefore, one might expect a substantial gender wage gap to be present already at labor market entry.

This paper's central message is that Korea's large aggregate gap does not originate at labor market entry. Using the Graduates Occupational Mobility Survey (GOMS) data, I establish four findings that build to this conclusion. First, large national gender gaps can coexist with small conditional entry-level gaps. Among recent college graduates, the raw entry-level wage gap remains sizable (around 12\%), but after controlling for educational background, age, and job characteristics, the conditional hourly wage gap is only 3.3\% over 2008--2019, and it falls across the period (from 3.4\% in 2008--2011 to 2.5\% in 2016--2019). Second, this conclusion survives flexible semiparametric selection corrections that do not rely on exclusion restrictions, all of which yield estimates within one log point of OLS. Third, the remaining entry-level gap is concentrated at the top of the wage distribution. Selection-corrected quantile estimates show a pronounced pattern by 2016--2019. The gap has effectively closed at the 10th percentile but remains around 5\% at the 90th, consistent with the hypothesis of \cite{goldin2014grand} about high-paying positions (``greedy jobs''). Fourth, because the aggregate gender gap combines within-cohort lifecycle divergence with differences across cohorts, the discrepancy between the small entry-level gap and the large aggregate gap suggests that gender disparity in Korea is generated largely after entry.\footnote{For instance, \cite{kleven2025child} estimate a child penalty in employment of 49\% for South Korea, one of the largest among advanced economies. In their framework, the child penalty measures the post-childbirth divergence in employment between women and men.} 

To examine this lifecycle interpretation, I apply the same estimation methods to the Korean Labor and Income Panel Study (KLIPS) and trace the cross-sectional age profile of the gap across the prime-age workforce. The selection-corrected gap more than doubles among college-educated workers, from 7\% at ages 24 to 35 to 16\% at ages 24 to 55, and more than triples across all workers, from 8\% to 26\%. The selection correction, negligible at entry, becomes first order once the female employment ``M-curve'' takes hold.\footnote{The ``M-curve'' describes the typical age profile of female labor supply. Participation rises in the early career, falls following marriage and childrearing, and recovers later.} 

The methodological challenge in establishing these findings is that wages are observed only for the employed. If the unobservable determinants of employment correlate with wages, OLS estimates conditional on employment are biased, and the direction of that bias is what differential selection would reveal. This selection problem is a central econometric concern in the gender wage gap literature \citep{mulligan2008selection, blau2024selection}. Because the missing wage offers cannot be recovered even by experimental variation, the sample selection framework is the natural econometric approach in this setting. The classical Heckman selection model \citep{heckman1979} imposes joint normality, and without exclusion restrictions its finite-sample performance can be poor.\footnote{See the related discussions in \cite{hay1984}, \cite{duan1984}, \cite{manning1987}, and \cite{leung1996}.} I therefore estimate the entry-level gap using the semiparametric selection model of \cite{KL2026}, which achieves point identification without exclusion restrictions and allows for heteroskedasticity. I then compare the results with those from debiased machine learning methods \citep{chernozhukov2018double,pan2024}, bounds approaches \citep{lee2009,honore2020selection}, and estimators using exclusion restrictions \citep{arellano2017quantile,biewen2020counterfactual}.

The paper contributes to the empirical literature on the gender wage gap and employment selection. The standard empirical framework is to estimate wage gaps conditional on observables and decompose them using the Oaxaca-Blinder framework \citep{oaxaca1973, blinder1973}. Among college-educated workers, differences in major and occupation explain a substantial share of the gender wage gap \citep{black2008gender, sloane2021college}. Because these decompositions condition on employment, selection into work can distort both the level and the trend of measured gaps \citep{fortin2011decomposition}. In the US literature, \cite{blau2017gender} review the broad evidence on long-run gender wage convergence, while \cite{goldin2014grand} emphasizes the role of temporal inflexibility in the remaining conditional gap. On selection, \cite{mulligan2008selection} attribute most of the measured narrowing to changes in female workforce composition. More recently, \cite{blau2024selection} show that the gender wage gap narrowed substantially even after correcting for selection using the Panel Study of Income Dynamics (PSID). Their preferred method uses longitudinal wage histories to impute missing wages. My approach instead estimates the gender wage gap using only repeated cross-sectional data.

The estimators I use are built on the econometric literature on sample selection. Semiparametric and nonparametric extensions of Heckman's model relax its distributional assumption but typically still require exclusion restrictions for identification \citep{ahn1993semiparametric, das2003, newey2009, machado2017}. \cite{chamberlain1986} characterizes what identification requires without them. Bounding methods \citep{blundell2007changes, lee2009, honore2020selection} weaken these requirements but generally yield partial identification. More recently, \cite{KL2026} show that point identification can be achieved without exclusion restrictions when the selection process is sufficiently nonlinear, and propose a semiparametric two-step sieve plug-in estimator. \cite{pan2024} develop a complementary debiased machine learning approach. I use these recent methods together with Lee and Honor\'e-Hu bounds to examine whether the paper's headline finding is robust to different identifying assumptions. 

The paper also relates to the literature on the gender wage gap at labor market entry. Studies that follow a cohort over time find a small gap at entry that widens sharply thereafter. Among US elite professionals, women and men start with nearly identical earnings yet diverge over their careers.\footnote{Among MBAs, the gender gap reaches almost 60 log points within a decade \citep{bertrand2010dynamics}. Among lawyers, initially small gaps widen substantially over time \citep{wood1993pay, noonan2005pay}. Differences in billable hours and client revenue help explain the gap among lawyers \citep{azmat2017gender}.} The pattern is not confined to elites. For US college graduates the gap expands by about 34 log points over the early career \citep{goldin2017expanding}. \cite{manning2008gender} also show that the gap is near zero among UK labor market entrants of all education levels and reaches roughly 25 log points after ten years. By contrast, studies measuring the gap among recent graduates often find it nontrivial already at entry.\footnote{See evidence for the United States \citep{weinberger1998race, joy2003salaries}, Germany \citep{francesconi2018early, sandner2025early}, Italy \citep{piazzalunga2018gender}, and Russia \citep{rozhkova2024dynamics}.} Almost all of this evidence conditions on the employed sample without selection correction. I extend this literature by estimating selection-corrected entry-level gender gaps in a nationally representative sample of Korean college graduates. I then apply the same methods to older workers to examine how the gap and selection bias vary with age.\footnote{Korean studies examine lifecycle gender wage gaps \citep{dynan2022korea, stansbury2024korea}, changes across the wage distribution \citep{tromp2019narrowing}, and the lasting effects of entering the labor market during the Asian financial crisis \citep{choi2020recession, choi2025recession}. Evidence on recent graduates includes \cite{tromp2022graduating}, \cite{cho2024entry}, \cite{kim2022taste}, and \cite{oh2024growth}. Studies accounting for employment selection use Heckman models \citep{cho2017heckman, moon2023selection} or imputation methods \citep{tromp2026selection}.}

The remainder of the paper proceeds as follows. Section~\ref{sec:framework} presents the econometric framework. Section~\ref{sec:data} describes the data. Section~\ref{sec:results} reports the main results and robustness analyses. Section~\ref{sec:klips} estimates the lifecycle evolution of the gender wage gap. Section~\ref{sec:conclusion} concludes. Additional empirical results are provided in the Online Appendix.

\section{Econometric Framework}\label{sec:framework}

\subsection{The Semiparametric Sample Selection Model}\label{sec:model}

The econometrician observes a random sample $\{(Y_i, D_i, X_i)\}_{i=1}^{n}$, where $Y_i = D_i Y_i^*$ is the observed outcome (log wage), $D_i \in \{0, 1\}$ is the selection indicator (employment status), and $X_i \in \mathbb{R}^{d_X}$ is a row vector of covariates. The latent wage equation is
\begin{equation}\label{eq:wage}
Y_i^* = \alpha_0 + X_i \beta_0 + V_i,
\end{equation}
where $\beta_0$ is the parameter vector and $V_i$ is an unobserved error. The selection mechanism takes the form $D_i = \mathbbm{1}\{h_0(X_i) \geq \varepsilon_i \},$
where $h_0(\cdot)$ is an unknown function and $\varepsilon_i$ is an unobserved selection error that may be correlated with $V_i$.\footnote{The threshold-crossing structure connects to conditional monotonicity \citep{semenova2023generalized}, which allows the direction of selection to differ across covariate values while requiring a common direction within each covariate cell. This relaxes the unconditional monotonicity assumption of \cite{lee2009}.} I leave both the functional form $h_0(\cdot)$ and the distribution of $\varepsilon_i$ unspecified, so identification cannot lean on a parametric selection rule. In \cite{KL2026} the primitive condition is the following control-function restriction:
\begin{equation}\label{eq:cond_mean}
E[Y_i \mid X_i = x, D_i = 1] = x\beta_0 + \lambda_0(p_0(x)),
\end{equation}
where $p_0(x) := P(D_i = 1 \mid X_i = x)$ is the conditional selection probability and $\lambda_0(\cdot)$ is an unknown function that captures the selection bias.\footnote{The outcome intercept $\alpha_0$ is not separately identified from the level of the selection bias function $\lambda_0$, so it is normalized to $0$. All coefficients reported in this paper exclude the intercept.} Selection bias enters the wage equation only through the scalar selection probability, which is what makes the model tractable for identification and estimation. This restriction is not a selection-on-observables assumption. It allows selection on unobservables by permitting correlation between $V_i$ and $\varepsilon_i$. 

The threshold-crossing structure is sufficient to imply the control-function restriction when $\varepsilon_i$ is independent of $X_i$ and $E[V_i \mid X_i, \varepsilon_i] = E[V_i \mid \varepsilon_i]$. Let $F_\varepsilon$ denote the continuous cumulative distribution function of $\varepsilon_i$ and define $U_i = F_\varepsilon(\varepsilon_i) \sim \text{Unif}(0,1)$. The selection rule then becomes $D_i = \mathbbm{1}\{p_0(X_i) \geq U_i\}$, and the conditional mean wage among employed workers becomes \eqref{eq:cond_mean}.

The restriction \eqref{eq:cond_mean} allows for heteroskedasticity, as it does not restrict the conditional variance of the wage error among the employed. A sufficient structure under the threshold-crossing rule with $\varepsilon_i$ independent of $X_i$ is
\begin{equation}\label{eq:hetero}
V_i = \nu(\varepsilon_i) + \sigma(X_i, \varepsilon_i)\,\xi_i, \qquad E[\xi_i \mid X_i, \varepsilon_i] = 0.
\end{equation}
Because employment is determined by $(X_i,\varepsilon_i)$, the second component has mean zero among the employed, leaving $E[V_i \mid X_i=x,D_i=1]=E[\nu(\varepsilon_i)\mid\varepsilon_i\le h_0(x)]=\lambda_0(p_0(x))$. This differs from the multiplicative location-scale specification $V_i=s(X_i)e_i$, with $(e_i,\varepsilon_i)$ independent of $X_i$, where the correction becomes $s(X_i)m(p_0(X_i))$ \citep{donald1995,chen2003heteroskedastic}. In such a specification, the same factor $s(X_i)$ scales both wage shocks and the mean selection effect. Under \eqref{eq:hetero}, on the other hand, workers with the same selection probability can have different wage dispersion while sharing the same mean selection correction. Neither structure generally nests the other.

The semiparametric model nests the Heckman model and the \cite{honore2020selection} framework as special cases: it preserves their threshold-crossing selection structure while relaxing both the linear-index restriction in the first stage and the joint normality of the unobserved heterogeneity terms. The identification challenge is to separate $\beta_0$ from $\lambda_0(\cdot)$. If $p_0(x)$ were a known linear index like $\Phi(x\gamma)$ (as in the Heckman model), the selection bias function $\lambda_0(\Phi(x\gamma))$ would be collinear with $x\beta_0$ in the absence of exclusion restrictions, making $\beta_0$ unidentified. \cite{KL2026} show that identification is restored when the selection probability $p_0(x)$ is a nonlinear function of $x$. The key conditions are:

\begin{assumption}\label{ass:id}
(i) At least one component $X_1$ of $X$ is continuously distributed. (ii) $p_0(X)$ is continuously differentiable with respect to $x_1$. (iii) $\partial p_0 / \partial x_1 \neq 0$ with probability one. (iv) $E[Y \mid X, D=1] = X\beta_0 + \lambda_0(p_0(X))$. (v) $\lambda_0(\cdot)$ is continuously differentiable almost everywhere. (vi) $X$ has no perfect multicollinearity (i.e., $E[X'X]$ has full rank).
\end{assumption}

Conditions (i)--(iii) require at least one continuous covariate that moves the selection probability smoothly and nontrivially. Condition (iv) is the control-function restriction. Condition (v) is a smoothness condition that allows this unknown selection bias function to be approximated by sieves. Condition (vi) is the standard rank condition needed for the slope coefficients in the wage equation. The central identification result is as follows:

\begin{prop}\label{prop:id}
Under Assumption~\ref{ass:id}, $\beta_0$ and $\lambda_0$ are identified if any one of the following conditions holds:
\begin{enumerate}
\item $X$ consists of a single continuously distributed covariate $X_1$, and there exist two distinct values $x'$ and $x''$ in the support of $X_1$ such that $p_0(x') = p_0(x'')$.
\item $X$ contains another continuously distributed covariate $X_2$, $p_0(X)$ is continuously differentiable with respect to $x_2$, $\partial p_0 / \partial x_2 \neq 0$ with probability one, and the ratio $(\partial p_0 / \partial x_1) / (\partial p_0 / \partial x_2)$ is not constant on the support of $X$.
\item $X$ contains $X_1$ and a binary covariate $X_2$. Holding the remaining covariates fixed, write $p_0(x_1, x_2)$ for the selection probability. For almost every $x_1$ in the support of $X_1$, there exists $\psi(x_1)$ in the support of $X_1$ such that $p_0(\psi(x_1), 1) = p_0(x_1, 0)$ and  $x_1 - \psi(x_1)$ is not constant in $x_1$. 
\end{enumerate}
\end{prop}

Proposition~\ref{prop:id} establishes point identification without an excluded variable in three cases. First, equal selection probabilities at two different values of a covariate suffice for identification. Second, when at least two continuous covariates are present, non-proportional marginal effects of the two covariates on selection provide identification. Lastly, identification is obtained when shifts in selection by a binary covariate cannot be absorbed by a constant adjustment in one continuous covariate. These conditions fail under the linear-index specification $p_0(x) = F(x\gamma)$. Crucially, nonlinearity in the selection equation is required for identification but is not imposed a priori in estimation. Instead, I assess its empirical relevance after estimating the selection probability. The estimation of such a model follows the two-step plug-in procedure of \cite{KL2026}.

\textbf{Step 1: Selection probability.} Using the full sample, estimate $p_0(X_i)$ using sieve maximum likelihood estimation:
\[
P(D_i = 1 \mid X_i) = F\Big(\sum_j f_j(X_{ij}) + \sum_{j<k} g_{jk}(X_{ij}, X_{ik}) + \sum_j \sum_l f_{jl}(X_{ij}) \cdot W_{il} + Z_i'\delta\Big),
\]
where $f_j(\cdot)$ are marginal sieve basis functions for the continuous covariates $X_j$, $g_{jk}(\cdot,\cdot)$ are tensor-product bases over the continuous covariates that capture flexible interactions between them, $f_{jl}(\cdot)$ are sieve interaction terms between the continuous covariates and the discrete covariates $W_l$, $F$ is the cdf of $\varepsilon_i$, and the vector $Z_i$ collects all the discrete covariates including $\{W_l\}$. In estimation, I use the standard normal cdf for $F$. This is an innocuous normalization, since composing any strictly increasing error distribution with the index is absorbed into the growing sieve expansion.

\textbf{Step 2: Partially linear regression.} Conditional on $D_i = 1$, estimate $\beta_0$ and $\lambda_0$ via partially linear regression. The unknown function $\lambda_0(p)$ is approximated by a sieve approximation: $Y_i = X_i \beta + \sum_{k=1}^{K} \gamma_k B_k(\hat{p}_i) + e_i$,
where $B_k(\cdot)$ are sieve basis functions for the estimated selection probability $\hat{p}_i$ from Step 1. The coefficient vector $\beta$ and sieve coefficients $\gamma_k$ are estimated by OLS. 

Both estimation steps can be implemented using standard regression routines. Under the regularity conditions in \cite{KL2026}, with the first stage estimated by sieve maximum likelihood on a fixed basis, $\hat{\beta}$ is $\sqrt{n}$-consistent and asymptotically normal. Writing $\tilde{X}_i = D_i\{X_i - E[X_i \mid p_0(X_i), D_i = 1]\}$ for the partialling-out residual, $e_i$ for the second-stage regression error above, $s_i$ and $J$ for the first-stage score and information matrix, and $G = E[\tilde{X}_i' \lambda_0'(p_0(X_i))\, \partial \hat{p}(X_i)/\partial \pi']$ for the cross-derivative of the selection correction term with respect to the first-stage coefficients, the estimator admits the influence function $A^{-1}\{\tilde{X}_i'e_i - G J^{-1} s_i\}$ with $A = E[\tilde{X}_i'\tilde{X}_i]$, so the asymptotic variance is $A^{-1}(\Omega + G J^{-1}\Sigma_s J^{-1} G')A^{-1}$, where $\Omega = E[\tilde{X}_i'\tilde{X}_i e_i^2]$ and $\Sigma_s = E[s_i s_i']$.\footnote{The term $GJ^{-1}s_i$ accounts for first-stage estimation error through the score and information matrix of the first-stage sieve MLE, following the usual two-step variance adjustment \citep{newey1994asymptotic,newey2009}. This computational approach also applies under growing-sieve asymptotics: under appropriate regularity conditions and rates of sieve growth, \cite{ackerberg2012practical} establish numerical equivalence between the conventional two-step sandwich, computed by treating the sieve coefficients as finite-dimensional parameters, and a consistent estimator of the semiparametric variance.} This asymptotic variance accounts for heteroskedasticity. 

\subsection{Alternative Methods}\label{sec:comparison}

For comparison, I estimate the gender wage gap using four additional methods. The first is ordinary least squares (OLS) on the employed subsample ($D_i = 1$), ignoring selection. The difference between $\hat{\beta}$ obtained by OLS and by selection correction methods is informative about the magnitude of selection bias. Second, I estimate the Heckman selection model, with joint normality of $(V_i, \varepsilon_i)$, using MLE. Third, two bounds approaches are considered: the fully nonparametric \cite{lee2009} bounds and the \cite{honore2020selection} bounds that tighten the Lee bounds using linear-index structures for both selection and outcome equations. Lastly, I implement the double/debiased machine learning (DML) estimator of \cite{pan2024}, in which the selection probability is estimated by a random forest with $K$-fold cross-fitting, and the second-stage partially linear moment \citep{robinson1988root} is Neyman-orthogonalized \citep{chernozhukov2018double}. I refer to this estimator as DML-RF. The original Pan-Zhang estimator uses both employed and non-employed graduates in the second stage, so all the covariates in the wage equation must be observed for both groups. Job characteristics such as firm size and industry, however, are observed only for workers. I therefore use a selected-sample extension that includes these controls. Table~\ref{tab:assumptions} summarizes the identifying assumptions these estimators impose.

\begin{table}[htbp]
	\centering
	\caption{Identifying Assumptions across Estimators}\label{tab:assumptions}
	{\footnotesize\setlength{\tabcolsep}{2pt}
		\begin{tabular}{lccccccc}
			\toprule
			Method
			& \shortstack{Linear\\outcome}
			& \shortstack{Linear\\selection}
			& \shortstack{Joint\\normality}
			& \shortstack{Random\\selection}
			& \shortstack{Selection\\monotonicity}
			& \shortstack{Nonlinear\\selection}
			& \shortstack{Assumption~\ref{ass:id}\\(iv)} \\
			\midrule
			Selected-sample OLS   & \checkmark &            &            & \checkmark &            &            & \\
			Heckman               & \checkmark & \checkmark & \checkmark &            & \checkmark &            & \checkmark \\
			Kim-Lee               & \checkmark &            &            &            & \checkmark & \checkmark & \checkmark \\
			Pan-Zhang DML        & \checkmark &            &            &            & \checkmark & \checkmark & \checkmark \\
			Lee bounds            &            &            &            &            & \checkmark &            & \\
			Honor\'e-Hu bounds   & \checkmark & \checkmark &            &            & \checkmark &            & \checkmark \\
			\bottomrule
	\end{tabular}}
	
	\vspace{0.5em}
	\parbox{\textwidth}{\scriptsize\textit{Notes:} A check mark indicates that the method relies on the assumption in that column, either directly or through its other assumptions.}
\end{table}

\section{Data}\label{sec:data}

The data come from the GOMS, administered annually by the Korea Employment Information Service. GOMS is a nationally representative repeated cross-sectional survey of graduates of postsecondary institutions (two-year colleges and four-year universities). Each cohort is surveyed approximately 18 months after graduation, and the survey collects detailed information on employment status, wages, working hours, job characteristics, educational background, and family background. I use 12 waves from 2008 to 2019.\footnote{The first wave of GOMS (2007) is excluded. It contains only 4,568 full-time wage workers, compared with more than 9,000 in each subsequent wave: 75\% of its respondents fall outside the full-time wage sample, against 30 to 48\% in later waves.} GOMS labels each wave by its graduation year: the 2019 wave, for example, covers graduates of August 2018 and February 2019 and was surveyed with a reference date of September 1, 2020. Each wave surveys approximately 18,000 graduates, and the sample is refreshed each year. The GOMS data are well suited for studying the entry-level gender wage gap because all respondents are at approximately the same career stage and have similar levels of labor market experience. Focusing on postsecondary graduates is especially informative in Korea because tertiary education is nearly universal among young adults: 71\% of Koreans aged 25--34 have completed tertiary education, the highest rate in the OECD and far above the OECD average of 48\% \citep{oecd2025eag}.

I construct the analysis sample from the 2008--2019 waves as follows. First, I drop 67 individuals with missing or implausible ages, leaving 217,649 observations. Second, I restrict the sample to graduates aged 35 or younger, which retains 95.5\% of these observations (207,918). The excluded graduates (ages above 35) are mostly non-traditional students from online universities and distance education programs: 70\% attended two-year colleges, 48\% studied social science (typically business or public administration), and their mean age is 46.5 years. These individuals are typically mid-career workers completing tertiary degrees while employed, so the gender wage gap among them reflects the lifecycle divergence. Third, I identify respondents who are employed in full-time wage work (excluding self-employment, unpaid family work, and part-time employment). This defines the selection indicator $D_i = 1$. Fourth, I construct the dependent variable as log hourly wage, computed by dividing monthly wage (in 10,000 KRW) by monthly working hours (including overtime). I drop 549 observations whose hourly wage is below half of the statutory minimum wage in force at the survey date or above 100,000 KRW, because these values likely reflect measurement error. Dropping 5,255 observations with missing GPA or other covariates yields the final analysis sample of 202,114 observations, of which 127,628 are employed in full-time wage work (63.1\%). GOMS provides sampling weights that align the sample composition with the graduate population. I report unweighted estimates of the conditional wage equations, which are consistent under correctly specified conditional means and exogenous survey sampling conditional on the covariates \citep{solon2015weighting}.\footnote{Applying the survey weights changes the pooled OLS female coefficient from $-0.0326$ to $-0.0365$.}

The primary variable of interest is \textit{female}, an indicator for the respondent being female. The female coefficient in the wage equation is my estimate of the gender wage gap, conditional on observable characteristics. The covariates are grouped into three categories:

\begin{itemize}
	\item Individual characteristics: age, GPA on a 0--100 scale, and marital status.
	\item Educational background: a four-year university dummy (versus a two-year college), school-area dummies (17 city/province categories), and a set of major field category dummies (social science, humanities, education, engineering, natural science, medical/health, and arts/sports).
	\item Job characteristics: indicators for employment at a large firm (300 or more employees) and in the public sector, a set of industry dummies (manufacturing, construction, wholesale/retail, IT/communication, finance, professional services, public administration, education, and health/social work), and workplace-region dummies (17 city/province categories).
\end{itemize}

\begin{table}[p!]
\centering
\caption{Summary Statistics by Gender and Employment Status}\label{tab:summary}
\begin{tabular}{lcccc}
\toprule
 & \multicolumn{2}{c}{Male} & \multicolumn{2}{c}{Female} \\
\cmidrule(lr){2-3} \cmidrule(lr){4-5}
 & Employed & Not employed & Employed & Not employed \\
\midrule
\multicolumn{5}{l}{\textit{Panel A: Demographics \& education}} \\
$N$ & 70,656 & 38,054 & 56,972 & 36,432 \\
Age & 27.3 & 27.0 & 24.9 & 24.9 \\
GPA (0--100) & 80.5 & 80.0 & 83.2 & 82.7 \\
4-year university (\%) & 75.4 & 76.8 & 72.9 & 75.5 \\
School in Seoul (\%) & 19.6 & 20.9 & 23.8 & 26.8 \\
Worked in college (\%) & 62.7 & 58.1 & 67.9 & 64.6 \\
Currently married (\%) & 5.8 & 2.3 & 2.5 & 4.9 \\
\midrule
\multicolumn{5}{l}{\textit{Panel B: Major field of study (\%)}} \\
Social science & 20.6 & 18.5 & 23.5 & 21.8 \\
Engineering & 44.6 & 38.2 & 11.3 & 10.7 \\
Humanities & 7.2 & 9.7 & 13.8 & 17.5 \\
Natural science & 11.2 & 16.0 & 12.9 & 16.6 \\
Education & 4.2 & 4.9 & 12.5 & 9.4 \\
Medical/health & 5.1 & 3.1 & 12.1 & 5.0 \\
Arts/sports & 7.2 & 9.5 & 13.9 & 19.1 \\
\midrule
\multicolumn{5}{l}{\textit{Panel C: Job characteristics (employed only)}} \\
Log hourly wage & \multicolumn{2}{c}{0.093} & \multicolumn{2}{c}{$-$0.029} \\
Monthly wage (10K KRW) & \multicolumn{2}{c}{232.9} & \multicolumn{2}{c}{192.4} \\
Large firm (\%) & \multicolumn{2}{c}{44.9} & \multicolumn{2}{c}{33.1} \\
Public sector (\%) & \multicolumn{2}{c}{15.6} & \multicolumn{2}{c}{26.7} \\
Manufacturing (\%) & \multicolumn{2}{c}{27.3} & \multicolumn{2}{c}{11.9} \\
Finance (\%) & \multicolumn{2}{c}{4.8} & \multicolumn{2}{c}{4.5} \\
Education sector (\%) & \multicolumn{2}{c}{8.7} & \multicolumn{2}{c}{20.5} \\
Health/social (\%) & \multicolumn{2}{c}{5.1} & \multicolumn{2}{c}{18.1} \\
\bottomrule
\end{tabular}

\vspace{0.5em}
\parbox{\textwidth}{\scriptsize\textit{Notes:} GOMS waves 2008--2019, restricted to graduates aged 35 or younger. ``Employed'' denotes full-time wage workers with a valid hourly wage ($N = 127{,}628$). Graduates with an hourly wage below half of the minimum wage at the survey date or above 100,000 KRW are dropped. ``Not employed'' includes all other graduates ($N = 74{,}486$). Log hourly wage is computed as log(monthly wage / (total weekly hours $\times$ 4.345)). Entries are means and percentages.}
\end{table}

Table~\ref{tab:summary} presents summary statistics by gender and employment status. Male graduates outnumber females in the full sample, and their employment rate is higher. Men are on average 2 years older than women, reflecting mandatory military service. Men are also much more likely to have studied engineering, whereas women are more likely to have studied education or medical and health fields. Women have higher average GPAs. Among those employed, women are less likely to work in large firms and more likely to work in the public sector, education, and health services. Women with medical and health degrees make up 12.1\% of employed women, more than twice their share among the non-employed (5.0\%). The opposite is true of humanities graduates and of arts and sports graduates. Arts and sports alone account for nearly one in five non-employed women, compared with 13.9\% of those employed. For men, engineering is more common and natural science is less common among the employed.

\section{Gender Wage Gap Estimation}\label{sec:results}

This section estimates the entry-level gender wage gap, accounting for selection into employment. I first report the first-stage estimates, which show strong nonlinearity in the selection process. I then present the main OLS, Heckman, Kim-Lee, and DML-RF estimates and investigate whether the change in the gap over 2008--2019 reflects changes in who is employed. An Oaxaca-Blinder decomposition and selection-corrected quantile regressions follow, and the section concludes with robustness analyses.

\subsection{First-Stage Estimation}\label{sec:first_stage}

The Kim-Lee first stage estimates the selection probability $p_0(X_i)$ by sieve probit maximum likelihood, using cubic B-spline expansions of age and GPA together with their tensor product and their interactions with gender, school type, school area, and major category ($K = 204$ basis functions in the pooled sample). The second stage estimates the partially linear wage equation by OLS, approximating the selection correction $\lambda_0(\hat{p})$ with a cubic B-spline with 5 degrees of freedom. Standard errors are computed using the first-stage-corrected sandwich formula.\footnote{The corrected variance treats observations as i.i.d. Clustering it at the school-area $\times$ wave level (199 clusters) moves the pooled standard error from 0.0027 to 0.0031. All conclusions are unchanged, so the independent-observation version is reported throughout.} Appendix~\ref{sec:appendix_flexible} shows that the estimates are insensitive to both the first-stage basis dimension (Table~\ref{tab:flexible_gam}) and the second-stage spline dimension (Table~\ref{tab:bs_df}).

\begin{table}[htbp]
	\centering
	\caption{First-Stage Sieve Probit: Wald Tests of Nonlinearity}\label{tab:first_stage}
	\begin{tabular}{lcccc}
\toprule
 & Pooled & 2008--2011 & 2012--2015 & 2016--2019 \\
\midrule
\multicolumn{5}{l}{\textit{Wald tests of sieve term blocks ($\chi^2$ / $p$-value)}} \\
Age spline & 16.1 / 0.007 & 15.1 / 0.010 & 24.0 / $<$0.001 & 4.2 / 0.527 \\
GPA spline & 20.0 / 0.001 & 4.2 / 0.515 & 9.4 / 0.094 & 1.8 / 0.879 \\
Age $\times$ GPA tensor product & 62.7 / $<$0.001 & 20.7 / 0.014 & 29.9 / $<$0.001 & 33.3 / $<$0.001 \\
Age spline $\times$ female & 182.7 / $<$0.001 & 148.3 / $<$0.001 & 82.9 / $<$0.001 & 10.3 / 0.066 \\
GPA spline $\times$ female & 18.7 / 0.002 & 2.3 / 0.803 & 7.1 / 0.213 & 16.5 / 0.006 \\
Age and GPA splines $\times$ 4-year & 42.5 / $<$0.001 & 26.3 / $<$0.001 & 25.2 / $<$0.001 & 4.4 / 0.619 \\
Age and GPA splines $\times$ school area & 327.8 / $<$0.001 & 130.5 / 0.003 & 199.1 / $<$0.001 & 200.3 / $<$0.001 \\
Age and GPA splines $\times$ major & 511.9 / $<$0.001 & 200.6 / $<$0.001 & 195.9 / $<$0.001 & 208.1 / $<$0.001 \\
\midrule
Sieve dimension $K$ & 204 & 189 & 196 & 196 \\
$N$ (full sample) & 202,114 & 66,008 & 66,995 & 69,111 \\
\midrule
\multicolumn{5}{l}{\textit{Model comparison (pooled)}} \\
\multicolumn{3}{l}{Likelihood ratio vs.\ linear probit} & \multicolumn{2}{c}{1,511.9 (df $=164$)} \\
\multicolumn{3}{l}{AIC difference (linear $-$ sieve)} & \multicolumn{2}{c}{1,183.9} \\
\bottomrule
\end{tabular}

	\vspace{0.5em}
\parbox{\textwidth}{\scriptsize\textit{Notes:} Each $\chi^2$ statistic tests that all coefficients of the indicated sieve block are jointly zero. The last three Wald statistics test the interactions of both splines with the indicated covariate group (6, 96, and 36 basis coefficients, respectively, in the pooled sample). The linear probit includes age, age$^2$, GPA, and all discrete covariates; AIC is the Akaike information criterion.}
\end{table}

Table~\ref{tab:first_stage} reports joint Wald tests of the first-stage sieve blocks, each testing whether all coefficients in the indicated block are zero. The results provide strong evidence of nonlinearity in the selection index. In the pooled sample, the age and GPA spline terms are significant, as is their interaction, indicating that the relationship between GPA and employment varies with age. The age and GPA profiles also differ by gender, school type, school area, and major. For example, the age spline interactions with gender are jointly significant ($\chi^2 = 182.7$, $p < 0.001$). Across subperiods, the significance of individual blocks varies, but the interactions with major remain significant in all three periods, and the age-by-gender interactions are significant in 2008--2011 and 2012--2015 ($p = 0.066$ in 2016--2019). The pooled likelihood-ratio test rejects a probit specification with age, age squared, GPA, and discrete covariates in favor of the sieve specification ($\chi^2 = 1{,}511.9$, 164 degrees of freedom). The sieve specification also lowers the AIC despite its additional parameters. Both specifications retain the probit link, so the comparison shows that a quadratic age term alone does not adequately capture the nonlinearity in the selection index.

\begin{figure}[b!]
\centering
\begin{subfigure}[t]{0.48\textwidth}
\centering
\includegraphics[width=\textwidth]{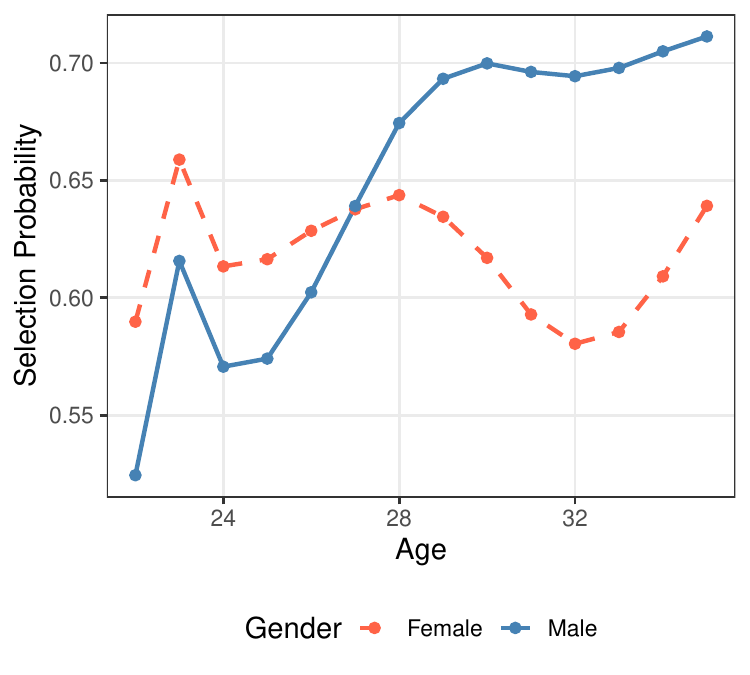}
\caption{Estimated selection probability by age}\label{fig:selection_prob}
\end{subfigure}
\hfill
\begin{subfigure}[t]{0.48\textwidth}
\centering
\includegraphics[width=\textwidth]{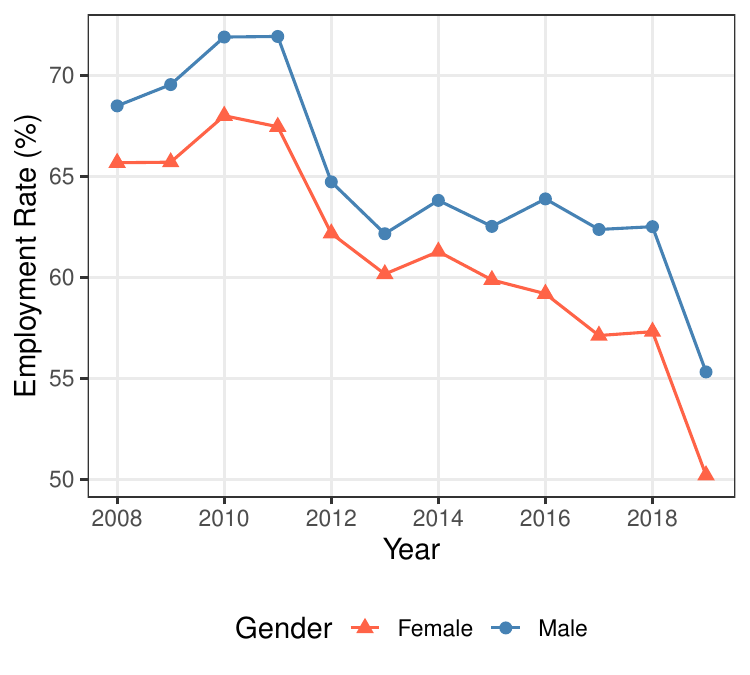}
\caption{Employment rates by year}\label{fig:selection_rates}
\end{subfigure}

\vspace{0.5em}
\caption{Selection into full-time wage employment in GOMS, 2008--2019. Panel (a) plots mean predicted probabilities by rounded age after setting the female indicator to zero or one for every observation while holding all other covariates fixed; panel (b) plots raw employment rates by gender and wave.}\label{fig:selection}
\end{figure}

Figure~\ref{fig:selection_prob} plots predicted employment probabilities by age and gender. For each graduate, I predict employment with the female indicator set to zero and then to one, holding all other covariates fixed, and average these predicted probabilities within age bins. The age profiles differ markedly. Men's employment probability rises sharply from the mid- to late twenties and remains relatively high thereafter. Women's probability rises between ages 24 and 28, then falls through their early thirties before recovering at ages 34--35. Women have a higher predicted probability through age 26, and men have a higher probability from age 27 onward. This crossover is consistent with men's later labor market entry due to mandatory military service. The flexible sieve specification captures this pattern as it allows the relationship between gender and employment to vary with age. Figure~\ref{fig:selection_rates} shows employment rates by gender and wave. Both rates are highest in 2010--2011, at about 72\% for men and 68\% for women, and fluctuate at lower levels thereafter before falling sharply in 2019 to about 55\% and 50\%, respectively. Men have a higher employment rate throughout the period.

\begin{table}[b!]
\centering
\caption{Anatomy of the Entry-Level Gender Wage Gap}\label{tab:anatomy}
\begin{tabular}{lrrrr}
\toprule
Specification & \multicolumn{1}{c}{Pooled} & \multicolumn{1}{c}{2008--2011} & \multicolumn{1}{c}{2012--2015} & \multicolumn{1}{c}{2016--2019} \\
\midrule
Raw gap (female + year dummies) & $-$0.125 & $-$0.146 & $-$0.126 & $-$0.100 \\
 & (0.002) & (0.004) & (0.004) & (0.003) \\
+ Education \& major & $-$0.119 & $-$0.140 & $-$0.115 & $-$0.098 \\
 & (0.002) & (0.004) & (0.004) & (0.003) \\
+ Job controls (firm, industry, region) & $-$0.095 & $-$0.114 & $-$0.096 & $-$0.072 \\
 & (0.002) & (0.004) & (0.004) & (0.003) \\
+ Demographics (age, age$^2$, GPA, married) & $-$0.033 & $-$0.034 & $-$0.037 & $-$0.025 \\
 & (0.003) & (0.005) & (0.005) & (0.004) \\
\bottomrule
\end{tabular}

\vspace{0.5em}
\parbox{\textwidth}{\scriptsize\textit{Notes:} OLS female coefficients on log hourly wages as controls are added cumulatively across rows. Heteroskedasticity-robust standard errors are in parentheses.}
\end{table}

\subsection{Main Estimates}\label{sec:main_estimates}

Employed male and female graduates differ in age, field of study, and job characteristics, so the raw wage gap conflates these differences. Before turning to the main estimates, I show how the OLS gap shrinks as these characteristics are controlled for. Table~\ref{tab:anatomy} adds the controls in three blocks. With only year effects, the gap is about 12\%. Adding education and major controls reduces it only slightly, to 11.9\%. Men are more likely to have studied engineering, but women are more likely to have studied education and medical/health fields, which also pay relatively well. These differences largely offset one another. Controlling for firm size, public-sector employment, industry, and workplace region reduces the gap further to 9.5\%. The largest reduction comes from adding age, age squared, GPA, and marital status, which brings the gap down to 3.3\%. Age is particularly important: employed men are on average 2.3 years older than employed women, reflecting mandatory military service, and wages are positively associated with age for both genders.

\begin{table}[b!]
\centering
\caption{Gender Wage Gap Estimates: Log Hourly Wage}\label{tab:main}

\begin{tabular}{lcccc}
\toprule
 & Pooled & 2008--2011 & 2012--2015 & 2016--2019 \\
\midrule
OLS & $-$0.033 & $-$0.034 & $-$0.037 & $-$0.025 \\
 & (0.003) & (0.005) & (0.005) & (0.004) \\
Heckman & $-$0.028 & $-$0.034 & $-$0.037 & $-$0.024 \\
 & (0.003) & (0.005) & (0.005) & (0.004) \\
Kim-Lee & $-$0.033 & $-$0.031 & $-$0.043 & $-$0.024 \\
 & (0.003) & (0.005) & (0.005) & (0.004) \\
DML-RF & $-$0.032 & $-$0.032 & $-$0.038 & $-$0.022 \\
 & (0.003) & (0.005) & (0.005) & (0.004) \\
\midrule
HH 95\% CI & [$-$0.043, $-$0.022] & [$-$0.049, $-$0.019] & [$-$0.064, $-$0.015] & [$-$0.035, $-$0.014] \\
Lee bounds & [$-$0.080, $+$0.002] & [$-$0.085, $+$0.015] & [$-$0.073, $-$0.015] & [$-$0.067, $+$0.007] \\
\midrule
Employed $N$ & 127,628 & 45,406 & 41,640 & 40,582 \\
\bottomrule
\end{tabular}

\vspace{0.5em}
\parbox{\textwidth}{\scriptsize\textit{Notes:} Female coefficients with the corresponding heteroskedasticity-robust standard errors in parentheses, except for Heckman, whose maximum likelihood standard errors do not allow for heteroskedasticity. Wage covariates: age, age$^2$, GPA, marriage indicator, 4-year university, raw school-area dummies, major category, large firm, public sector, industry, raw workplace-region dummies, and year dummies. ``Lee bounds'' are the age-conditional \cite{lee2009} bounds and ``HH 95\% CI'' is the \cite{honore2020selection} confidence interval described in the text.}
\end{table}

Selection correction has little effect on the estimated conditional gap. Table~\ref{tab:main} reports female coefficients from OLS, Heckman MLE (with age squared included in both the selection and wage equations), Kim-Lee (KL), and DML-RF for the pooled 2008--2019 sample and three subperiods, together with age-conditional Lee bounds and Honor\'e-Hu confidence intervals. All four estimators yield negative and statistically significant coefficients in every period. In the pooled sample, OLS, Kim-Lee, and DML-RF place the conditional gap at about 3.3 log points, roughly one-quarter of the raw gap. The Kim-Lee and DML-RF estimates remain within 0.6 log points of OLS in every period, suggesting that employment selection has little effect on the estimated gender wage gap among recent graduates. Heckman produces a somewhat smaller pooled gap of 2.8 log points, compared with 3.3 under OLS, but its subperiod estimates are close to OLS.

\begin{figure}[t!]
	\centering
	\includegraphics[width=0.85\textwidth]{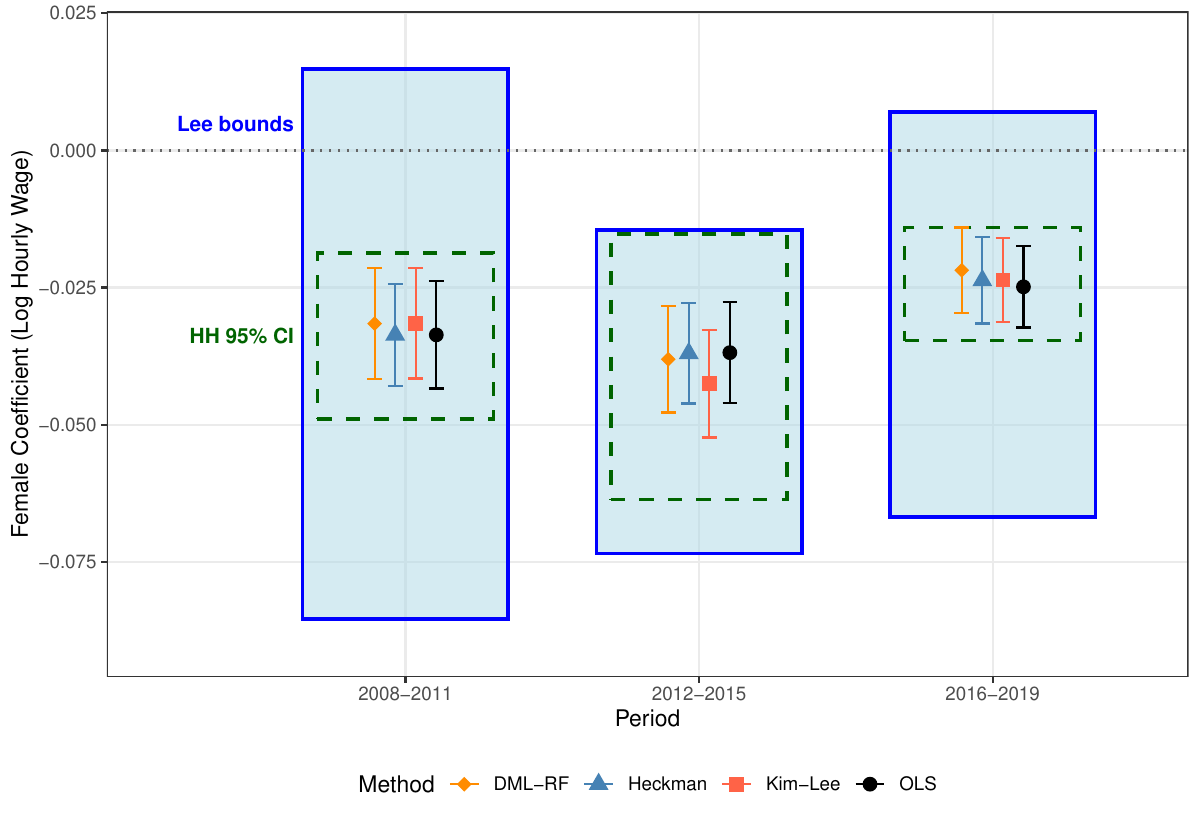}
	\caption{Gender wage gap estimates by method and subperiod. Point estimates with 95\% confidence intervals based on the standard errors in Table~\ref{tab:main}. The blue boxes are the age-conditional Lee bounds and the dashed green boxes the Honor\'e-Hu 95\% confidence intervals.}\label{fig:comparison}
\end{figure}

The unconditional \cite{lee2009} bounds assume that gender shifts employment in the same direction for every graduate, which does not hold here, as shown in Figure~\ref{fig:selection_prob}. I therefore use \emph{age-conditional} Lee bounds, allowing the direction of selection to differ across age bins while maintaining a common direction within each bin.\footnote{Log wages are residualized on all controls except gender and age, the trimming is applied within five age bins (19--24, 25--26, 27--28, 29--31, and 32--35), and the bin-specific bounds are integrated over the age distribution. Appendix~\ref{sec:appendix_semenova} reports the unconditional Lee bounds and generalized Lee bounds \citep{semenova2023generalized} in which the direction of selection varies with all pre-market covariates.} The estimated interval excludes zero only in 2012--2015, $[-0.073, -0.015]$. The intervals for the pooled sample, $[-0.080, 0.002]$, for 2008--2011, $[-0.085, 0.015]$, and for 2016--2019, $[-0.067, 0.007]$, include zero. Conditioning on both age and major gives a wider pooled interval, $[-0.096, -0.010]$, which excludes zero. The Kim-Lee and DML-RF estimates fall within the age-conditional intervals in every period. 

The \cite{honore2020selection} (HH) bounds impose linear-index structures on the selection and wage equations, with errors independent of the covariates. In each of the three subperiods, no parameter value satisfies all of the sample moment inequalities, so the sample analogue of the identified set is empty, and in the pooled sample it is $[-0.032, -0.031]$. \cite{honore2020selection} obtain an estimated set by allowing each inequality to be violated by up to a tolerance they arbitrarily choose. I instead use their subsampling inference method and report 95\% confidence intervals.\footnote{Each step uses 1,000 subsamples of one-eighth of the sample, close to HH's original implementation (15,000 out of about 120,000).} The HH 95\% confidence interval of the female coefficient is non-empty in every sample, so the HH inequalities are not rejected. It contains both the OLS and the Kim-Lee estimates and excludes zero in every sample. Figure~\ref{fig:comparison} plots the point estimates, the Lee bounds, and the HH confidence intervals by subperiod.

\begin{figure}[h!]
\centering
\includegraphics[width=\textwidth]{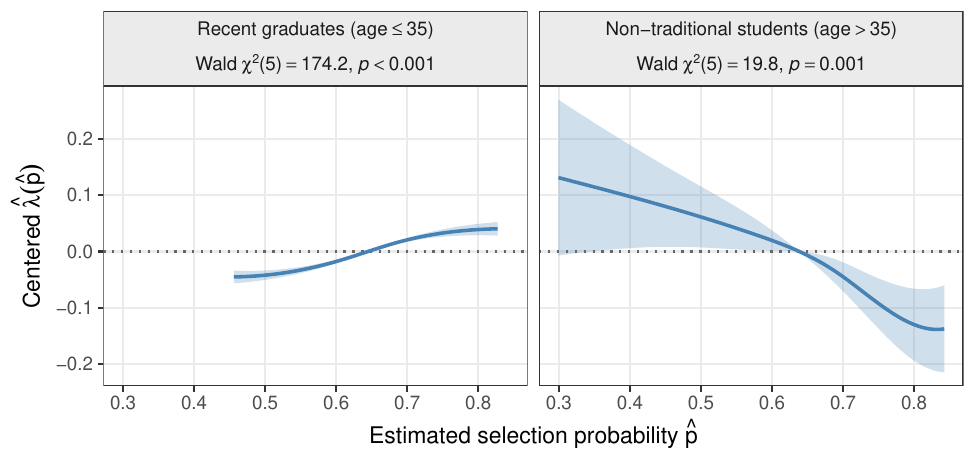}
\caption{Estimated selection correction function $\hat{\lambda}(\hat{p})$ from the pooled (2008--2019) Kim-Lee estimator, normalized to zero at the median $\hat{p}$ of employed graduates, with 95\% pointwise confidence bands (shaded). The bands and the joint Wald test of no selection bias in the panel headers use the heteroskedasticity-robust second-stage covariance, without the first-stage correction.}\label{fig:lambda}
\end{figure}

Although selection correction is negligible for the female coefficient, it changes several other wage coefficients (Appendix Table~\ref{tab:full_coefs}), reducing the medical/health-major coefficient from 0.143 to 0.110 and the coefficient on the linear age term from 0.069 to 0.054. The left panel of Figure~\ref{fig:lambda} plots the estimated selection correction function $\hat{\lambda}(\hat{p})$ with 95\% pointwise confidence bands. The estimated correction function varies by about 9 log points over the plotted range, and a joint Wald test strongly rejects the null hypothesis that it is constant ($p < 0.001$). The right panel shows the estimated correction function for graduates older than 35, who are excluded from the analysis sample. The direction of selection correction is opposite to that of the analysis sample. The entry-level correction is about a third as large in terms of magnitude but far more precisely estimated. 

\begin{figure}[t!]
	\centering
	\includegraphics[width=0.82\textwidth]{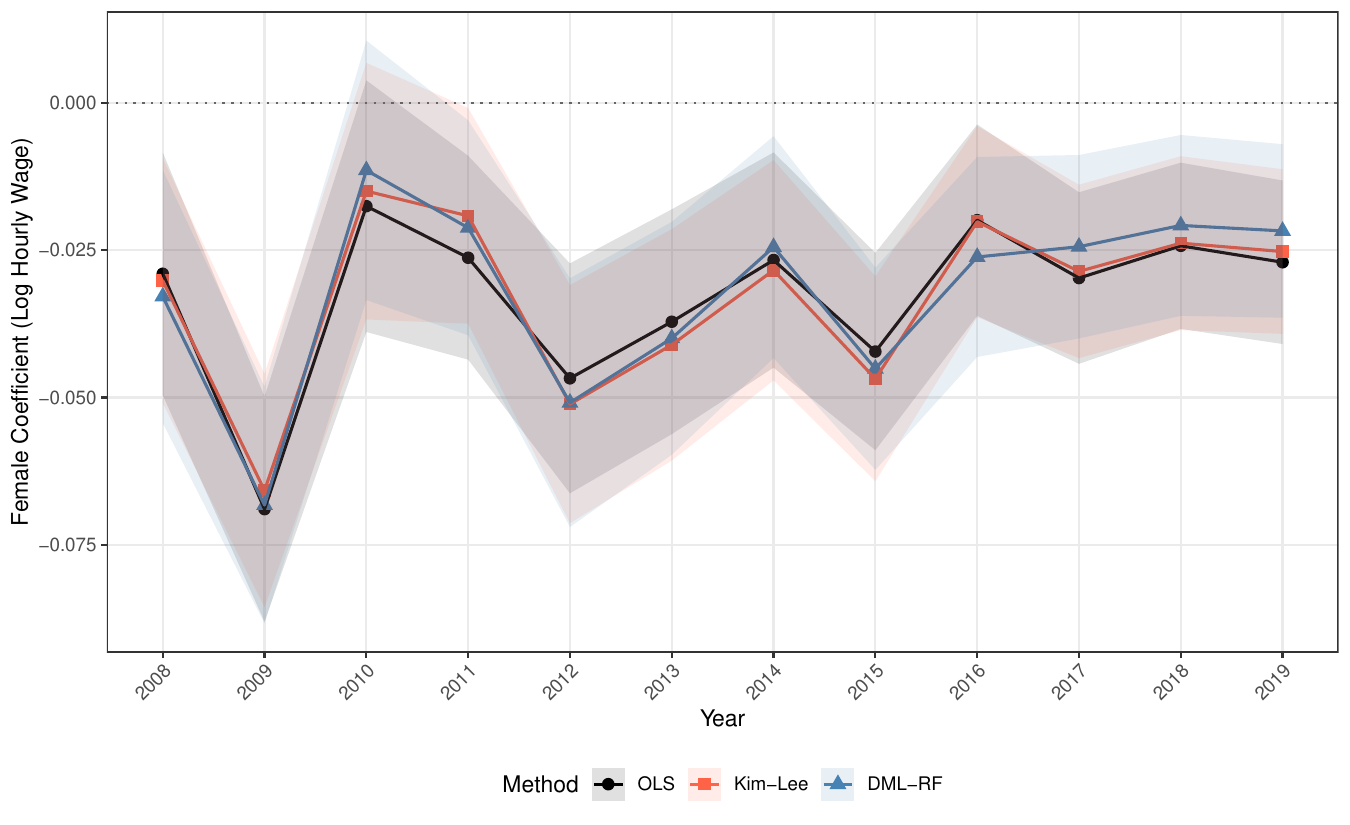}
	\caption{Year-by-year estimates of the female coefficient on log hourly wages for the entry-level sample, by wave. Shaded bands denote 95\% confidence intervals.}\label{fig:trend}
\end{figure}

I also estimate each model separately for each wave, allowing the selection and wage coefficients to change annually. Figure~\ref{fig:trend} plots the OLS, Kim-Lee, and DML-RF estimates, and Appendix Table~\ref{tab:yearly} reports the annual estimates and standard errors for all four methods. The yearly estimates fluctuate considerably and do not decline steadily. All four methods give the largest gaps in 2009, 2012, 2013, and 2015, while their estimates for 2008, 2010, 2011, and 2014 are similar to those for 2016--2019. The Kim-Lee estimate is 6.6 log points in 2009 and between 2.0 and 2.9 log points in each year from 2016 to 2019. The Kim-Lee and DML-RF estimates remain close to OLS throughout.

\subsection{Employment Decline and the Composition of the Employed}\label{sec:emp_decline}

The conditional gap narrowed as graduate employment fell sharply over the sample period. How much of this change reflects changes in who is employed, rather than changes in relative wages among otherwise comparable graduates? As Table~\ref{tab:emp_activity} shows, the share of graduates in full-time wage work fell by 9 percentage points for men and 11 for women between 2008--2011 and 2016--2019. From the 2010 peak to 2019, the overall decline was 17.4 percentage points. Among non-working graduates, job search became more common for both men and women, while the share pursuing further education fell.

\begin{table}[ht!]
\centering
\caption{Employment Rates and Non-Employment Activities by Gender}\label{tab:emp_activity}
\begin{tabular}{lcccccc}
\toprule
 & \multicolumn{2}{c}{Employment rate (\%)} & \multicolumn{4}{c}{Non-working graduates (\%)} \\
\cmidrule(lr){2-3} \cmidrule(lr){4-7}
Period & Male & Female & \multicolumn{2}{c}{Job searching} & \multicolumn{2}{c}{Further education} \\
 & & & Male & Female & Male & Female \\
\midrule
2008--2011 & 70.4 & 66.8 & 47.7 & 44.2 & 41.3 & 35.1 \\
2012--2015 & 63.3 & 60.9 & 48.2 & 48.1 & 40.1 & 33.3 \\
2016--2019 & 61.2 & 55.9 & 56.1 & 58.0 & 32.1 & 25.4 \\
\bottomrule
\end{tabular}

\vspace{0.2em}
\parbox{\textwidth}{\scriptsize\textit{Notes:} Job searching includes employment preparation and training, and further education includes university or graduate-school enrollment and preparation for further study.}
\end{table}

The gender gap need not be the same across ages, majors, or jobs, so changes in the composition of employed graduates can affect the estimate. Using the reweighting method of \cite{dinardo1996labor}, I give the 2016--2019 employed sample the covariate distribution of the 2008--2011 sample while holding its wage structure fixed. Table~\ref{tab:trend_decomp} attributes 0.011 (SE 0.003) to compositional change, more than the total change of 0.009 (SE 0.006), and $-0.002$ (SE 0.007) to the wage structure. The composition component is precisely estimated, whereas the total change and the wage-structure component are not statistically distinguishable from zero. At the 2008--2011 composition, the 2016--2019 gap would be 3.5 log points, so at a fixed composition the gap did not narrow. Reweighting one covariate group at a time, I find that individual characteristics (age, GPA, education, marriage, and school area) account for 0.006 (SE 0.002) and field of study accounts for 0.0029 (SE 0.0005). Job characteristics contribute little.

\begin{table}[htbp]
	\centering
	\caption{Decomposition of the Change in the Conditional Gender Gap}\label{tab:trend_decomp}
		\begin{tabular}{lc}
		\toprule
		& Female coefficient \\
		\midrule
		2008--2011 (OLS) & $-$0.034 (0.005) \\
		2016--2019 (OLS) & $-$0.025 (0.004) \\
		2016--2019 reweighted to 2008--2011 composition & $-$0.035 (0.005) \\
		\midrule
		Total change & $+$0.009 (0.006) \\
		\quad Composition & $+$0.011 (0.003) \\
		\quad Wage structure & $-$0.002 (0.007) \\
		\bottomrule
	\end{tabular}

\vspace{0.2em}
\parbox{\textwidth}{\scriptsize\textit{Notes:} Weights are constructed from a logit model predicting whether an employed graduate belongs to the 2016--2019 or 2008--2011 sample, using all wage covariates except gender and year dummies. Standard errors in parentheses are from a bootstrap (400 replications) that resamples employed graduates within each subperiod and re-estimates the weights.}
\end{table}

\begin{figure}[h!]
\centering
\includegraphics[width=0.85\textwidth]{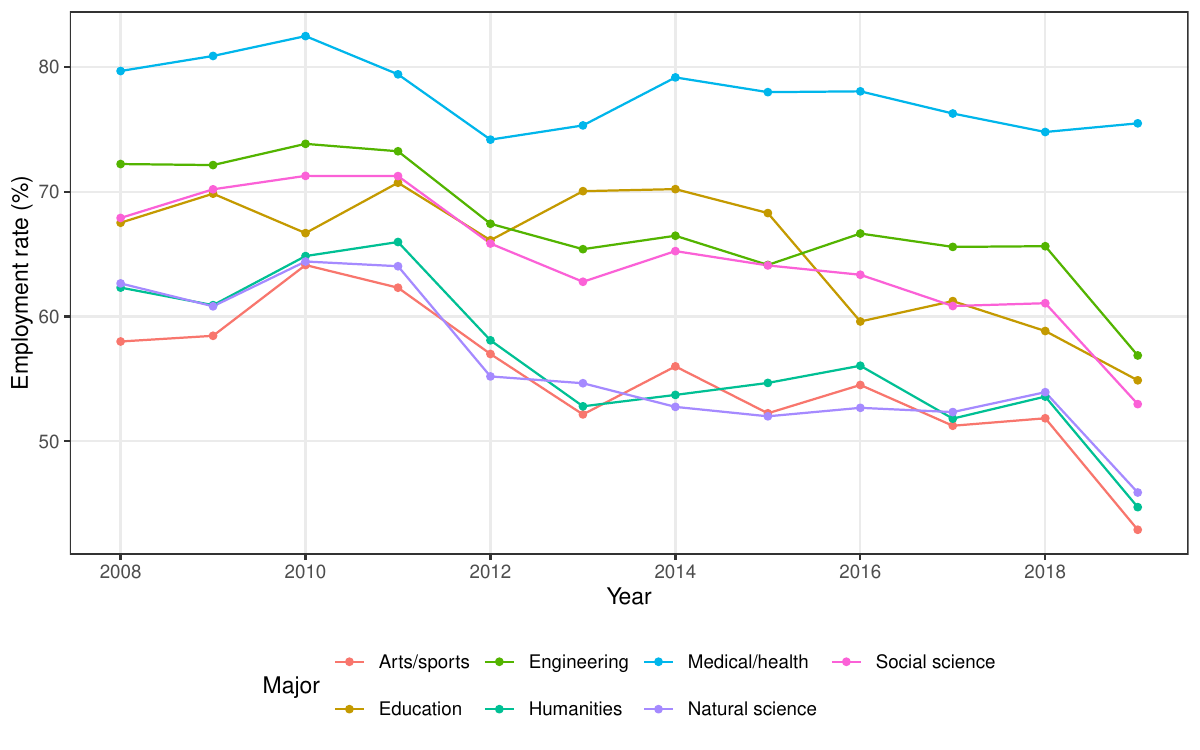}
\caption{Employment rate by major category, 2008--2019. Each line plots the annual share of graduates in full-time wage employment with valid wage data, pooling men and women within each major category.}\label{fig:emp_major}
\end{figure}

Employment rates fell in most fields, as shown in Figure~\ref{fig:emp_major}. Rates among medical/health graduates stayed between 74\% and 82\%, but in other fields they fell by 16 to 21 percentage points from their early-2010s peaks to 2019. Social science graduates accounted for 20.2\% of employed graduates in 2016--2019, down from 24.6\% in 2008--2011. This field had the largest conditional gender gap in the later period, at 4.6 log points (Appendix Table~\ref{tab:major_gap_composition} reports the conditional gender wage gap in each field of study). The smaller share of social science graduates and larger share of medical/health graduates, whose estimated gap was close to zero, would both tend to reduce the overall gender wage gap. But workers also became more concentrated in engineering, where the conditional gap was 2.9 log points, and less concentrated in education, where the gap was small. These changes partly offset one another, so field of study accounts for about a quarter of the composition component.

\subsection{Decomposition of the Gender Wage Gap}\label{sec:oaxaca}

To examine how much of the raw gender wage gap is explained by differences in observed characteristics, I apply the \cite{oaxaca1973}--\cite{blinder1973} decomposition. I estimate separate semiparametric wage equations for men and women and use the pooled Kim-Lee coefficients as the reference, following \cite{oaxaca1994identification}. The endowments component measures differences in observed characteristics between men and women, evaluated at these reference coefficients. The residual is the part of the raw gap not explained by these differences.\footnote{The intercept is not separately identified from the level of the selection correction, so the coefficients and selection components cannot be separately interpreted. I therefore report their combined residual, calculated as the raw gap minus the endowments component.}

Table~\ref{tab:oaxaca} shows that observed characteristics explain nearly two-thirds of the raw wage gap. The explained share is 71\% in 2008--2011 and 66\% in 2016--2019. Between these periods the raw gap falls from 0.143 to 0.099, mostly in the endowments component (0.102 to 0.065), while the residual component falls from 0.041 to 0.034. Age makes the largest contribution: 0.058 in the pooled sample, or 48\% of the raw gap. Male graduates are on average 2 years older because of mandatory military service, and wages rise with age for both genders. In the separate Kim-Lee wage equations, the marginal effect of age at each gender's mean age is 0.021 for men and 0.022 for women in the pooled sample and is positive for both genders in every subperiod. The age contribution therefore reflects the older age of male graduates combined with a positive age gradient found among women as well as men. Differences in firm size, public-sector employment, industry, and workplace region contribute another 0.042. These contributions are partly offset by women's higher GPAs ($-0.006$) and differences in education and major ($-0.015$), including employed women's greater representation in the relatively well-paid education and medical/health fields.

\begin{table}[t!]
\centering
\caption{Oaxaca-Blinder Decomposition with Kim-Lee Selection Correction}\label{tab:oaxaca}
\begin{tabular}{lcccc}
\toprule
 & Pooled & 2008--2011 & 2012--2015 & 2016--2019 \\
\midrule
Raw gap ($\bar{Y}_M - \bar{Y}_F$) & 0.122 & 0.143 & 0.126 & 0.099 \\
\midrule
Endowments (characteristics) & 0.077 & 0.102 & 0.075 & 0.065 \\
 & [63.3\%] & [71.0\%] & [59.7\%] & [65.8\%] \\
\quad Age (incl.\ age$^2$) & 0.058 & 0.082 & 0.054 & 0.043 \\
\quad GPA & $-$0.006 & $-$0.008 & $-$0.007 & $-$0.004 \\
\quad Married & 0.002 & 0.003 & 0.002 & 0.001 \\
\quad Education \& major & $-$0.015 & $-$0.016 & $-$0.011 & $-$0.017 \\
\quad Job characteristics & 0.042 & 0.043 & 0.037 & 0.043 \\
\quad Year & $-$0.004 & $-$0.003 & 0.000 & $-$0.001 \\
\midrule
Residual (unexplained + selection) & 0.045 & 0.041 & 0.051 & 0.034 \\
 & [36.7\%] & [29.0\%] & [40.3\%] & [34.2\%] \\
\bottomrule
\end{tabular}

\vspace{0.5em}
\parbox{\textwidth}{\scriptsize\textit{Notes:} Reference coefficients are the pooled Kim-Lee estimates. The residual component is the raw gap minus the endowments component. Percentages in brackets report the endowments and residual components as shares of the raw gap.}
\end{table}

The results are similar when the sample is restricted to graduates aged 30 or younger or to four-year university graduates (Appendix Table~\ref{tab:subsample_oaxaca}). In both samples, observed characteristics explain about 60\% of the pooled raw gap, with age and job characteristics making the largest contributions. Among graduates older than 35, however, observed characteristics explain only 33\% of a much larger raw gap of 47 log points. The age component is slightly negative in this sample, where women are slightly older than men.

\subsection{Heterogeneity across the Wage Distribution}\label{sec:quantile}

\begin{figure}[ht]
\centering
\includegraphics[width=\textwidth]{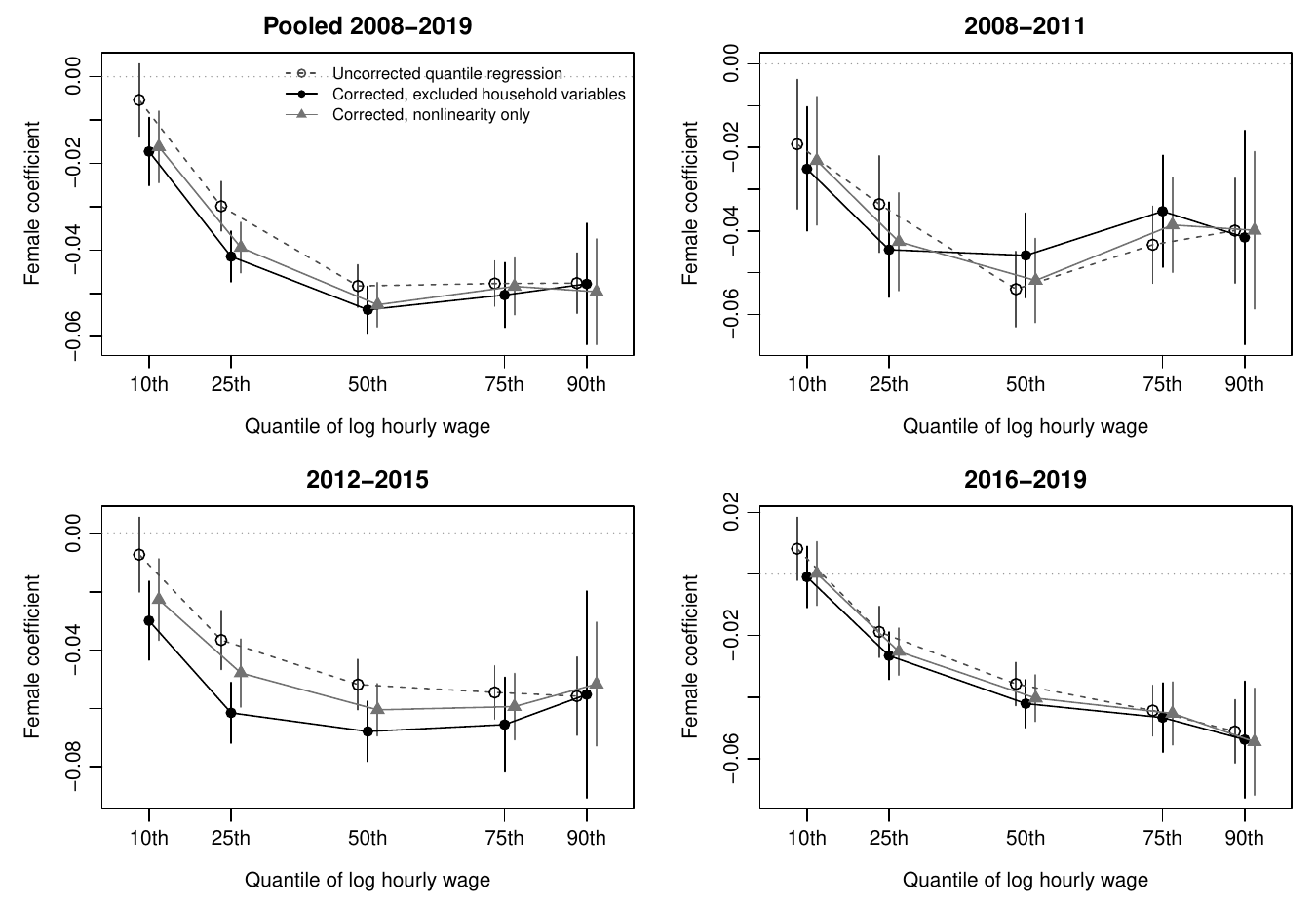}
\caption{Selection-corrected female coefficient by quantile, GOMS. Corrected estimates are obtained by the \cite{arellano2017quantile} estimator. Vertical bars are 95\% confidence intervals, from a bootstrap over graduates (200 replications) for the corrected estimates and from heteroskedasticity-robust standard errors for uncorrected quantile regression.}\label{fig:quantile}
\end{figure}

I use the selection-corrected quantile estimator of \cite{arellano2017quantile} to examine whether the entry-level gender gap is larger at the top or bottom of the wage distribution. The model allows wage coefficients to differ across quantiles and uses a Frank copula to model dependence between the wage rank and the selection error. Figure~\ref{fig:quantile} reports the female coefficient at the 10th, 25th, 50th, 75th, and 90th percentiles for the pooled sample and each subperiod. I use co-residence with parents and household size as excluded variables that enter only the selection equation.\footnote{In the Arellano-Bonhomme quantile selection model, using only the family-background variables places the estimated copula parameter at or near the boundary of the search range (Appendix~\ref{sec:appendix_iv}).} For comparison, I also estimate the model without excluded variables, relying instead on the nonlinearity of the sieve selection probability (as in the Kim-Lee estimator), and report uncorrected quantile regression estimates \citep{koenker1978regression}. All specifications include the full set of wage covariates. 

With the excluded household variables, the selection-corrected gap in the pooled sample is 1.7 log points at the 10th percentile and about 5 log points at the median and above. The gender gap varies most across quantiles in 2016--2019. The female coefficient at the 10th percentile is close to zero and statistically insignificant ($-0.001$, SE $=0.005$), whereas the coefficient at the 90th percentile remains $-0.054$ (SE $=0.010$). Over the sample period, the gap falls at the lower quantiles but increases at the upper quantiles. The small change in the mean gap therefore conceals opposite changes at the two ends of the distribution.

The estimates with excluded household variables and those relying on nonlinearity differ by less than one log point at every reported quantile in the pooled sample, 2008--2011, and 2016--2019. The differences are larger in 2012--2015, reaching 1.4 log points at the 25th percentile. Relative to uncorrected quantile regression, the median adjustments are below two log points in every period (Appendix Table~\ref{tab:iv_goms_q}). The largest adjustment across the reported quantiles is 2.5 log points at the 25th percentile in 2012--2015. Both selection-corrected specifications and uncorrected quantile regression show a small gap at the bottom and a substantial gap at the top in 2016--2019.

Recentered influence function (RIF) regressions \citep{firpo2009unconditional} also show a larger gender gap at the top of the unconditional wage distribution among observed wage workers. The pooled female effect is $+0.015$ at the 10th percentile, $-0.035$ at the median, and $-0.079$ at the 90th. In 2016--2019 it is $+0.030$ at the 10th percentile and $-0.073$ at the 90th. Distribution regressions on a nominal wage grid \citep{chernozhukov2013inference,biewen2026distribution} show a similar pattern in the pooled sample ($+0.008$ at the 10th percentile and $-0.075$ at the 90th) and in 2016--2019 ($+0.037$ and $-0.055$). Appendix~\ref{sec:appendix_dr} reports these estimates, differences between the observed and counterfactual wage distributions, and sensitivity to the RIF bandwidth. These unconditional estimates are not selection-corrected.

\FloatBarrier
\subsection{Robustness Analyses}\label{sec:robustness}

I examine how sensitive the estimated gender gap is to the assumptions used to account for selection into employment. I include standard excluded variables used in the literature and allow the selection correction to differ across observed groups, and then explore alternative wage measures, job controls, sample definitions, and first-stage specifications of the Heckman selection model.

\textit{Exclusion restrictions.}\phantomsection\label{sec:instruments} The baseline selection-corrected estimates in Table~\ref{tab:main} do not rely on exclusion restrictions. I further examine whether adding excluded variables commonly used in the literature, such as household resources and composition \citep{arellano2017quantile,biewen2020counterfactual}, alters the gender gap estimates. GOMS provides parental income and education, co-residence with parents, household size, spouse income, and dependent children (Appendix~\ref{sec:appendix_iv}). The exclusion restriction assumes that, conditional on the wage covariates, these variables affect employment through the reservation wage but do not directly affect wages. Family background may nevertheless capture ability and networks, while household structure may itself depend on employment. I therefore compare estimates across different sets of excluded variables. Both sets of excluded variables are jointly significant in the Kim-Lee sieve probit first stage ($p<0.001$), with household size accounting for most of the explanatory power of the household-structure variables (Appendix Tables~\ref{tab:iv_firststage} and~\ref{tab:iv_wald}).

Table~\ref{tab:iv_goms} reports Kim-Lee estimates and Heckman maximum likelihood estimates with the excluded variables added to the first stage. With co-residence and household size as excluded variables, the pooled estimates are $-0.034$ for Kim-Lee and $-0.034$ for Heckman, compared with $-0.033$ for Kim-Lee without exclusions. Both remain within half a log point of the corresponding baseline Kim-Lee estimate in every period. The results are similar with the full set of candidate excluded variables. With the family-background variables as the only excluded variables, the Kim-Lee estimates stay within half a log point of their no-exclusion counterparts on the same observations, whereas the Heckman estimates are $-0.026$ in the pooled sample and $-0.021$ in 2012--2015. The Heckman estimates are therefore more sensitive to the choice of excluded variables than the Kim-Lee estimates.

\begin{table}[tp]
\centering
\caption{Exclusion-Based Estimates of the Entry-Level Gender Gap, GOMS}\label{tab:iv_goms}
{\footnotesize\setlength{\tabcolsep}{4pt}
\begin{tabular}{lrrrr}
\toprule
 & \multicolumn{1}{c}{Pooled} & \multicolumn{1}{c}{2008--2011} & \multicolumn{1}{c}{2012--2015} & \multicolumn{1}{c}{2016--2019} \\
\midrule
\multicolumn{5}{l}{\textit{Panel A: Kim-Lee with excluded variables}} \\
Household structure & $-$0.034 & $-$0.032 & $-$0.045 & $-$0.024 \\
 & (0.003) & (0.005) & (0.005) & (0.004) \\
Family background & $-$0.033 & $-$0.033 & $-$0.038 & $-$0.026 \\
 & (0.003) & (0.005) & (0.005) & (0.004) \\
All candidate excluded variables & $-$0.034 & $-$0.033 & $-$0.041 & $-$0.026 \\
 & (0.003) & (0.005) & (0.005) & (0.004) \\
\midrule
\multicolumn{5}{l}{\textit{Panel B: Heckman maximum likelihood with excluded variables}} \\
Household structure & $-$0.034 & $-$0.033 & $-$0.040 & $-$0.025 \\
 & (0.003) & (0.005) & (0.005) & (0.004) \\
Family background & $-$0.026 & $-$0.033 & $-$0.021 & $-$0.026 \\
 & (0.003) & (0.005) & (0.005) & (0.004) \\
All candidate excluded variables & $-$0.034 & $-$0.033 & $-$0.038 & $-$0.026 \\
 & (0.003) & (0.005) & (0.005) & (0.004) \\
\bottomrule
\end{tabular}
}

\vspace{0.5em}
\parbox{\textwidth}{\scriptsize\textit{Notes:} Excluded variables enter the selection equation only: ``household structure'' is co-residence with parents and household size; ``family background'' is parental income at admission and at the survey, an indicator for deceased parents, and father's and mother's education; ``all candidate excluded variables'' combines the two and adds spouse income and the number of dependent children. First-stage-corrected standard errors for Kim-Lee and MLE standard errors for Heckman estimates are in parentheses.}
\end{table}

\textit{Heterogeneous selection.}\phantomsection\label{sec:hetero_selection} The control-function restriction in Assumption~\ref{ass:id}(iv) requires workers with the same selection probability to share the same selection correction in the conditional mean. The test of \cite{huber2015test} examines the stronger restriction that the wage error is independent of the covariates conditional on the selection probability, which also rules out differences in wage-error variance at a given selection probability. The test compares the covariate slopes across quantile regressions of log wages that also control for a flexible function of $\hat{p}$. Table~\ref{tab:hm} reports rejections of equal slopes across deciles in the pooled sample, both jointly and for the female coefficient alone ($p<0.01$), as found in \cite{biewen2020counterfactual} for Germany. This rejection does not by itself rule out the control-function restriction. As Equation~(\ref{eq:hetero}) shows, wage dispersion can vary across workers even when their selection correction is the same.

\begin{table}[t!]
\centering
\caption{Huber-Melly Test of Conditional Independence}\label{tab:hm}
{\footnotesize\setlength{\tabcolsep}{4pt}
\begin{tabular}{lrrrr}
\toprule
 & \multicolumn{1}{c}{Pooled} & \multicolumn{1}{c}{2008--2011} & \multicolumn{1}{c}{2012--2015} & \multicolumn{1}{c}{2016--2019} \\
\midrule
\multicolumn{5}{l}{\textit{All slopes}} \\
Kolmogorov-Smirnov & 16.60 & 11.13 & 9.69 & 8.81 \\
 & [$<$0.01] & [$<$0.01] & [$<$0.01] & [$<$0.01] \\
Cram\'er-von Mises & 9,022 & 2,270 & 2,106 & 2,716 \\
 & [$<$0.01] & [$<$0.01] & [$<$0.01] & [$<$0.01] \\
\multicolumn{5}{l}{\textit{Female coefficient}} \\
Kolmogorov-Smirnov & 9.55 & 4.67 & 5.44 & 6.98 \\
 & [$<$0.01] & [$<$0.01] & [$<$0.01] & [$<$0.01] \\
Cram\'er-von Mises & 226 & 58 & 51 & 127 \\
 & [$<$0.01] & [$<$0.01] & [$<$0.01] & [$<$0.01] \\
\multicolumn{5}{l}{\textit{Female coefficient by decile, given $\hat{p}$}} \\
First decile & $-$0.007 & $-$0.014 & $-$0.016 & $+$0.007 \\
Median & $-$0.050 & $-$0.053 & $-$0.059 & $-$0.035 \\
Ninth decile & $-$0.047 & $-$0.039 & $-$0.057 & $-$0.050 \\
\bottomrule
\end{tabular}
}

\vspace{0.5em}
\parbox{\textwidth}{\scriptsize\textit{Notes:} Quantile regressions of log hourly wages on the full wage controls and the cubic B-spline in the baseline $\hat{p}$ at the deciles, on the employed sample. The statistics are the maximum absolute value (Kolmogorov-Smirnov) and the sum of squares (Cram\'er-von Mises) of the slope differences relative to the median, standardized by their bootstrap standard errors, over all slope-decile pairs. $p$-values in brackets from 199 replications of a bootstrap over graduates that re-estimates the sieve probit in each replication and recenters the process at the sample estimates.}
\end{table}

To test the common selection correction directly, I interact the second-stage spline in $\hat{p}$ with school type, marital status, and major. The joint test of the 40 interaction terms strongly rejects a common correction ($p<0.001$ in all four samples). I estimate a single wage regression with a common female coefficient, allowing the selection correction to vary additively across the covariate cells. The correction remains common to men and women with the same characteristics, preserving the within-cell identifying argument of Proposition~\ref{prop:id}. Table~\ref{tab:cellcf} reports the estimates with standard errors. Allowing the correction to differ across these groups changes the female coefficient by at most half a log point in every period and in the age and school-type subsamples.

\begin{table}[t!]
\centering
\caption{Kim-Lee Estimates of Female Coefficients with Cell-Specific Selection Corrections}\label{tab:cellcf}
{\footnotesize\setlength{\tabcolsep}{4pt}
\begin{tabular}{lrrrrrr}
\toprule
 & \multicolumn{1}{c}{Pooled} & \multicolumn{1}{c}{2008--2011} & \multicolumn{1}{c}{2012--2015} & \multicolumn{1}{c}{2016--2019} & \multicolumn{1}{c}{Age $\le 30$} & \multicolumn{1}{c}{Four-year} \\
\midrule
Single correction (Table~\ref{tab:main}) & $-$0.033 & $-$0.031 & $-$0.043 & $-$0.024 & $-$0.033 & $-$0.030 \\
 & (0.003) & (0.005) & (0.005) & (0.004) & (0.003) & (0.003) \\
Cell-specific corrections & $-$0.028 & $-$0.026 & $-$0.046 & $-$0.021 & $-$0.029 & $-$0.028 \\
 & (0.003) & (0.006) & (0.005) & (0.004) & (0.003) & (0.003) \\
Wald test of the interaction terms & 496 & 136 & 103 & 92 & 458 & 104 \\
\quad restrictions & 40 & 40 & 40 & 40 & 40 & 35 \\
Employed $N$ & 127,628 & 45,406 & 41,640 & 40,582 & 121,854 & 94,806 \\
\bottomrule
\end{tabular}
}

\vspace{0.5em}
\parbox{\textwidth}{\scriptsize\textit{Notes:} ``Cell-specific corrections'' interact the cubic B-spline in $\hat{p}$ (5 degrees of freedom) with the four-year, married, and major indicators (married and major only in the four-year subsample). First-stage-corrected robust standard errors in parentheses; the Wald test uses the same corrected covariance.}
\end{table}

\textit{Alternative dependent variable.} Using log monthly wages and controlling separately for regular and overtime hours gives a larger pooled Kim-Lee gap of 6.5 log points, compared with 3.3 for hourly wages. Men report more overtime, but monthly earnings rise less than proportionally with hours worked, so dividing earnings by total hours reduces men's relative wages. Both specifications show a smaller gender gap in 2016--2019 than in 2008--2011, and OLS and Kim-Lee estimates remain close in every period (Appendix Table~\ref{tab:lwage_dynamics}).

\textit{Job sorting.} Omitting controls for firm size, public-sector status, industry, and workplace region increases the pooled Kim-Lee gap from 3.3 to 5.1 log points. OLS, Kim-Lee, and DML-RF \citep{pan2024} give similar estimates with or without job controls (Appendix Table~\ref{tab:nojob}), so the small impact of selection correction does not depend on conditioning on job characteristics.

\textit{Sample definition.} I vary the age cutoff for the analysis sample to investigate the sensitivity of the gender gap estimate. When mature non-traditional students (age $> 35$) are included, the OLS and Kim-Lee estimates move to $-0.055$ and $-0.050$, reflecting the much larger wage gaps among mid-career graduates (Appendix Table~\ref{tab:age_sensitivity}). Analyzed separately, the age $> 35$ subsample shows a far larger conditional gap (25 to 37 log points) with qualitatively different selection behavior (Appendix Table~\ref{tab:subsample_robust}). Restricting the sample to four-year university graduates leaves the OLS and Kim-Lee estimates within 0.6 log points of each other in all periods.

\textit{Heckman with a nonlinear first stage.} The baseline Heckman specification in Table~\ref{tab:main} already includes age squared in both the selection and wage equations. I add GPA squared and age $\times$ GPA to the selection equation, along with interactions of the female indicator with age, age squared, GPA, GPA squared, and age $\times$ GPA. These additional terms allow the age and GPA profiles of employment to differ between men and women, while leaving the wage equation and the joint normality assumption unchanged. The pooled maximum likelihood estimate of the female coefficient moves from $-0.028$ to $-0.033$, close to Kim-Lee's $-0.033$. The estimates are within 0.6 log points of Kim-Lee's in every period (Appendix Table~\ref{tab:heckman_fs}).

\section{The Gender Gap over the Lifecycle: Evidence from KLIPS}\label{sec:klips}

Given the small entry-level gap, the 29\% aggregate gap suggests that most of the gender wage inequality is likely generated after labor market entry. This section examines how the gender gap and the effect of selection correction vary with age. I apply the same estimation methods to the 2008--2019 waves of the Korean Labor and Income Panel Study (KLIPS), restricting the sample to ages 24 to 55. The pooled sample contains 100,116 person-year observations on 17,606 individuals, is 50\% female, and has a mean age of 40. The share married is 72\%, against about 4\% in GOMS. Hourly wages are monthly earnings divided by monthly hours worked, trimmed by the same rule as in GOMS. The selection indicator equals one for wage and salary workers with a valid wage observation. I investigate four groups: the pooled sample (ages 24 to 55), a young subsample (24 to 35) that brackets the GOMS entry ages, and the college-educated counterparts of each. College graduates aged 24 to 35 are the most comparable to the GOMS sample in age and education, although they are not all recent graduates. The gender difference in employment is much larger than in GOMS. Men's employment rate is 22.1 percentage points higher than women's in the pooled sample and 18.7 percentage points higher among the young.

The specifications closely follow the GOMS analysis. Both equations include a female indicator, age, age$^2$, years of education, indicators for a two-year and a four-year degree, a marriage indicator, residence-region dummies, field-of-study dummies, and year dummies.\footnote{The field of study comes from the KLIPS education histories, and each person-year is assigned the undergraduate field of the degree held in that year (Appendix~\ref{sec:appendix_klips_adjusted}).} As in GOMS, the wage equation includes the job characteristics: indicators for a large firm and for public-sector employment, industry dummies, and workplace-region dummies. GPA is omitted as it is collected only in occasional supplementary modules and cannot be included consistently across the analysis sample. I pool the person-year observations and cluster all standard errors by individual.

\begin{table}[t!]
\centering
\caption{Gender Wage Gap over the Lifecycle: KLIPS, 2008--2019}\label{tab:klips}
\scriptsize
\renewcommand{\arraystretch}{0.92}
\setlength{\tabcolsep}{4.5pt}
\begin{tabular}{l rrrr rrrr}
\toprule
 & \multicolumn{4}{c}{All workers} & \multicolumn{4}{c}{College graduates} \\
\cmidrule(lr){2-5} \cmidrule(lr){6-9}
 & \multicolumn{1}{c}{OLS} & \multicolumn{1}{c}{Heckman} & \multicolumn{1}{c}{Kim-Lee} & \multicolumn{1}{c}{DML-RF}
 & \multicolumn{1}{c}{OLS} & \multicolumn{1}{c}{Heckman} & \multicolumn{1}{c}{Kim-Lee} & \multicolumn{1}{c}{DML-RF} \\
\midrule
\multicolumn{9}{l}{\textit{Panel A: By age}} \\
Pooled, 24--55 & $-$0.342 & $-$0.311 & $-$0.260 & $-$0.232 & $-$0.229 & $-$0.225 & $-$0.156 & $-$0.135 \\
 & (0.009) & (0.014) & (0.012) & (0.011) & (0.011) & (0.012) & (0.014) & (0.013) \\
\quad HH 95\% CI & \multicolumn{4}{c}{empty} & \multicolumn{4}{c}{[$-$0.331, $-$0.131]} \\
\quad Lee bounds & \multicolumn{4}{c}{[$-$0.476, $-$0.070]} & \multicolumn{4}{c}{[$-$0.372, $+$0.030]} \\
Young, 24--35 & $-$0.151 & $-$0.147 & $-$0.078 & $-$0.090 & $-$0.120 & $-$0.121 & $-$0.070 & $-$0.072 \\
 & (0.010) & (0.010) & (0.012) & (0.011) & (0.011) & (0.011) & (0.013) & (0.012) \\
\quad HH 95\% CI & \multicolumn{4}{c}{[$-$0.219, $-$0.067]} & \multicolumn{4}{c}{[$-$0.183, $+$0.006]} \\
\quad Lee bounds & \multicolumn{4}{c}{[$-$0.301, $+$0.070]} & \multicolumn{4}{c}{[$-$0.263, $+$0.094]} \\
\midrule
\multicolumn{9}{l}{\textit{Panel B: By subperiod (ages 24--55)}} \\
2008--2011 & $-$0.375 & $-$0.297 & $-$0.277 & $-$0.246 & $-$0.222 & $-$0.105 & $-$0.156 & $-$0.135 \\
 & (0.011) & (0.017) & (0.017) & (0.016) & (0.017) & (0.023) & (0.022) & (0.020) \\
\quad Lee bounds & \multicolumn{4}{c}{[$-$0.541, $-$0.079]} & \multicolumn{4}{c}{[$-$0.393, $+$0.044]} \\
2012--2015 & $-$0.365 & $-$0.306 & $-$0.313 & $-$0.252 & $-$0.246 & $-$0.153 & $-$0.197 & $-$0.163 \\
 & (0.012) & (0.021) & (0.019) & (0.016) & (0.016) & (0.026) & (0.025) & (0.019) \\
\quad Lee bounds & \multicolumn{4}{c}{[$-$0.493, $-$0.085]} & \multicolumn{4}{c}{[$-$0.385, $+$0.021]} \\
2016--2019 & $-$0.297 & $-$0.284 & $-$0.230 & $-$0.193 & $-$0.219 & $-$0.215 & $-$0.133 & $-$0.110 \\
 & (0.009) & (0.010) & (0.013) & (0.013) & (0.012) & (0.012) & (0.016) & (0.014) \\
\quad Lee bounds & \multicolumn{4}{c}{[$-$0.407, $-$0.053]} & \multicolumn{4}{c}{[$-$0.345, $+$0.027]} \\
\midrule
\multicolumn{9}{l}{\textit{Panel C: Young (24--35) by subperiod}} \\
2008--2011 & $-$0.157 & $-$0.157 & $-$0.077 & $-$0.080 & $-$0.113 & $-$0.123 & $-$0.057 & $-$0.059 \\
 & (0.015) & (0.016) & (0.018) & (0.017) & (0.017) & (0.018) & (0.022) & (0.020) \\
\quad Lee bounds & \multicolumn{4}{c}{[$-$0.340, $+$0.100]} & \multicolumn{4}{c}{[$-$0.288, $+$0.133]} \\
2012--2015 & $-$0.165 & $-$0.112 & $-$0.117 & $-$0.111 & $-$0.139 & $-$0.089 & $-$0.099 & $-$0.094 \\
 & (0.015) & (0.024) & (0.021) & (0.017) & (0.017) & (0.020) & (0.022) & (0.019) \\
\quad Lee bounds & \multicolumn{4}{c}{[$-$0.316, $+$0.069]} & \multicolumn{4}{c}{[$-$0.283, $+$0.089]} \\
2016--2019 & $-$0.126 & $-$0.121 & $-$0.067 & $-$0.061 & $-$0.098 & $-$0.097 & $-$0.056 & $-$0.045 \\
 & (0.013) & (0.013) & (0.016) & (0.015) & (0.014) & (0.014) & (0.017) & (0.017) \\
\quad Lee bounds & \multicolumn{4}{c}{[$-$0.234, $+$0.045]} & \multicolumn{4}{c}{[$-$0.209, $+$0.069]} \\
\bottomrule
\end{tabular}

\vspace{0.5em}
\parbox{\textwidth}{\scriptsize\textit{Notes:} Female coefficients on log hourly wages, with standard errors clustered by individual in parentheses. ``Lee bounds'' are the age-conditional \cite{lee2009} bounds, and ``HH 95\% CI'' is the \cite{honore2020selection} confidence interval computed by subsampling. HH confidence intervals are not reported for the subperiods because of small sample sizes; ``empty'' means that HH's 95\% confidence set is empty.}
\end{table}

Table~\ref{tab:klips} reports much larger wage gaps than at entry, and selection correction now matters. The OLS estimate is $-0.342$ for the pooled sample and $-0.151$ for the younger group. Kim-Lee and DML-RF reduce the pooled gap to $-0.26$ and $-0.23$ and the gap in the young sample to $-0.08$ and $-0.09$. This larger correction is in line with the female employment M-curve, as many married women are out of wage employment in this age range. Kim-Lee and DML-RF are closely aligned in the young groups, within about $0.01$. In the prime-age samples they differ by $0.02$ to $0.03$ in the pooled sample and by up to about $0.06$ in some subperiods. Nonetheless, both methods substantially reduce the estimated gender gap. The Heckman corrections vary widely across cells: below $0.005$ in most young and college-graduate cells but larger than the Kim-Lee corrections among college graduates aged 24 to 55 in 2008--2011 and 2012--2015. The age-conditional Lee bounds exclude zero for all workers aged 24 to 55, in the pooled sample and in every subperiod, but not for college graduates or the young groups. For all workers aged 24 to 55, HH's 95\% confidence set is empty, so the HH inequalities (and the underlying model specification) are rejected.\footnote{As in the GOMS subperiods, no parameter value satisfies all of the sample moment inequalities in any KLIPS group.} In the other three groups the HH confidence intervals contain the Kim-Lee estimates.

The pooled gap is smaller in 2016--2019 than in 2008--2011. The Kim-Lee estimate changes from $-0.277$ to $-0.230$, with the largest gap ($-0.313$) in the intervening period. This comparison partly reflects the aging of the panel. Evaluated at a common age, the narrowing is larger, from $-0.325$ to $-0.247$, and the 2012--2015 estimate, $-0.316$, is close to that of 2008--2011 (Appendix Table~\ref{tab:klips_adjusted}). Within the young 24--35 band, whose age composition is nearly stationary across subperiods (Appendix Figure~\ref{fig:age_dist}), the corrected gap stays small with no clear trend.

\begin{figure}[t!]
\centering
\includegraphics[width=0.95\textwidth]{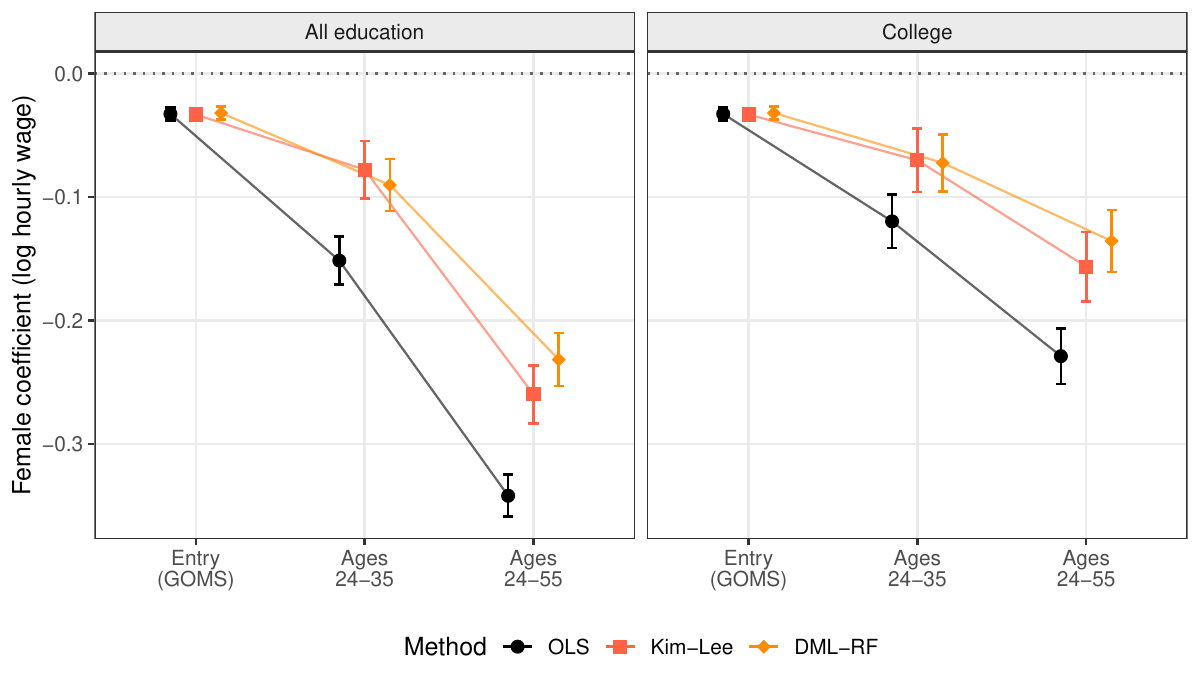}
\caption{The gender wage gap over the lifecycle. Female coefficient in a log hourly wage regression, with 95\% confidence intervals.}\label{fig:klips_lifecycle}
\end{figure}

Figure~\ref{fig:klips_lifecycle} compares the entry-level estimates with the KLIPS age groups. Among college graduates aged 24 to 35, the Kim-Lee gap is about 7 log points, against 3.3 at entry in GOMS, and the selection correction reduces the estimated gap by about two-fifths.\footnote{Dropping GPA, which KLIPS regressions do not include, leaves the GOMS OLS and Kim-Lee estimates at $-0.027$ and $-0.030$, within $0.003$ of each other, against a $0.082$ wedge in the pooled KLIPS sample.} The corrected gap more than doubles among the college-educated, from 7 log points at ages 24 to 35 to 16 at ages 24 to 55, and more than triples across all workers, from about 8 to 26. The selection correction also increases, from 4.9 to 7.3 among the college-educated and from 7.4 to 8.2 across all workers.

\begin{figure}[htbp]
\centering
\includegraphics[width=0.85\textwidth]{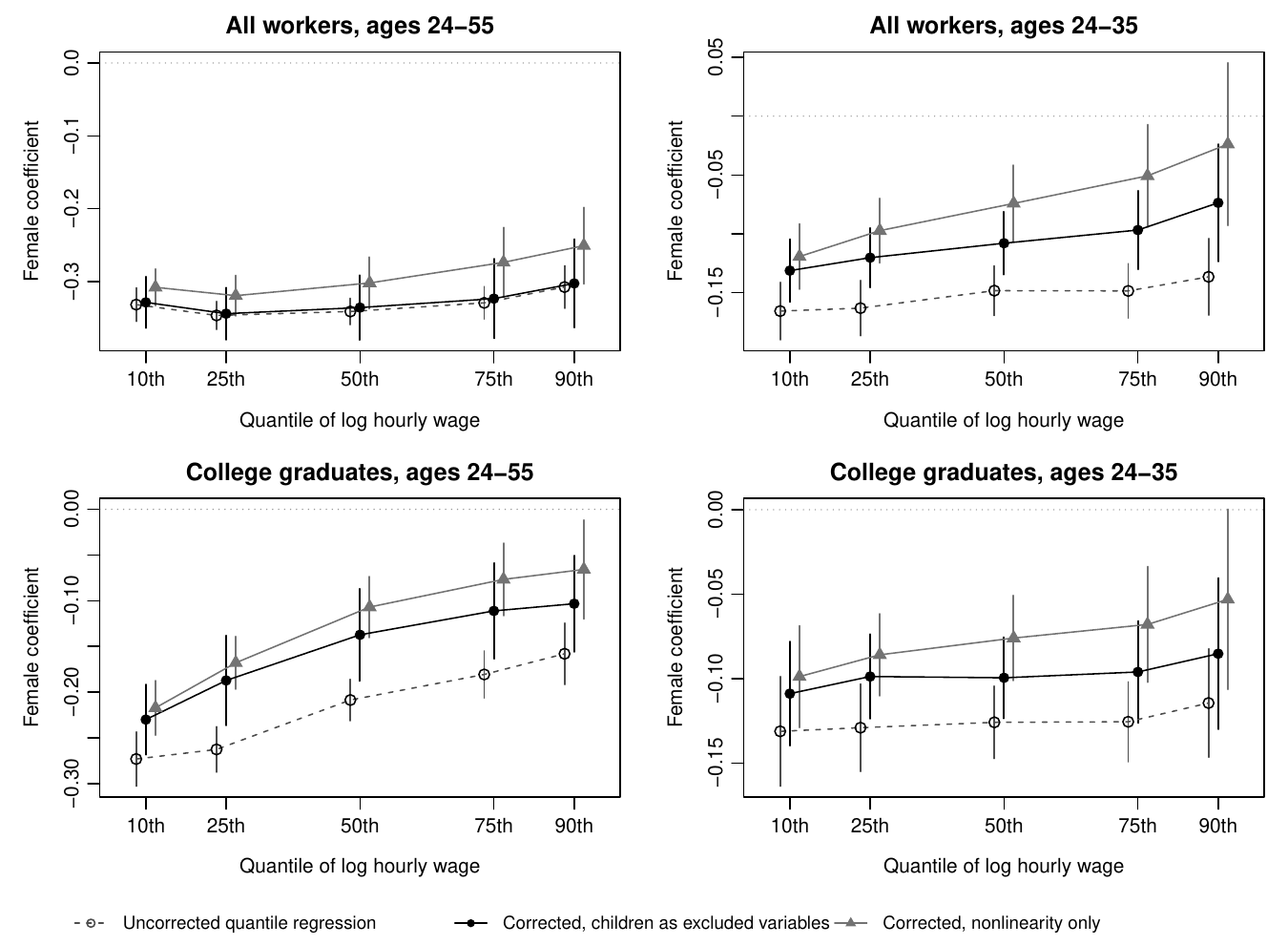}
\caption{Selection-corrected female coefficient by quantile, KLIPS, 2008--2019. Vertical bars are 95\% confidence intervals from a bootstrap over persons (200 replications).}\label{fig:klips_quantile}
\end{figure}

I also apply the selection-corrected quantile estimator to KLIPS, using the numbers of children by age group as excluded variables or relying on the nonlinearity of the sieve selection probability alone. Figure~\ref{fig:klips_quantile} compares both sets of estimates with uncorrected quantile regression. In contrast to the entry-level pattern, the corrected gap at the 90th percentile is not significantly larger than at the 10th in any group, and in several it is significantly smaller. Among college graduates aged 24 to 55, the corrected gap is 22 to 23 log points at the 10th percentile and 7 to 10 at the 90th, and uncorrected quantile regression shows the same decline. Among young college graduates the corrected gap is relatively flat across quantiles, whereas among all young workers it falls from 12 to 13 log points at the 10th percentile to 2 to 7 at the 90th. In the prime-age pool the estimates that use the children counts as excluded variables are 2 to 5 log points larger in absolute value than the nonlinearity-based estimates, but none of these differences is statistically significant.

In contrast to the small entry-level gap, I obtain much larger gender gaps among workers at later career stages, even after selection correction. The adjustment for employment selection is also much larger than at entry. I further present an Oaxaca-Blinder decomposition for KLIPS and two additional robustness analyses.

\textit{Decomposition.} A two-part Oaxaca-Blinder decomposition of the pooled gap, using the pooled Kim-Lee coefficients as the reference, as for GOMS, attributes $0.090$ of the $0.426$ raw gap, about 21\%, to differences in observed endowments, against 63\% explained for recent graduates. Job characteristics account for the largest share of the endowments component (Appendix~\ref{sec:appendix_klips_adjusted}). For the broad prime-age sample, most of the gap therefore reflects differences in returns to characteristics and employment selection.

\textit{Heterogeneous selection.} Selection corrections that differ by education, marital status, and field of study shift the Kim-Lee estimates by up to four log points: toward zero in the prime-age groups, from $-0.26$ to $-0.22$ among all workers and from $-0.16$ to $-0.14$ among college graduates, and away from zero among young college graduates, from $-0.07$ to $-0.08$. By comparison, allowing these differences in GOMS changes the entry-level estimate by less than one log point. The Huber-Melly test rejects conditional independence for all slopes in every group, and for the female coefficient in the prime-age groups but not in the young groups. Appendix Table~\ref{tab:klips_hetero} reports these estimates and tests.

\textit{Exclusion-based estimates.} The excluded variables differ from those used for GOMS, where few graduates are married or have children and the excluded variables describe the parental household. For KLIPS I use the variables common in the literature \citep{arellano2017quantile,biewen2020counterfactual}: children counts by age group, spouse earnings, household asset income, and transfer income. Each is interacted with gender in the selection equation to allow its relationship with employment to differ between men and women. Appendix Table~\ref{tab:iv_klips} reports Kim-Lee and Heckman estimates with these excluded variables. Adding them to the Kim-Lee first stage changes the estimates by at most $0.03$, and the children counts alone move them by at most $0.011$, relative to Kim-Lee without them on the same observations. The Heckman estimates with the excluded variables correct less in the prime-age pool, $-0.30$ to $-0.32$ against Kim-Lee's $-0.26$ and DML-RF's $-0.23$. All mean corrections with the excluded variables move the gap toward zero relative to OLS. For the young groups, whose life stage is closest to that of recent graduates, I also use the GOMS excluded variables, co-residence with parents and household size. They move the Kim-Lee estimates by at most $0.002$. In the quantile model, using all four groups of excluded variables gives a corrected pooled median of $-0.29$, close to the nonlinearity-based $-0.30$ (Appendix Table~\ref{tab:iv_klips_q}).

\section{Conclusion}\label{sec:conclusion}

This paper provides evidence that the large aggregate Korean gender wage gap does not originate at labor market entry. The semiparametric selection correction methods yield virtually identical estimates to OLS, and the age-conditional Lee bounds and HH confidence intervals contain them. Allowing heterogeneous corrections or using excluded variables changes the semiparametric mean estimate by less than a log point. Selection-corrected quantiles preserve the concentration of the remaining gap at the top. Differential selection into full-time wage employment does not materially change the entry-level mean gap for this population. These results suggest that most of Korea's gender wage disparity is generated after labor market entry. The lifecycle analysis of KLIPS demonstrates much larger gender gaps among older prime-age workers. 

These results imply that policy efforts should focus more on women's employment and pay growth after labor market entry. Measures that act on the hiring stage alone address a small part of Korea's gender gap. Keeping women employed after childbirth is a priority. Korean women's employment falls by almost half relative to men's after the birth of a first child \citep{kleven2025child}. Korea already offers each parent up to 18 months of parental leave, but the replacement rate is relatively low \citep{oecd2026family}. Among eligible parents of children born in 2024, 72\% of mothers and only 10\% of fathers took leave in the year of birth \citep{mods2025parental}. Raising benefit ceilings would reduce the income loss for the higher-earning parent. Requiring fathers to take part of their leave would also make fathers' participation in early childcare more common. Helping mothers stay with their current employer, where they have accumulated seniority, is also important. Flexible working-hour arrangements can help mothers remain in the same job. Under current law, however, reduced hours are limited to one year plus twice any unused parental leave, so a parent who has taken a full year of leave can work reduced hours for only one more year. Policy should make flexible working arrangements more widely available so that parents can remain employed while raising children. Encouraging fathers to use reduced hours as well would help share childcare and reduce the pressure on mothers to interrupt their careers.

\clearpage
\runtitle{Online Appendix: Gender wage gap at labor market entry}
\markboth{DONGWOO KIM}{ONLINE APPENDIX: GENDER WAGE GAP AT LABOR MARKET ENTRY}
\thispagestyle{empty}
\begin{center}
\vspace*{1.3in}
{\large\bfseries ONLINE APPENDIX\par}
\vspace{1.25em}
{\LARGE\bfseries Does the Gender Wage Gap Originate at Labor Market Entry?\par}
\vspace{0.4em}
{\LARGE\bfseries Evidence from South Korea\par}
\vspace{2em}
{\large Dongwoo Kim\par}
\vspace{0.4em}
{\normalsize Simon Fraser University and Korea University\par}
\end{center}
\vfill
\clearpage

\setcounter{footnote}{0}
\makeatletter
\setlength{\@fptop}{0pt}
\setlength{\@fpsep}{18pt}
\setlength{\@fpbot}{0pt plus 1fil}
\renewcommand{\floatpagefraction}{0.8}
\makeatother

\appendix
\section{Additional Tables}\label{sec:appendix}

This appendix collects additional empirical results for Section~\ref{sec:results}. Table~\ref{tab:full_coefs} reports the full pooled 2008--2019 coefficients of OLS, the Heckman MLE (wage and selection equations), and Kim-Lee. Table~\ref{tab:yearly} reports the year-by-year female coefficients of all four estimators. Figure~\ref{fig:trend} plots the OLS, Kim-Lee, and DML-RF series. Table~\ref{tab:major_gap_composition} reports the major shares of employed graduates and the conditional gender gap within each major, which Section~\ref{sec:emp_decline} uses to interpret the change in the major composition of the employed sample. Tables~\ref{tab:robust_lwage} and \ref{tab:robust_nojob} report the full coefficient vectors of the log monthly wage and no-job-control specifications of Appendix~\ref{sec:appendix_robustness}.

\newcommand{\AfterFullCoefficientTable}{%
\begin{table}[p]
	\centering
	\caption{Year-by-Year Gender Wage Gap, 2008--2019}\label{tab:yearly}
	\begin{tabular}{lcccc}
	\toprule
	Year & \multicolumn{1}{c}{OLS} & \multicolumn{1}{c}{Heckman} & \multicolumn{1}{c}{Kim-Lee} & \multicolumn{1}{c}{DML-RF} \\
	\midrule
	2008 & $-$0.029 (0.011) & $-$0.023 (0.010) & $-$0.030 (0.011) & $-$0.033 (0.011) \\
	2009 & $-$0.069 (0.010) & $-$0.068 (0.010) & $-$0.066 (0.010) & $-$0.068 (0.010) \\
	2010 & $-$0.018 (0.011) & $-$0.018 (0.010) & $-$0.015 (0.011) & $-$0.011 (0.011) \\
	2011 & $-$0.026 (0.009) & $-$0.026 (0.009) & $-$0.019 (0.009) & $-$0.021 (0.009) \\
	2012 & $-$0.047 (0.010) & $-$0.047 (0.010) & $-$0.051 (0.010) & $-$0.051 (0.011) \\
	2013 & $-$0.037 (0.010) & $-$0.037 (0.010) & $-$0.041 (0.010) & $-$0.040 (0.010) \\
	2014 & $-$0.027 (0.009) & $-$0.020 (0.010) & $-$0.028 (0.010) & $-$0.024 (0.010) \\
	2015 & $-$0.042 (0.009) & $-$0.042 (0.008) & $-$0.047 (0.009) & $-$0.045 (0.009) \\
	2016 & $-$0.020 (0.008) & $-$0.018 (0.008) & $-$0.020 (0.008) & $-$0.026 (0.009) \\
	2017 & $-$0.030 (0.007) & $-$0.030 (0.008) & $-$0.029 (0.008) & $-$0.024 (0.008) \\
	2018 & $-$0.024 (0.007) & $-$0.018 (0.008) & $-$0.024 (0.008) & $-$0.021 (0.008) \\
	2019 & $-$0.027 (0.007) & $-$0.030 (0.008) & $-$0.025 (0.007) & $-$0.022 (0.008) \\
	\bottomrule
\end{tabular}

	\vspace{0.5em}
	\parbox{\textwidth}{\scriptsize\textit{Notes:} Female coefficients on log hourly wages from separate regressions for each wave (approximately 8,800--11,900 wage workers per year) with the Table~\ref{tab:main} covariates except the year dummies. Heteroskedasticity-robust standard errors are in parentheses, except for Heckman, whose maximum likelihood standard errors do not allow for heteroskedasticity.}
\end{table}

\begin{table}[p]
\centering
\caption{Major Composition and Within-Major Gender Wage Gaps}\label{tab:major_gap_composition}
{\renewcommand{\arraystretch}{0.9}
\begin{tabular}{lrrc}
\toprule
 & \multicolumn{2}{c}{Share of wage sample (\%)} & Female coefficient \\
\cmidrule(lr){2-3}
Major & 2008--2011 & 2016--2019 & 2016--2019 (log points) \\
\midrule
Social science & 24.6 & 20.2 & $-$4.60 (0.60) \\
Engineering & 28.4 & 31.4 & $-$2.87 (0.63) \\
Natural science & 11.7 & 12.3 & $-$2.79 (0.85) \\
Arts/sports & 10.3 & 9.1 & $-$2.24 (1.14) \\
Humanities & 9.6 & 11.0 & $-$1.46 (0.88) \\
Education & 7.7 & 7.1 & $-$0.53 (1.08) \\
Medical/health & 7.7 & 8.8 & $+$1.23 (1.08) \\
\midrule
$N$ & 45,406 & 40,582 & 40,582 \\
\bottomrule
\end{tabular}
}

\vspace{0.5em}
\parbox{\textwidth}{\scriptsize\textit{Notes:} Shares pool male and female employed graduates. Female coefficients are from one OLS regression of log hourly wages on the 2016--2019 sample that interacts the female indicator with the major dummies and keeps common coefficients on the other Table~\ref{tab:main} controls. Heteroskedasticity-robust standard errors in parentheses.}
\end{table}}
\newcommand{\OmitExclusionTable}{}

\begin{table}[p]
\centering
\caption{Full Regression Coefficients: Log Hourly Wage (Pooled 2008--2019)}\label{tab:full_coefs}
\scriptsize
\setlength{\tabcolsep}{4pt}
\renewcommand{\arraystretch}{0.78}
\begin{tabular}{lcccc}
\toprule
 & & \multicolumn{2}{c}{Heckman} & \\
\cmidrule(lr){3-4}
 & OLS & Wage eq. & Selection eq. & Kim-Lee \\
\midrule
\multicolumn{5}{l}{\textit{Personal characteristics}} \\
Female & $-$0.033 (0.003) & $-$0.028 (0.003) & 0.017 (0.007) & $-$0.033 (0.003) \\
Age & 0.069 (0.008) & 0.102 (0.008) & 0.197 (0.024) & 0.054 (0.008) \\
Age$^2$ ($\times 100$) & $-$0.076 (0.015) & $-$0.128 (0.015) & $-$0.293 (0.043) & $-$0.056 (0.015) \\
GPA (0--100 scale) & 0.003 (0.000) & 0.003 (0.000) & 0.003 (0.000) & 0.002 (0.000) \\
Married & 0.064 (0.005) & 0.062 (0.005) & $-$0.004 (0.015) & 0.063 (0.005) \\
4-year university & 0.120 (0.003) & 0.108 (0.003) & $-$0.063 (0.009) & 0.126 (0.003) \\
School-area dummies & Yes & Yes & Yes & Yes \\
[4pt]
\multicolumn{5}{l}{\textit{Major (ref: Social science)}} \\
Humanities & $-$0.028 (0.003) & $-$0.061 (0.004) & $-$0.191 (0.010) & $-$0.004 (0.004) \\
Education & 0.178 (0.005) & 0.194 (0.005) & 0.040 (0.012) & 0.172 (0.005) \\
Engineering & 0.026 (0.003) & 0.032 (0.003) & 0.070 (0.009) & 0.020 (0.003) \\
Natural science & $-$0.025 (0.003) & $-$0.061 (0.004) & $-$0.197 (0.010) & 0.000 (0.004) \\
Medical/health & 0.143 (0.005) & 0.194 (0.005) & 0.409 (0.014) & 0.110 (0.006) \\
Arts/sports & $-$0.057 (0.004) & $-$0.100 (0.004) & $-$0.194 (0.010) & $-$0.029 (0.005) \\
[4pt]
\multicolumn{5}{l}{\textit{Job characteristics}} \\
Large firm ($\geq$ 300 emp.) & 0.139 (0.002) & 0.134 (0.002) &  & 0.138 (0.002) \\
Public sector & $-$0.020 (0.003) & $-$0.022 (0.003) &  & $-$0.019 (0.003) \\
[4pt]
\multicolumn{5}{l}{\textit{Industry (ref: Other services)}} \\
Manufacturing & 0.116 (0.003) & 0.125 (0.003) &  & 0.115 (0.003) \\
Construction & 0.067 (0.005) & 0.074 (0.005) &  & 0.066 (0.005) \\
Wholesale/retail & 0.049 (0.004) & 0.052 (0.004) &  & 0.049 (0.004) \\
IT/communication & 0.035 (0.004) & 0.043 (0.004) &  & 0.035 (0.004) \\
Finance/insurance & 0.206 (0.005) & 0.209 (0.005) &  & 0.205 (0.005) \\
Professional services & 0.037 (0.004) & 0.042 (0.004) &  & 0.036 (0.004) \\
Public administration & $-$0.098 (0.005) & $-$0.093 (0.005) &  & $-$0.094 (0.005) \\
Education services & $-$0.025 (0.005) & $-$0.037 (0.004) &  & $-$0.024 (0.005) \\
Health/social & 0.003 (0.005) & 0.007 (0.004) &  & $-$0.000 (0.005) \\
Workplace-region dummies & Yes & Yes &  & Yes \\
[4pt]
Year dummies & Yes & Yes & Yes & Yes \\
\midrule
$\hat{\rho}$ & & \multicolumn{2}{c}{0.715} & \\
$\hat{\sigma}$ & & \multicolumn{2}{c}{0.384} & \\
$N$ (selected) & 127,628 & \multicolumn{2}{c}{127,628} & 127,628 \\
$N$ (total) & & \multicolumn{2}{c}{202,114} & \\
\bottomrule
\end{tabular}

\vspace{0.5em}
\parbox{\textwidth}{\scriptsize\textit{Notes:} Heckman is estimated by maximum likelihood; $\hat{\rho}$ is its error correlation and $\hat{\sigma}$ the standard deviation of the wage error. Standard errors in parentheses: heteroskedasticity-robust for OLS, maximum likelihood for Heckman, and first-stage-corrected for the Kim-Lee estimator of \cite{KL2026}. Age$^2$ coefficients are multiplied by 100. Intercepts are not reported (see the text); year dummies (reference 2008) are included but not shown.}
\end{table}

\ifdefined%
\begin{table}[p]
	\centering
	\caption{Year-by-Year Gender Wage Gap, 2008--2019}\label{tab:yearly}
	\begin{tabular}{lcccc}
	\toprule
	Year & \multicolumn{1}{c}{OLS} & \multicolumn{1}{c}{Heckman} & \multicolumn{1}{c}{Kim-Lee} & \multicolumn{1}{c}{DML-RF} \\
	\midrule
	2008 & $-$0.029 (0.011) & $-$0.023 (0.010) & $-$0.030 (0.011) & $-$0.033 (0.011) \\
	2009 & $-$0.069 (0.010) & $-$0.068 (0.010) & $-$0.066 (0.010) & $-$0.068 (0.010) \\
	2010 & $-$0.018 (0.011) & $-$0.018 (0.010) & $-$0.015 (0.011) & $-$0.011 (0.011) \\
	2011 & $-$0.026 (0.009) & $-$0.026 (0.009) & $-$0.019 (0.009) & $-$0.021 (0.009) \\
	2012 & $-$0.047 (0.010) & $-$0.047 (0.010) & $-$0.051 (0.010) & $-$0.051 (0.011) \\
	2013 & $-$0.037 (0.010) & $-$0.037 (0.010) & $-$0.041 (0.010) & $-$0.040 (0.010) \\
	2014 & $-$0.027 (0.009) & $-$0.020 (0.010) & $-$0.028 (0.010) & $-$0.024 (0.010) \\
	2015 & $-$0.042 (0.009) & $-$0.042 (0.008) & $-$0.047 (0.009) & $-$0.045 (0.009) \\
	2016 & $-$0.020 (0.008) & $-$0.018 (0.008) & $-$0.020 (0.008) & $-$0.026 (0.009) \\
	2017 & $-$0.030 (0.007) & $-$0.030 (0.008) & $-$0.029 (0.008) & $-$0.024 (0.008) \\
	2018 & $-$0.024 (0.007) & $-$0.018 (0.008) & $-$0.024 (0.008) & $-$0.021 (0.008) \\
	2019 & $-$0.027 (0.007) & $-$0.030 (0.008) & $-$0.025 (0.007) & $-$0.022 (0.008) \\
	\bottomrule
\end{tabular}

	\vspace{0.5em}
	\parbox{\textwidth}{\scriptsize\textit{Notes:} Female coefficients on log hourly wages from separate regressions for each wave (approximately 8,800--11,900 wage workers per year) with the Table~\ref{tab:main} covariates except the year dummies. Heteroskedasticity-robust standard errors are in parentheses, except for Heckman, whose maximum likelihood standard errors do not allow for heteroskedasticity.}
\end{table}

\begin{table}[p]
\centering
\caption{Major Composition and Within-Major Gender Wage Gaps}\label{tab:major_gap_composition}
{\renewcommand{\arraystretch}{0.9}
\begin{tabular}{lrrc}
\toprule
 & \multicolumn{2}{c}{Share of wage sample (\%)} & Female coefficient \\
\cmidrule(lr){2-3}
Major & 2008--2011 & 2016--2019 & 2016--2019 (log points) \\
\midrule
Social science & 24.6 & 20.2 & $-$4.60 (0.60) \\
Engineering & 28.4 & 31.4 & $-$2.87 (0.63) \\
Natural science & 11.7 & 12.3 & $-$2.79 (0.85) \\
Arts/sports & 10.3 & 9.1 & $-$2.24 (1.14) \\
Humanities & 9.6 & 11.0 & $-$1.46 (0.88) \\
Education & 7.7 & 7.1 & $-$0.53 (1.08) \\
Medical/health & 7.7 & 8.8 & $+$1.23 (1.08) \\
\midrule
$N$ & 45,406 & 40,582 & 40,582 \\
\bottomrule
\end{tabular}
}

\vspace{0.5em}
\parbox{\textwidth}{\scriptsize\textit{Notes:} Shares pool male and female employed graduates. Female coefficients are from one OLS regression of log hourly wages on the 2016--2019 sample that interacts the female indicator with the major dummies and keeps common coefficients on the other Table~\ref{tab:main} controls. Heteroskedasticity-robust standard errors in parentheses.}
\end{table}%

\begin{table}[p]
\centering
\caption{Major Composition and Within-Major Gender Wage Gaps}\label{tab:major_gap_composition}
{\renewcommand{\arraystretch}{0.9}
\begin{tabular}{lrrc}
\toprule
 & \multicolumn{2}{c}{Share of wage sample (\%)} & Female coefficient \\
\cmidrule(lr){2-3}
Major & 2008--2011 & 2016--2019 & 2016--2019 (log points) \\
\midrule
Social science & 24.6 & 20.2 & $-$4.60 (0.60) \\
Engineering & 28.4 & 31.4 & $-$2.87 (0.63) \\
Natural science & 11.7 & 12.3 & $-$2.79 (0.85) \\
Arts/sports & 10.3 & 9.1 & $-$2.24 (1.14) \\
Humanities & 9.6 & 11.0 & $-$1.46 (0.88) \\
Education & 7.7 & 7.1 & $-$0.53 (1.08) \\
Medical/health & 7.7 & 8.8 & $+$1.23 (1.08) \\
\midrule
$N$ & 45,406 & 40,582 & 40,582 \\
\bottomrule
\end{tabular}
}

\vspace{0.5em}
\parbox{\textwidth}{\scriptsize\textit{Notes:} Shares pool male and female employed graduates. Female coefficients are from one OLS regression of log hourly wages on the 2016--2019 sample that interacts the female indicator with the major dummies and keeps common coefficients on the other Table~\ref{tab:main} controls. Heteroskedasticity-robust standard errors in parentheses.}
\end{table}\fi

\begin{table}[p]
\centering
\caption{Full Regression Coefficients: Log Hourly Wage vs.\ Log Monthly Wage}\label{tab:robust_lwage}
\scriptsize
\setlength{\tabcolsep}{1.7pt}
\renewcommand{\arraystretch}{0.9}
\begin{tabular}{lcccccc}
\toprule
 & \multicolumn{3}{c}{Log hourly wage} & \multicolumn{3}{c}{Log monthly wage} \\
\cmidrule(lr){2-4} \cmidrule(lr){5-7}
 & OLS & Heckman & Kim-Lee & OLS & Heckman & Kim-Lee \\
\midrule
\multicolumn{7}{l}{\textit{Personal characteristics}} \\
Female & $-$0.033 (0.003) & $-$0.028 (0.003) & $-$0.033 (0.003) & $-$0.064 (0.002) & $-$0.071 (0.002) & $-$0.065 (0.003) \\
Age & 0.069 (0.008) & 0.102 (0.008) & 0.054 (0.008) & 0.065 (0.007) & 0.031 (0.007) & 0.044 (0.008) \\
Age$^2$ ($\times 100$) & $-$0.076 (0.015) & $-$0.128 (0.015) & $-$0.056 (0.015) & $-$0.072 (0.013) & $-$0.019 (0.014) & $-$0.041 (0.014) \\
GPA (0--100 scale) & 0.003 (0.000) & 0.003 (0.000) & 0.002 (0.000) & 0.002 (0.000) & 0.002 (0.000) & 0.002 (0.000) \\
Married & 0.064 (0.005) & 0.062 (0.005) & 0.063 (0.005) & 0.059 (0.004) & 0.061 (0.005) & 0.059 (0.005) \\
4-year university & 0.120 (0.003) & 0.108 (0.003) & 0.126 (0.003) & 0.109 (0.002) & 0.118 (0.003) & 0.116 (0.003) \\
School-area dummies & Yes & Yes & Yes & Yes & Yes & Yes \\
[4pt]
\multicolumn{7}{l}{\textit{Major (ref: Social science)}} \\
Humanities & $-$0.028 (0.003) & $-$0.061 (0.004) & $-$0.004 (0.004) & $-$0.041 (0.003) & $-$0.008 (0.003) & $-$0.011 (0.004) \\
Education & 0.178 (0.005) & 0.194 (0.005) & 0.172 (0.005) & 0.222 (0.004) & 0.205 (0.004) & 0.214 (0.005) \\
Engineering & 0.026 (0.003) & 0.032 (0.003) & 0.020 (0.003) & 0.029 (0.002) & 0.023 (0.003) & 0.021 (0.003) \\
Natural science & $-$0.025 (0.003) & $-$0.061 (0.004) & 0.000 (0.004) & $-$0.022 (0.003) & 0.015 (0.003) & 0.012 (0.004) \\
Medical/health & 0.143 (0.005) & 0.194 (0.005) & 0.110 (0.006) & 0.167 (0.004) & 0.123 (0.004) & 0.127 (0.006) \\
Arts/sports & $-$0.057 (0.004) & $-$0.100 (0.004) & $-$0.029 (0.005) & $-$0.105 (0.003) & $-$0.063 (0.003) & $-$0.068 (0.004) \\
[4pt]
\multicolumn{7}{l}{\textit{Job characteristics}} \\
Large firm ($\geq$ 300 emp.) & 0.139 (0.002) & 0.134 (0.002) & 0.138 (0.002) & 0.153 (0.002) & 0.159 (0.002) & 0.152 (0.002) \\
Public sector & $-$0.020 (0.003) & $-$0.022 (0.003) & $-$0.019 (0.003) & $-$0.035 (0.003) & $-$0.031 (0.002) & $-$0.033 (0.003) \\
[4pt]
\multicolumn{7}{l}{\textit{Industry (ref: Other services)}} \\
Manufacturing & 0.116 (0.003) & 0.125 (0.003) & 0.115 (0.003) & 0.151 (0.003) & 0.137 (0.003) & 0.150 (0.003) \\
Construction & 0.067 (0.005) & 0.074 (0.005) & 0.066 (0.005) & 0.126 (0.004) & 0.115 (0.004) & 0.124 (0.004) \\
Wholesale/retail & 0.049 (0.004) & 0.052 (0.004) & 0.049 (0.004) & 0.080 (0.004) & 0.075 (0.003) & 0.079 (0.004) \\
IT/communication & 0.035 (0.004) & 0.043 (0.004) & 0.035 (0.004) & 0.056 (0.003) & 0.043 (0.003) & 0.056 (0.003) \\
Finance/insurance & 0.206 (0.005) & 0.209 (0.005) & 0.205 (0.005) & 0.219 (0.005) & 0.220 (0.004) & 0.217 (0.005) \\
Professional services & 0.037 (0.004) & 0.042 (0.004) & 0.036 (0.004) & 0.064 (0.003) & 0.056 (0.003) & 0.063 (0.003) \\
Public administration & $-$0.098 (0.005) & $-$0.093 (0.005) & $-$0.094 (0.005) & $-$0.085 (0.004) & $-$0.095 (0.004) & $-$0.080 (0.004) \\
Education services & $-$0.025 (0.005) & $-$0.037 (0.004) & $-$0.024 (0.005) & $-$0.120 (0.005) & $-$0.108 (0.003) & $-$0.118 (0.005) \\
Health/social & 0.003 (0.005) & 0.007 (0.004) & $-$0.000 (0.005) & 0.007 (0.004) & $-$0.008 (0.004) & 0.003 (0.004) \\
Workplace-region dummies & Yes & Yes & Yes & Yes & Yes & Yes \\
[4pt]
\midrule
Year dummies & \multicolumn{6}{c}{Yes} \\
$N$ & \multicolumn{3}{c}{127,628} & \multicolumn{3}{c}{127,628} \\
\bottomrule
\end{tabular}

\vspace{0.5em}
\parbox{\textwidth}{\scriptsize\textit{Notes:} Pooled 2008--2019. The log monthly wage columns also control for regular and overtime weekly hours (not shown). Standard errors in parentheses: heteroskedasticity-robust for OLS, maximum likelihood for Heckman, and first-stage-corrected for Kim-Lee. Age$^2$ coefficients are multiplied by 100; intercepts and year dummies are not shown.}
\end{table}

\begin{table}[p]
\centering
\caption{Full Regression Coefficients with and without Job Controls}\label{tab:robust_nojob}
\scriptsize
\setlength{\tabcolsep}{1.7pt}
\renewcommand{\arraystretch}{0.9}
\begin{tabular}{lcccccc}
\toprule
 & \multicolumn{3}{c}{With job controls} & \multicolumn{3}{c}{Without job controls} \\
\cmidrule(lr){2-4} \cmidrule(lr){5-7}
 & OLS & Heckman & Kim-Lee & OLS & Heckman & Kim-Lee \\
\midrule
\multicolumn{7}{l}{\textit{Personal characteristics}} \\
Female & $-$0.033 (0.003) & $-$0.028 (0.003) & $-$0.033 (0.003) & $-$0.049 (0.003) & $-$0.050 (0.003) & $-$0.051 (0.003) \\
Age & 0.069 (0.008) & 0.102 (0.008) & 0.054 (0.008) & 0.110 (0.008) & 0.106 (0.008) & 0.083 (0.009) \\
Age$^2$ ($\times 100$) & $-$0.076 (0.015) & $-$0.128 (0.015) & $-$0.056 (0.015) & $-$0.148 (0.015) & $-$0.143 (0.015) & $-$0.108 (0.017) \\
GPA (0--100 scale) & 0.003 (0.000) & 0.003 (0.000) & 0.002 (0.000) & 0.003 (0.000) & 0.003 (0.000) & 0.003 (0.000) \\
Married & 0.064 (0.005) & 0.062 (0.005) & 0.063 (0.005) & 0.065 (0.005) & 0.065 (0.005) & 0.065 (0.006) \\
4-year university & 0.120 (0.003) & 0.108 (0.003) & 0.126 (0.003) & 0.125 (0.003) & 0.127 (0.003) & 0.136 (0.003) \\
School-area dummies & Yes & Yes & Yes & Yes & Yes & Yes \\
[4pt]
\multicolumn{7}{l}{\textit{Major (ref: Social science)}} \\
Humanities & $-$0.028 (0.003) & $-$0.061 (0.004) & $-$0.004 (0.004) & $-$0.042 (0.004) & $-$0.039 (0.004) & $-$0.004 (0.005) \\
Education & 0.178 (0.005) & 0.194 (0.005) & 0.172 (0.005) & 0.079 (0.004) & 0.078 (0.004) & 0.071 (0.004) \\
Engineering & 0.026 (0.003) & 0.032 (0.003) & 0.020 (0.003) & 0.045 (0.003) & 0.044 (0.003) & 0.035 (0.003) \\
Natural science & $-$0.025 (0.003) & $-$0.061 (0.004) & 0.000 (0.004) & $-$0.024 (0.003) & $-$0.020 (0.004) & 0.017 (0.005) \\
Medical/health & 0.143 (0.005) & 0.194 (0.005) & 0.110 (0.006) & 0.121 (0.004) & 0.115 (0.005) & 0.068 (0.006) \\
Arts/sports & $-$0.057 (0.004) & $-$0.100 (0.004) & $-$0.029 (0.005) & $-$0.093 (0.004) & $-$0.089 (0.004) & $-$0.047 (0.005) \\
[4pt]
\multicolumn{7}{l}{\textit{Job characteristics}} \\
Large firm ($\geq$ 300 emp.) & 0.139 (0.002) & 0.134 (0.002) & 0.138 (0.002) &  &  &  \\
Public sector & $-$0.020 (0.003) & $-$0.022 (0.003) & $-$0.019 (0.003) &  &  &  \\
[4pt]
\multicolumn{7}{l}{\textit{Industry (ref: Other services)}} \\
Manufacturing & 0.116 (0.003) & 0.125 (0.003) & 0.115 (0.003) &  &  &  \\
Construction & 0.067 (0.005) & 0.074 (0.005) & 0.066 (0.005) &  &  &  \\
Wholesale/retail & 0.049 (0.004) & 0.052 (0.004) & 0.049 (0.004) &  &  &  \\
IT/communication & 0.035 (0.004) & 0.043 (0.004) & 0.035 (0.004) &  &  &  \\
Finance/insurance & 0.206 (0.005) & 0.209 (0.005) & 0.205 (0.005) &  &  &  \\
Professional services & 0.037 (0.004) & 0.042 (0.004) & 0.036 (0.004) &  &  &  \\
Public administration & $-$0.098 (0.005) & $-$0.093 (0.005) & $-$0.094 (0.005) &  &  &  \\
Education services & $-$0.025 (0.005) & $-$0.037 (0.004) & $-$0.024 (0.005) &  &  &  \\
Health/social & 0.003 (0.005) & 0.007 (0.004) & $-$0.000 (0.005) &  &  &  \\
Workplace-region dummies & Yes & Yes & Yes & No & No & No \\
[4pt]
\midrule
Year dummies & \multicolumn{6}{c}{Yes} \\
$N$ & \multicolumn{3}{c}{127,628} & \multicolumn{3}{c}{127,628} \\
\bottomrule
\end{tabular}

\vspace{0.5em}
\parbox{\textwidth}{\scriptsize\textit{Notes:} Pooled 2008--2019. Job controls are the large-firm, public-sector, industry, and workplace-region dummies. Standard errors in parentheses: heteroskedasticity-robust for OLS, maximum likelihood for Heckman, and first-stage-corrected for Kim-Lee. Age$^2$ coefficients are multiplied by 100; intercepts and year dummies are not shown.}
\end{table}

\ifdefined\OmitExclusionTable\else
\begin{table}[p]
\centering
\caption{Robustness with and without Exclusion Restrictions}\label{tab:robust_excl_full}
\scriptsize
\setlength{\tabcolsep}{1.7pt}
\renewcommand{\arraystretch}{0.9}
\begin{tabular}{lcccccc}
\toprule
 & \multicolumn{3}{c}{With exclusion restrictions} & \multicolumn{3}{c}{Without exclusion restrictions} \\
\cmidrule(lr){2-4} \cmidrule(lr){5-7}
 & OLS & Heckman & Kim-Lee & OLS & Heckman & Kim-Lee \\
\midrule
\multicolumn{7}{l}{\textit{Personal characteristics}} \\
Female & $-$0.032 (0.003) & $-$0.027 (0.003) & $-$0.033 (0.003) & $-$0.032 (0.003) & $-$0.027 (0.003) & $-$0.034 (0.003) \\
Age & 0.070 (0.008) & 0.104 (0.009) & 0.059 (0.009) & 0.070 (0.008) & 0.104 (0.009) & 0.053 (0.009) \\
Age$^2$ ($\times 100$) & $-$0.078 (0.015) & $-$0.132 (0.016) & $-$0.062 (0.016) & $-$0.078 (0.015) & $-$0.132 (0.016) & $-$0.053 (0.016) \\
GPA (0--100 scale) & 0.003 (0.000) & 0.003 (0.000) & 0.002 (0.000) & 0.003 (0.000) & 0.003 (0.000) & 0.002 (0.000) \\
Married & 0.064 (0.005) & 0.062 (0.005) & 0.064 (0.005) & 0.064 (0.005) & 0.063 (0.005) & 0.064 (0.005) \\
4-year university & 0.120 (0.003) & 0.108 (0.003) & 0.124 (0.003) & 0.120 (0.003) & 0.107 (0.003) & 0.127 (0.003) \\
School-area dummies & Yes & Yes & Yes & Yes & Yes & Yes \\
[4pt]
\multicolumn{7}{l}{\textit{Major (ref: Social science)}} \\
Humanities & $-$0.027 (0.004) & $-$0.060 (0.004) & $-$0.013 (0.004) & $-$0.027 (0.004) & $-$0.059 (0.004) & $-$0.003 (0.004) \\
Education & 0.182 (0.005) & 0.197 (0.005) & 0.179 (0.005) & 0.182 (0.005) & 0.197 (0.005) & 0.177 (0.005) \\
Engineering & 0.026 (0.003) & 0.031 (0.003) & 0.023 (0.003) & 0.026 (0.003) & 0.032 (0.003) & 0.020 (0.003) \\
Natural science & $-$0.026 (0.003) & $-$0.062 (0.004) & $-$0.010 (0.004) & $-$0.026 (0.003) & $-$0.062 (0.004) & 0.002 (0.004) \\
Medical/health & 0.142 (0.005) & 0.191 (0.005) & 0.125 (0.006) & 0.142 (0.005) & 0.191 (0.005) & 0.111 (0.006) \\
Arts/sports & $-$0.058 (0.004) & $-$0.100 (0.004) & $-$0.040 (0.005) & $-$0.058 (0.004) & $-$0.100 (0.004) & $-$0.028 (0.005) \\
[4pt]
\multicolumn{7}{l}{\textit{Job characteristics}} \\
Large firm ($\geq$ 300 emp.) & 0.139 (0.002) & 0.134 (0.002) & 0.139 (0.002) & 0.139 (0.002) & 0.135 (0.002) & 0.138 (0.002) \\
Public sector & $-$0.019 (0.003) & $-$0.021 (0.003) & $-$0.018 (0.003) & $-$0.019 (0.003) & $-$0.021 (0.003) & $-$0.018 (0.003) \\
[4pt]
\multicolumn{7}{l}{\textit{Industry (ref: Other services)}} \\
Manufacturing & 0.118 (0.003) & 0.127 (0.003) & 0.118 (0.003) & 0.118 (0.003) & 0.127 (0.003) & 0.117 (0.003) \\
Construction & 0.069 (0.005) & 0.076 (0.005) & 0.069 (0.005) & 0.069 (0.005) & 0.076 (0.005) & 0.068 (0.005) \\
Wholesale/retail & 0.052 (0.004) & 0.055 (0.004) & 0.052 (0.004) & 0.052 (0.004) & 0.055 (0.004) & 0.052 (0.004) \\
IT/communication & 0.035 (0.004) & 0.042 (0.004) & 0.035 (0.004) & 0.035 (0.004) & 0.042 (0.004) & 0.034 (0.004) \\
Finance/insurance & 0.209 (0.005) & 0.211 (0.005) & 0.208 (0.005) & 0.209 (0.005) & 0.212 (0.005) & 0.207 (0.005) \\
Professional services & 0.038 (0.004) & 0.043 (0.004) & 0.038 (0.004) & 0.038 (0.004) & 0.044 (0.004) & 0.038 (0.004) \\
Public administration & $-$0.096 (0.005) & $-$0.092 (0.005) & $-$0.093 (0.005) & $-$0.096 (0.005) & $-$0.091 (0.005) & $-$0.092 (0.005) \\
Education services & $-$0.029 (0.005) & $-$0.042 (0.004) & $-$0.028 (0.005) & $-$0.029 (0.005) & $-$0.041 (0.004) & $-$0.028 (0.005) \\
Health/social & 0.003 (0.005) & 0.007 (0.005) & 0.001 (0.005) & 0.003 (0.005) & 0.007 (0.005) & $-$0.001 (0.005) \\
Workplace-region dummies & Yes & Yes & Yes & Yes & Yes & Yes \\
[4pt]
\midrule
Year dummies & \multicolumn{6}{c}{Yes} \\
$N$ (selected) & \multicolumn{6}{c}{119,157} \\
$N$ (total) & \multicolumn{6}{c}{186,513} \\
\bottomrule
\end{tabular}

\vspace{0.5em}
\parbox{\textwidth}{\scriptsize\textit{Notes:} Wage equation coefficients with and without exclusion restrictions in the selection equation, estimated on the same subsample of observations with valid exclusion restriction variables. The ``with'' specification adds parental income (middle tercile), father's education, and mother's education to the selection equation. The Age$^2$ coefficient is multiplied by 100. Standard errors in parentheses. The intercept and year dummies are suppressed.}
\end{table}
\fi

\section{Details of the Robustness Analyses}\label{sec:appendix_robustness}

This appendix reports the details of three robustness analyses summarized in Section~\ref{sec:robustness}: the log monthly wage as the dependent variable, the omission of job controls, and a Heckman model with a nonlinear first stage.

\subsection{Alternative Dependent Variable}\label{sec:lwage}

The baseline dependent variable is the log hourly wage, log(monthly wage / (total weekly hours $\times$ 4.345)), where total weekly hours include regular and overtime hours. As an alternative, I use the log monthly wage and control for regular and overtime weekly hours separately. This specification compares the monthly earnings of men and women who report the same regular and overtime hours, and it allows the two types of hours to have different coefficients.

The monthly wage gap is larger than the hourly wage gap in every period and for every estimator reported in Table~\ref{tab:lwage_dynamics}. The pooled Kim-Lee estimates are $-0.065$ (monthly) and $-0.033$ (hourly). Men report more overtime than women, 5.0 against 3.3 hours per week on average, and 54\% of men against 44\% of women report any overtime. Monthly earnings rise less than proportionally with hours, with an elasticity of roughly $0.2$, so dividing earnings by total hours reduces men's relative wages.

Both gaps are smaller in 2016--2019 than in 2008--2011, the monthly gap by more. The Kim-Lee estimates fall from $-0.031$ (hourly) and $-0.072$ (monthly) in 2008--2011 to $-0.024$ and $-0.047$ in 2016--2019, by less than one log point on the hourly measure and by two and a half on the monthly measure. OLS and Kim-Lee remain close in every period for both dependent variables, so selection correction has little effect under either measure. 

\begin{table}[htbp]
\centering
\caption{Gender Wage Gap: Log Hourly Wage vs.\ Log Monthly Wage}\label{tab:lwage_dynamics}
\begin{tabular}{lrrrrrr}
\toprule
 & \multicolumn{3}{c}{Log hourly wage} & \multicolumn{3}{c}{Log monthly wage} \\
\cmidrule(lr){2-4} \cmidrule(lr){5-7}
Period & \multicolumn{1}{c}{OLS} & \multicolumn{1}{c}{Kim-Lee} & \multicolumn{1}{c}{DML-RF} & \multicolumn{1}{c}{OLS} & \multicolumn{1}{c}{Kim-Lee} & \multicolumn{1}{c}{DML-RF} \\
\midrule
Pooled & $-$0.033 & $-$0.033 & $-$0.032 & $-$0.064 & $-$0.065 & $-$0.063 \\
 & (0.003) & (0.003) & (0.003) & (0.002) & (0.003) & (0.002) \\
2008--2011 & $-$0.034 & $-$0.031 & $-$0.032 & $-$0.075 & $-$0.072 & $-$0.074 \\
 & (0.005) & (0.005) & (0.005) & (0.004) & (0.005) & (0.005) \\
2012--2015 & $-$0.037 & $-$0.043 & $-$0.038 & $-$0.066 & $-$0.074 & $-$0.069 \\
 & (0.005) & (0.005) & (0.005) & (0.004) & (0.005) & (0.004) \\
2016--2019 & $-$0.025 & $-$0.024 & $-$0.022 & $-$0.049 & $-$0.047 & $-$0.045 \\
 & (0.004) & (0.004) & (0.004) & (0.004) & (0.004) & (0.004) \\
\bottomrule
\end{tabular}

\vspace{0.5em}
\parbox{\textwidth}{\scriptsize\textit{Notes:} Female coefficients with the Table~\ref{tab:main} controls; the log monthly wage columns add regular and overtime weekly hours as separate controls. DML-RF is the selected-sample estimator of Section~\ref{sec:comparison} with a random-forest first stage. Standard errors in parentheses: heteroskedasticity-robust for OLS, first-stage-corrected for Kim-Lee, and orthogonal-moment for DML-RF.}
\end{table}

\subsection{Job Sorting}\label{sec:nojob}

Job characteristics such as firm size, public-sector status, and industry may themselves reflect gender differences in sorting across jobs. With job controls, the female coefficient compares men and women with the same observed job characteristics. Without them, it includes the part of the gap associated with job sorting. Table~\ref{tab:nojob} reports both specifications. Omitting the job controls widens the pooled Kim-Lee estimate from $-0.033$ to $-0.051$ and the OLS estimate from $-0.033$ to $-0.049$, so conditioning on job characteristics reduces the gap by about 0.017 to 0.018, roughly one-third of the no-job estimate. Within subperiods the reduction ranges from roughly 0.011 to 0.021. OLS and Kim-Lee are virtually identical in the pooled sample with and without job controls and remain close in every subperiod, so selection correction has little effect in either specification.

\begin{table}[htbp]
\centering
\caption{Gender Wage Gap with and without Job Controls}\label{tab:nojob}
\begin{tabular}{lcccc}
\toprule
 & Pooled & 2008--2011 & 2012--2015 & 2016--2019 \\
\midrule
\multicolumn{5}{l}{\textit{With job controls (large firm, public sector, industry dummies, and workplace-region dummies)}} \\
\quad OLS     & $-$0.033 (0.003) & $-$0.034 (0.005) & $-$0.037 (0.005) & $-$0.025 (0.004) \\
\quad Kim-Lee & $-$0.033 (0.003) & $-$0.031 (0.005) & $-$0.043 (0.005) & $-$0.024 (0.004) \\
\quad DML-RF  & $-$0.032 (0.003) & $-$0.032 (0.005) & $-$0.038 (0.005) & $-$0.022 (0.004) \\
\addlinespace
\multicolumn{5}{l}{\textit{Without job controls (age, GPA, education, school area, major, married, and year dummies)}} \\
\quad OLS     & $-$0.049 (0.003) & $-$0.055 (0.005) & $-$0.048 (0.005) & $-$0.040 (0.004) \\
\quad Kim-Lee & $-$0.051 (0.003) & $-$0.051 (0.005) & $-$0.057 (0.005) & $-$0.039 (0.004) \\
\quad DML-RF  & $-$0.052 (0.003) & $-$0.055 (0.005) & $-$0.053 (0.005) & $-$0.041 (0.004) \\
\midrule
$N$ & 127{,}628 & 45{,}406 & 41{,}640 & 40{,}582 \\
\bottomrule
\end{tabular}

\vspace{0.5em}
\parbox{\textwidth}{\scriptsize\textit{Notes:} Female coefficients on log hourly wages; the panel headings list the controls. DML-RF is the \cite{pan2024} estimator with a random-forest first stage, in the selected-sample extension of Section~\ref{sec:comparison} when job controls are included. Standard errors in parentheses: heteroskedasticity-robust for OLS, first-stage-corrected for Kim-Lee, and orthogonal-moment for DML-RF.}
\end{table}

The last row of each panel reports the debiased machine-learning estimator of \cite{pan2024} with a random-forest first stage (DML-RF), whose orthogonalized moment accommodates a fully nonparametric, non-smooth first stage. Without job controls every covariate is observed for all graduates, which is the whole-sample setting of \cite{pan2024}. With job controls I use the selected-sample extension defined in Section~\ref{sec:comparison}. DML-RF tracks OLS and Kim-Lee closely in both panels: $-0.052$ against $-0.049$ and $-0.051$ without job controls, and $-0.032$ against $-0.033$ and $-0.033$ with them. It differs from OLS by at most 0.005 in any cell (2012--2015 without job controls), within sampling error, and shows no systematic wedge across subperiods. The agreement between DML-RF, with its random-forest first stage, and Kim-Lee, with its sieve-probit first stage, supports the conclusion that selection correction has little effect on the entry-level gap.

\subsection{Heckman with a Nonlinear First Stage}\label{sec:appendix_heckman_fs}

Table~\ref{tab:heckman_fs} compares the Heckman MLE of Table~\ref{tab:main}, whose selection index is linear apart from age squared, with a Heckman MLE whose index is quadratic in age and GPA, with every term also interacted with the female indicator. Both specifications retain joint normality. With the quadratic index the pooled estimate moves from $-0.028$ to $-0.033$, and the estimates are within 0.6 log points of Kim-Lee's in every period. The estimated error correlation with the linear index is 0.71 in the pooled sample and 0.70 in 2016--2019 but $-0.02$ in 2008--2011 and $-0.01$ in 2012--2015, and it lies between $-0.16$ and $-0.03$ with the quadratic index in every sample.

\begin{table}[htbp]
\centering
\caption{Heckman Estimates by First-Stage Specification}\label{tab:heckman_fs}
{\footnotesize\setlength{\tabcolsep}{4pt}
\begin{tabular}{lrrrr}
\toprule
First-stage index & \multicolumn{1}{c}{Pooled} & \multicolumn{1}{c}{2008--2011} & \multicolumn{1}{c}{2012--2015} & \multicolumn{1}{c}{2016--2019} \\
\midrule
\multicolumn{5}{l}{\textit{Panel A: Female coefficient}} \\
Heckman, linear (Table~\ref{tab:main}) & $-$0.028 & $-$0.034 & $-$0.037 & $-$0.024 \\
 & (0.003) & (0.005) & (0.005) & (0.004) \\
Heckman, quadratic with female interactions & $-$0.033 & $-$0.033 & $-$0.037 & $-$0.025 \\
 & (0.003) & (0.005) & (0.005) & (0.004) \\
Kim-Lee, sieve probit (Table~\ref{tab:main}) & $-$0.033 & $-$0.031 & $-$0.043 & $-$0.024 \\
 & (0.003) & (0.005) & (0.005) & (0.004) \\
\midrule
\multicolumn{5}{l}{\textit{Panel B: Heckman error correlation $\rho$}} \\
Linear & $+$0.71 & $-$0.02 & $-$0.01 & $+$0.70 \\
Quadratic with female interactions & $-$0.12 & $-$0.11 & $-$0.03 & $-$0.16 \\
\bottomrule
\end{tabular}
}

\vspace{0.5em}
\parbox{\textwidth}{\scriptsize\textit{Notes:} Female coefficients on log hourly wages with the Table~\ref{tab:main} controls; Heckman rows are maximum likelihood estimates. The selection index contains the wage controls except the job characteristics. The quadratic index replaces age, age$^2$, and GPA with age, GPA, their squares, and their product, each also interacted with the female indicator. Standard errors in parentheses: maximum likelihood for Heckman and first-stage-corrected for Kim-Lee.}
\end{table}

\section{Sensitivity to the Sieve Dimensions}\label{sec:appendix_flexible}

To check sensitivity to the first-stage dimension, I drop the interactions of the age and GPA splines with school type, school area, and major from the baseline sieve probit of Section~\ref{sec:first_stage}, keeping the marginal splines, their tensor product, and their interactions with gender. This reduces the pooled basis dimension from 204 to 66 and moves the female coefficient by at most 0.002 in any period (Table~\ref{tab:flexible_gam}).

\begin{table}[htbp]
\centering
\caption{Kim-Lee Estimates: Baseline vs.\ Parsimonious First Stage}\label{tab:flexible_gam}
\begin{tabular}{lrr}
\toprule
Period & \multicolumn{1}{c}{Baseline ($K = 204$)} & \multicolumn{1}{c}{Parsimonious ($K = 66$)} \\
\midrule
Pooled & $-$0.033 (0.003) & $-$0.035 (0.003) \\
2008--2011 & $-$0.031 (0.005) & $-$0.031 (0.005) \\
2012--2015 & $-$0.043 (0.005) & $-$0.043 (0.005) \\
2016--2019 & $-$0.024 (0.004) & $-$0.024 (0.004) \\
\bottomrule
\end{tabular}

\vspace{0.5em}
\parbox{\textwidth}{\scriptsize\textit{Notes:} Kim-Lee female coefficients on log hourly wages with the Table~\ref{tab:main} controls. ``Baseline'' is the sieve probit first stage of Section~\ref{sec:first_stage}; ``Parsimonious'' drops its interactions with school type, school area, and major. $K$ is the number of basis functions in the pooled sample (smaller in the subperiods). First-stage-corrected standard errors in parentheses.}
\end{table}

\begin{table}[htbp]
\centering
\caption{Sensitivity to Second-Stage B-Spline Flexibility}\label{tab:bs_df}
\begin{tabular}{lrrrr}
\toprule
 & \multicolumn{4}{c}{B-spline degrees of freedom in $\hat{p}$} \\
\cmidrule(lr){2-5}
Period & \multicolumn{1}{c}{5 (baseline)} & \multicolumn{1}{c}{7} & \multicolumn{1}{c}{10} & \multicolumn{1}{c}{15} \\
\midrule
Pooled     & $-$0.033 (0.003) & $-$0.033 (0.003) & $-$0.033 (0.003) & $-$0.033 (0.003) \\
2008--2011 & $-$0.031 (0.005) & $-$0.031 (0.005) & $-$0.031 (0.005) & $-$0.032 (0.005) \\
2012--2015 & $-$0.043 (0.005) & $-$0.042 (0.005) & $-$0.042 (0.005) & $-$0.042 (0.005) \\
2016--2019 & $-$0.024 (0.004) & $-$0.024 (0.004) & $-$0.024 (0.004) & $-$0.024 (0.004) \\
\bottomrule
\end{tabular}

\vspace{0.5em}
\parbox{\textwidth}{\scriptsize\textit{Notes:} Kim-Lee female coefficients on log hourly wages with the Table~\ref{tab:main} controls and the baseline sieve probit first stage. Columns vary the degrees of freedom of the second-stage cubic B-spline in $\hat{p}$. First-stage-corrected standard errors in parentheses.}
\end{table}

The second-stage spline dimension is also immaterial. The baseline approximates $\lambda_0(\hat{p})$ with a cubic B-spline with 5 degrees of freedom, with two interior knots at the 33rd and 67th percentiles of $\hat{p}$. Table~\ref{tab:bs_df} raises the dimension to 7, 10, and 15 degrees of freedom. The Kim-Lee female coefficient moves by less than 0.001 in every period.

\section{Sensitivity to the Age Cutoff}\label{sec:appendix_age}

The analysis sample is restricted to graduates aged 35 or younger. Table~\ref{tab:age_sensitivity} re-estimates the pooled OLS and Kim-Lee female coefficients with upper age cutoffs of 30, 32, and 35 and with no cutoff. Raising the cutoff widens the OLS gap from $-0.029$ at a cutoff of 30 to $-0.055$ with no cutoff, about two and a half log points, most of it when graduates older than 35 enter the sample. The difference between OLS and Kim-Lee stays within $\pm 0.005$ at every cutoff, so the cutoff changes the estimated gap through sample composition rather than through selection.

\begin{table}[htbp]
\centering
\caption{Sensitivity to the Age Cutoff}\label{tab:age_sensitivity}
\begin{tabular}{lrrrrr}
\toprule
Age cutoff & \multicolumn{1}{c}{$N$ (total)} & \multicolumn{1}{c}{$N$ (wage)} & \multicolumn{1}{c}{OLS} & \multicolumn{1}{c}{Kim-Lee} & \multicolumn{1}{c}{OLS $-$ KL} \\
\midrule
$\leq 30$ & 193,506 & 121,854 & $-$0.029 (0.003) & $-$0.033 (0.003) & $+$0.004 \\
$\leq 32$ & 199,568 & 125,887 & $-$0.031 (0.003) & $-$0.032 (0.003) & $+$0.001 \\
$\leq 35$ & 202,114 & 127,628 & $-$0.033 (0.003) & $-$0.033 (0.003) & 0.000 \\
Unrestricted & 211,439 & 133,025 & $-$0.055 (0.003) & $-$0.050 (0.003) & $-$0.004 \\
\bottomrule
\end{tabular}

\vspace{0.5em}
\parbox{\textwidth}{\scriptsize\textit{Notes:} Pooled 2008--2019 female coefficients on log hourly wages with the controls of Table~\ref{tab:main}; each row applies the upper age cutoff shown, and ``Unrestricted'' applies none. $N$ (wage) counts full-time wage workers with valid wages. Standard errors in parentheses: heteroskedasticity-robust for OLS and first-stage-corrected for Kim-Lee.}
\end{table}

\section{Subsample Robustness}\label{sec:appendix_subsamples}

Table~\ref{tab:subsample_robust} reports OLS and Kim-Lee estimates, pooled and by subperiod, for three subsamples: graduates aged 30 or younger, graduates older than 35 (non-traditional mature students), and four-year university graduates.

\begin{table}[htbp]
\centering
\caption{Subsample Robustness: OLS and Kim-Lee Estimates}\label{tab:subsample_robust}
\small
\begin{tabular}{lrrrrr}
\toprule
Subsample / Period & \multicolumn{1}{c}{$N$} & \multicolumn{1}{c}{$N$ (wage)} & \multicolumn{1}{c}{OLS} & \multicolumn{1}{c}{Kim-Lee} & \multicolumn{1}{c}{OLS $-$ KL} \\
\midrule
\multicolumn{6}{l}{\textit{Age $\leq$ 30}} \\
Pooled     & 193,506 & 121,854 & $-$0.029 & $-$0.033 & $+$0.004 \\
2008--2011 &  62,458 &  42,923 & $-$0.026 & $-$0.026 & $+$0.000 \\
2012--2015 &  64,126 &  39,745 & $-$0.037 & $-$0.047 & $+$0.010 \\
2016--2019 &  66,922 &  39,186 & $-$0.023 & $-$0.023 & $-$0.000 \\
\midrule
\multicolumn{6}{l}{\textit{Age $>$ 35 (non-traditional)}} \\
Pooled     &   9,325 &   5,397 & $-$0.323 & $-$0.372 & $+$0.050 \\
2008--2011 &   4,564 &   2,773 & $-$0.336 & $-$0.357 & $+$0.021 \\
2012--2015 &   2,809 &   1,577 & $-$0.342 & $-$0.351 & $+$0.009 \\
2016--2019 &   1,952 &   1,047 & $-$0.251 & $-$0.261 & $+$0.010 \\
\midrule
\multicolumn{6}{l}{\textit{Four-year university only}} \\
Pooled     & 151,567 &  94,806 & $-$0.030 & $-$0.030 & $-$0.000 \\
2008--2011 &  47,595 &  32,420 & $-$0.026 & $-$0.024 & $-$0.002 \\
2012--2015 &  49,109 &  30,188 & $-$0.037 & $-$0.043 & $+$0.006 \\
2016--2019 &  54,863 &  32,198 & $-$0.025 & $-$0.023 & $-$0.003 \\
\bottomrule
\end{tabular}

\vspace{0.5em}
\parbox{\textwidth}{\scriptsize\textit{Notes:} Female coefficients on log hourly wages, estimated separately for each subsample and period with the wage controls of Table~\ref{tab:main}. For the age $> 35$ subsample, the Kim-Lee first stage omits the covariate-by-spline interaction block (see text). $N$ counts all graduates and $N$ (wage) full-time wage workers with valid wages.}
\end{table}

In the age $\leq 30$ and four-year subsamples, OLS and Kim-Lee differ by at most $0.005$ in every period except 2012--2015, where the difference is $0.010$ among graduates aged 30 or younger and $0.006$ among four-year graduates, so selection correction remains small under both restrictions. The four-year estimates are within half a log point of the baseline estimates of Table~\ref{tab:main} in the pooled sample and the two later subperiods but about 0.8 log points smaller in 2008--2011, so among four-year graduates the gap does not narrow between 2008--2011 and 2016--2019. 

The age $> 35$ subsample is the exception and needs care. It is small, with as few as about one thousand wage workers in a subperiod, so I drop the covariate-by-spline interactions and keep the marginal age and GPA splines, their tensor product, and their interactions with gender, which holds the sieve dimension well below the sample size.\footnote{With the full sieve used for the large samples, the first-stage probit has roughly 190 parameters, about five employed observations per parameter in the smallest subperiod. Dropping the covariate-by-spline interactions raises the ratio to roughly twenty employed observations per parameter while preserving the marginal nonlinearity that identifies the correction.} Under this specification, OLS and Kim-Lee differ by $0.050$ in the pooled mature subsample, a nontrivial correction, but by $0.021$ in 2008--2011 and by only $0.009$ and $0.010$ in the two later subperiods. The correction is also sensitive to first-stage flexibility, ranging from about $0.01$ under the saturated sieve to about $0.07$ under coarser marginal splines, possibly due to the small sample size.

Table~\ref{tab:subsample_oaxaca} repeats the Oaxaca-Blinder decomposition of Section~\ref{sec:oaxaca} in each subsample. The age $\leq 30$ and four-year decompositions resemble the baseline: endowments explain about 60\% of the raw gap, and age and job characteristics are the largest components. The age $> 35$ subsample differs. Its pooled raw gap is 47 log points, of which observables explain only 33\%, and its age component is negative because women in this subsample are slightly older than men. This pattern is consistent with the lifecycle accumulation of gender disparities among older workers.

\begin{table}[htbp]
\centering
\caption{Oaxaca-Blinder Decomposition by Subsample}\label{tab:subsample_oaxaca}
\small
\setlength{\tabcolsep}{4.5pt}
\begin{tabular}{lrrrrrrr}
\toprule
Subsample / Period & \multicolumn{1}{c}{Raw gap} & \multicolumn{1}{c}{Endow} & \multicolumn{1}{c}{Endow\%} & \multicolumn{1}{c}{Age} & \multicolumn{1}{c}{GPA} & \multicolumn{1}{c}{Educ./major} & \multicolumn{1}{c}{Job} \\
\midrule
\multicolumn{8}{l}{\textit{Age $\leq 30$}} \\
Pooled     & 0.116 & 0.072 & 62.1\% & $+$0.054 & $-$0.005 & $-$0.016 & $+$0.042 \\
2008--2011 & 0.136 & 0.103 & 75.6\% & $+$0.085 & $-$0.008 & $-$0.017 & $+$0.043 \\
2012--2015 & 0.121 & 0.067 & 55.4\% & $+$0.046 & $-$0.006 & $-$0.011 & $+$0.037 \\
2016--2019 & 0.094 & 0.060 & 64.1\% & $+$0.039 & $-$0.004 & $-$0.017 & $+$0.043 \\
\midrule
\multicolumn{8}{l}{\textit{Age $> 35$ (non-traditional)}} \\
Pooled     & 0.471 & 0.155 & 32.9\% & $-$0.009 & $-$0.006 & $+$0.056 & $+$0.112 \\
2008--2011 & 0.480 & 0.141 & 29.4\% & $-$0.012 & $+$0.002 & $+$0.043 & $+$0.111 \\
2012--2015 & 0.490 & 0.154 & 31.4\% & $-$0.008 & $-$0.007 & $+$0.066 & $+$0.106 \\
2016--2019 & 0.397 & 0.157 & 39.6\% & $-$0.016 & $-$0.006 & $+$0.076 & $+$0.097 \\
\midrule
\multicolumn{8}{l}{\textit{Four-year university only}} \\
Pooled     & 0.114 & 0.067 & 59.1\% & $+$0.052 & $-$0.007 & $-$0.020 & $+$0.048 \\
2008--2011 & 0.139 & 0.103 & 73.6\% & $+$0.077 & $-$0.010 & $-$0.022 & $+$0.056 \\
2012--2015 & 0.116 & 0.067 & 57.9\% & $+$0.047 & $-$0.010 & $-$0.015 & $+$0.043 \\
2016--2019 & 0.096 & 0.062 & 64.0\% & $+$0.042 & $-$0.006 & $-$0.017 & $+$0.043 \\
\midrule
\multicolumn{8}{l}{\textit{Baseline (age $\leq 35$, all)}} \\
Pooled     & 0.122 & 0.077 & 63.3\% & $+$0.058 & $-$0.006 & $-$0.015 & $+$0.042 \\
\bottomrule
\end{tabular}

\vspace{0.5em}
\parbox{\textwidth}{\scriptsize\textit{Notes:} Two-part decomposition as in Table~\ref{tab:oaxaca}, estimated separately for each subsample and period with that sample's pooled Kim-Lee coefficients as the reference. ``Endow'' is the endowments component and ``Endow\%'' its share of the raw gap; ``Age,'' ``GPA,'' ``Educ./major,'' and ``Job'' are its components by variable group, and the married and year components are included in ``Endow'' but not shown. The baseline row repeats the pooled column of Table~\ref{tab:oaxaca}.}
\end{table}

\section{Generalized Lee Bounds}\label{sec:appendix_semenova}

The age-conditional Lee bounds of Section~\ref{sec:main_estimates} allow the direction of selection to vary with a single covariate. Here I let it vary with the entire pre-market covariate vector, using the conditional-monotonicity partition of \cite{semenova2023generalized}. This exercise investigates whether both selection directions remain present under this richer partition and how the resulting intervals compare with the unconditional and age-conditional bounds.

I estimate the selection equation by post-lasso logit. The lasso estimator selects regressors from about eighty predetermined terms and their interactions with the female indicator. The terms include age, age$^2$, GPA, GPA$^2$, four-year status, and school-area, major, and year dummies, together with interactions of age, age$^2$, GPA, and four-year status with major, of age with school area, and of age with four-year status and GPA. Marital status is excluded because it is measured at the survey date and can change after labor market entry. For each graduate, the ratio $\hat{p}_0(X) = \hat{s}_0(X)/\hat{s}_1(X)$ of the predicted employment probabilities of men and women classifies the cell as one where being female raises employment ($\hat{p}_0 \le 1$, ``female helps'') or lowers it ($\hat{p}_0 > 1$, ``female hurts''). Within each direction group, I compute the \cite{lee2009} bounds, trimming the over-represented gender by the group's own selection rates. I then aggregate the two group bounds by their always-taker \emph{mass} over the full covariate distribution. The always-taker probability in a cell is $\min(\hat{s}_0(X),\hat{s}_1(X))$, and a group's weight is the sum of these probabilities over its members.\footnote{Weighting by always-taker mass accounts for the relative size of the two regions. This matters when the regions are unbalanced. In 2012--2015, where three-quarters of graduates fall in ``female helps'' cells, mass weighting moves the lower bound from $-0.197$ to $-0.159$.} The 95\% confidence regions come from a bootstrap over graduates (200 replications) that re-estimates the lasso in every replication. I compare the bounds with the no-job Kim-Lee estimates of Section~\ref{sec:nojob}. That specification includes the marriage indicator, but only about 4\% of recent graduates are married, so the reference estimates are insensitive to the difference. Table~\ref{tab:semenova} reports the results.

Both selection directions are present in every period. The share of graduates in ``female helps'' cells ranges from 0.19 to 0.76 and is never close to zero or one. A single monotonicity direction, which the unconditional bounds of \cite{lee2009} require, therefore does not hold even after conditioning on the full pre-market covariate vector, so its failure is not specific to the age margin. The generalized bounds are also not nested within the unconditional bounds. Their upper bounds lie above the unconditional upper bounds in every period, by $0.003$ in 2012--2015 and by as much as $0.086$ in 2008--2011, the same non-nesting that \cite{semenova2023generalized} reports in her Job Corps application. The full-covariate partition is less stable near balanced-propensity cells ($\hat{p}_0 \approx 1$), where small changes in the employment probabilities can change the estimated trimming direction.

\begin{table}[htbp]
\centering
\caption{Generalized Lee Bounds with a High-Dimensional Lasso First Stage}\label{tab:semenova}
{\footnotesize\setlength{\tabcolsep}{3.5pt}
\begin{tabular}{lrrrr}
\toprule
 & \multicolumn{1}{c}{Pooled} & \multicolumn{1}{c}{2008--2011} & \multicolumn{1}{c}{2012--2015} & \multicolumn{1}{c}{2016--2019} \\
\midrule
Unconditional Lee                   & $[-0.175, -0.071]$ & $[-0.191, -0.094]$ & $[-0.160, -0.090]$ & $[-0.156, -0.038]$ \\
Generalized (lasso)                 & $[-0.169, -0.055]$ & $[-0.131, -0.008]$ & $[-0.159, -0.087]$ & $[-0.160, -0.030]$ \\
\quad 95\% CI                       & $(-0.182, -0.039)$ & $(-0.149, +0.020)$ & $(-0.187, -0.066)$ & $(-0.187, -0.016)$ \\
Age-conditional (Table~\ref{tab:main}) & $[-0.080, +0.002]$ & $[-0.085, +0.015]$ & $[-0.073, -0.015]$ & $[-0.067, +0.007]$ \\
\midrule
No-job Kim-Lee                      & $-0.051$ & $-0.051$ & $-0.057$ & $-0.039$ \\
Share ``female helps''              & $0.47$ & $0.40$ & $0.76$ & $0.19$ \\
\bottomrule
\end{tabular}
}

\vspace{0.5em}
\parbox{\textwidth}{\scriptsize\textit{Notes:} Bounds on the male-female contrast in log hourly wages, no-job specification. The unconditional and generalized bounds follow \cite{lee2009}; the generalized bounds use the lasso partition described in the text. Brackets are estimated bounds and parentheses 95\% bootstrap confidence regions for the generalized bounds. Share ``female helps'' is the fraction of graduates in cells where being female raises the fitted employment probability ($\hat{p}_0(X) \le 1$).}
\end{table}

\section{Supplementary KLIPS Results}\label{sec:appendix_klips_adjusted}

\textit{Field of study.} Each person-year is assigned the undergraduate field of the degree held in that year, using the KLIPS education histories. Department codes recorded in the 1998--2000 waves are converted to fields with the official classification of the Korean Educational Development Institute distributed with KLIPS, and a program the person enrolled in after the observation year is never used to assign that year's major field. The field is identified for 94\% of college-educated person-years. The remaining 6\% enter with a separate indicator, as do the 2.5\% whose field lies outside the GOMS groups. Dropping the unidentified person-years changes the Kim-Lee estimates in Panel A of Table~\ref{tab:klips} by at most $0.004$.

\begin{figure}[htbp]
\centering
\includegraphics[width=\textwidth]{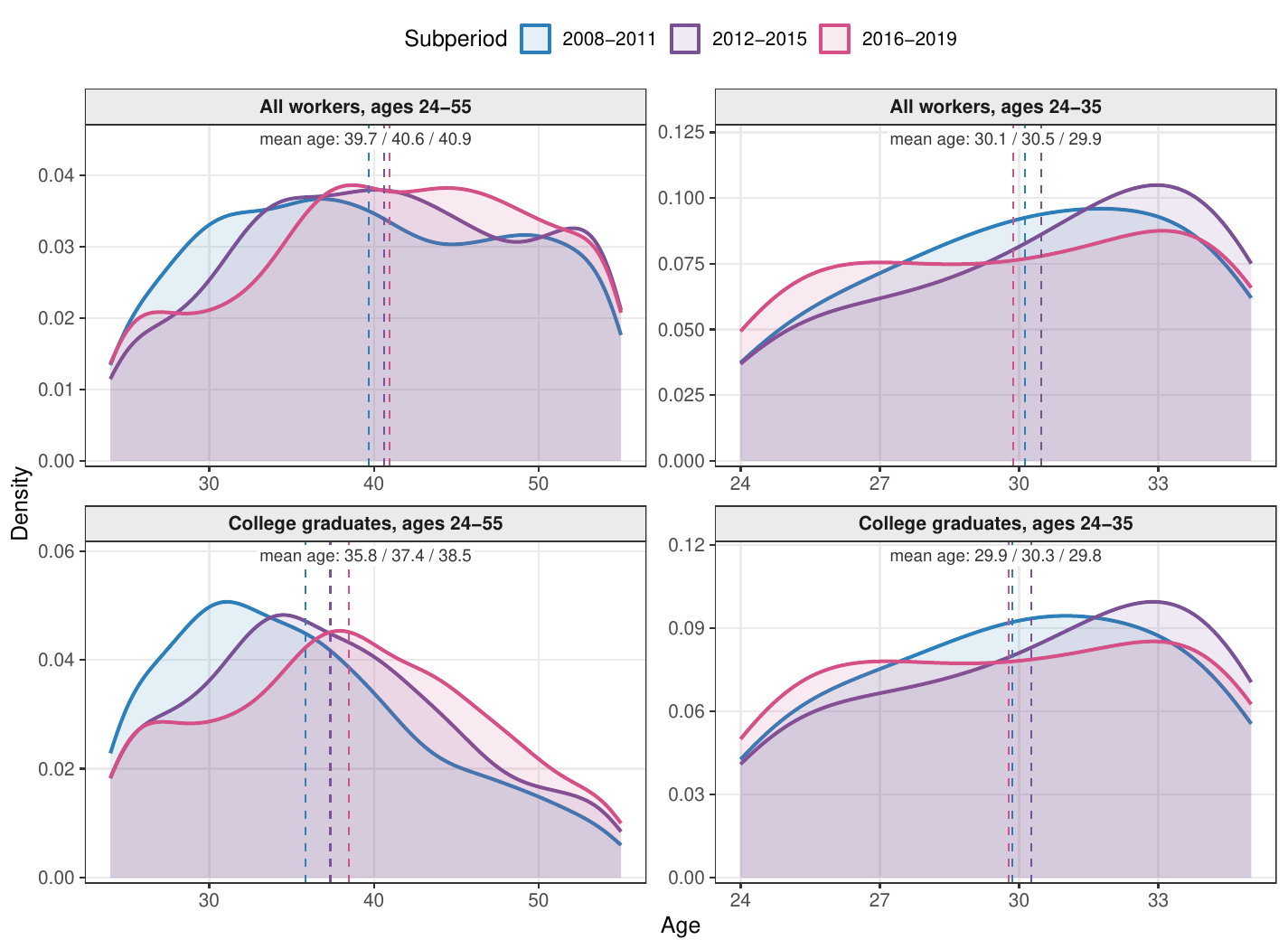}
\caption{Age distribution by subperiod, KLIPS 2008--2019. Kernel densities of age for all workers (top row) and college graduates (bottom row), ages 24 to 55 (left) and 24 to 35 (right). Dashed vertical lines mark each subperiod's mean age.}\label{fig:age_dist}
\end{figure}

\textit{Subperiod estimates at a common age.} Panel B of Table~\ref{tab:klips} reports the KLIPS subperiod gaps with the female indicator entering the wage equation without an age interaction. The KLIPS sample grows older over the survey window (Figure~\ref{fig:age_dist}), so the subperiod comparison in Panel B partly reflects the changing age composition of the prime-age sample. I interact age with the female indicator and report the gap at a common age, the pooled mean of 40.4, in every subperiod (Table~\ref{tab:klips_adjusted}). Under a linear age profile of the gap, this equals the gap averaged over the pooled age distribution. Evaluated at this age, the gap among all workers narrows across the three subperiods under all four estimators, from $-0.325$ in 2008--2011 to $-0.247$ in 2016--2019 for Kim-Lee. I report the adjustment for all four estimators. Among all workers, DML-RF with the female-age interaction added to its wage equation stays within about $0.04$ of Kim-Lee in every subperiod.

\begin{table}[htbp]
\centering
\caption{KLIPS Gender Wage Gap by Subperiod, Evaluated at Age 40.4}\label{tab:klips_adjusted}
\scriptsize
\setlength{\tabcolsep}{4.5pt}
\renewcommand{\arraystretch}{0.95}
\begin{tabular}{l rrrr rrrr}
\toprule
 & \multicolumn{4}{c}{All workers} & \multicolumn{4}{c}{College graduates} \\
\cmidrule(lr){2-5} \cmidrule(lr){6-9}
 & \multicolumn{1}{c}{OLS} & \multicolumn{1}{c}{Heckman} & \multicolumn{1}{c}{Kim-Lee} & \multicolumn{1}{c}{DML-RF}
 & \multicolumn{1}{c}{OLS} & \multicolumn{1}{c}{Heckman} & \multicolumn{1}{c}{Kim-Lee} & \multicolumn{1}{c}{DML-RF} \\
\midrule
2008--2011 & $-$0.402 & $-$0.320 & $-$0.325 & $-$0.299 & $-$0.316 & $-$0.319 & $-$0.269 & $-$0.232 \\
 & (0.012) & (0.018) & (0.017) & (0.016) & (0.024) & (0.024) & (0.033) & (0.030) \\
2012--2015 & $-$0.368 & $-$0.300 & $-$0.316 & $-$0.275 & $-$0.302 & $-$0.209 & $-$0.272 & $-$0.228 \\
 & (0.011) & (0.018) & (0.018) & (0.015) & (0.019) & (0.028) & (0.030) & (0.024) \\
2016--2019 & $-$0.295 & $-$0.288 & $-$0.247 & $-$0.212 & $-$0.255 & $-$0.253 & $-$0.195 & $-$0.163 \\
 & (0.009) & (0.010) & (0.013) & (0.011) & (0.013) & (0.013) & (0.020) & (0.016) \\
\bottomrule
\end{tabular}

\vspace{0.5em}
\parbox{\textwidth}{\scriptsize\textit{Notes:} Female coefficients on log hourly wages evaluated at age 40.4, the pooled mean age, from wage equations that interact age with the female indicator; KLIPS, ages 24--55. Other covariates and samples as in Panel B of Table~\ref{tab:klips}. Standard errors clustered by individual in parentheses.}
\end{table}

\textit{Oaxaca-Blinder decomposition.} The two-part decomposition of the pooled KLIPS gap uses the pooled Kim-Lee coefficients as the reference. Differences in observed endowments account for $0.090$ of the $0.426$ raw gap, about 21\%: job characteristics contribute $0.044$, education $0.040$, age $0.009$, marriage $0.006$, residence region $0.002$, field of study $-0.005$, and survey year $-0.006$.

\section{Unconditional Quantile Effects}\label{sec:appendix_dr}

\begin{figure}[!htbp]
\centering
\includegraphics[width=\textwidth]{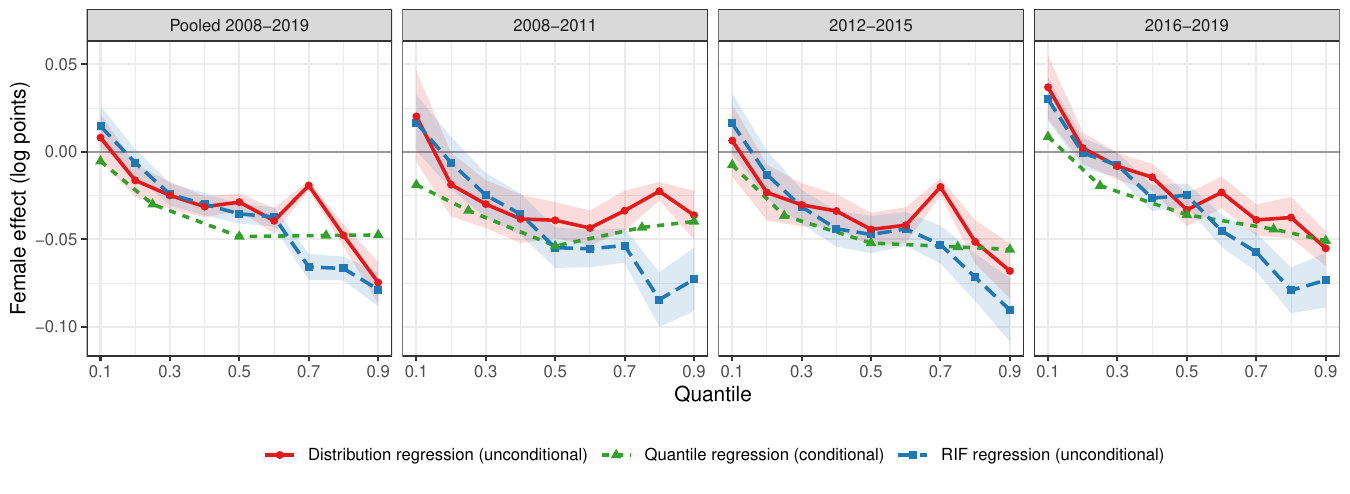}
\caption{Unconditional and conditional gender gaps across the wage distribution, GOMS. Distribution regression: women's unconditional quantile minus the counterfactual quantile with the female indicator switched off. RIF regression: unconditional quantile partial effect of the female indicator. Quantile regression: uncorrected conditional coefficient. All three use the Table~\ref{tab:main} covariates. Shaded bands around the distribution-regression and RIF estimates: 95\% confidence intervals from a bootstrap over graduates (200 replications).}\label{fig:unconditional_dr}
\end{figure}

Figure~\ref{fig:unconditional_dr} reports three sets of estimates, pooled and by subperiod: the RIF estimates summarized in Section~\ref{sec:quantile}, distribution-regression differences between women's quantiles and their counterfactual quantiles, and uncorrected conditional quantile regression coefficients. None of the three is selection-corrected. All describe the wage distribution among observed workers.

\begin{figure}[!htbp]
\centering
\includegraphics[width=\textwidth]{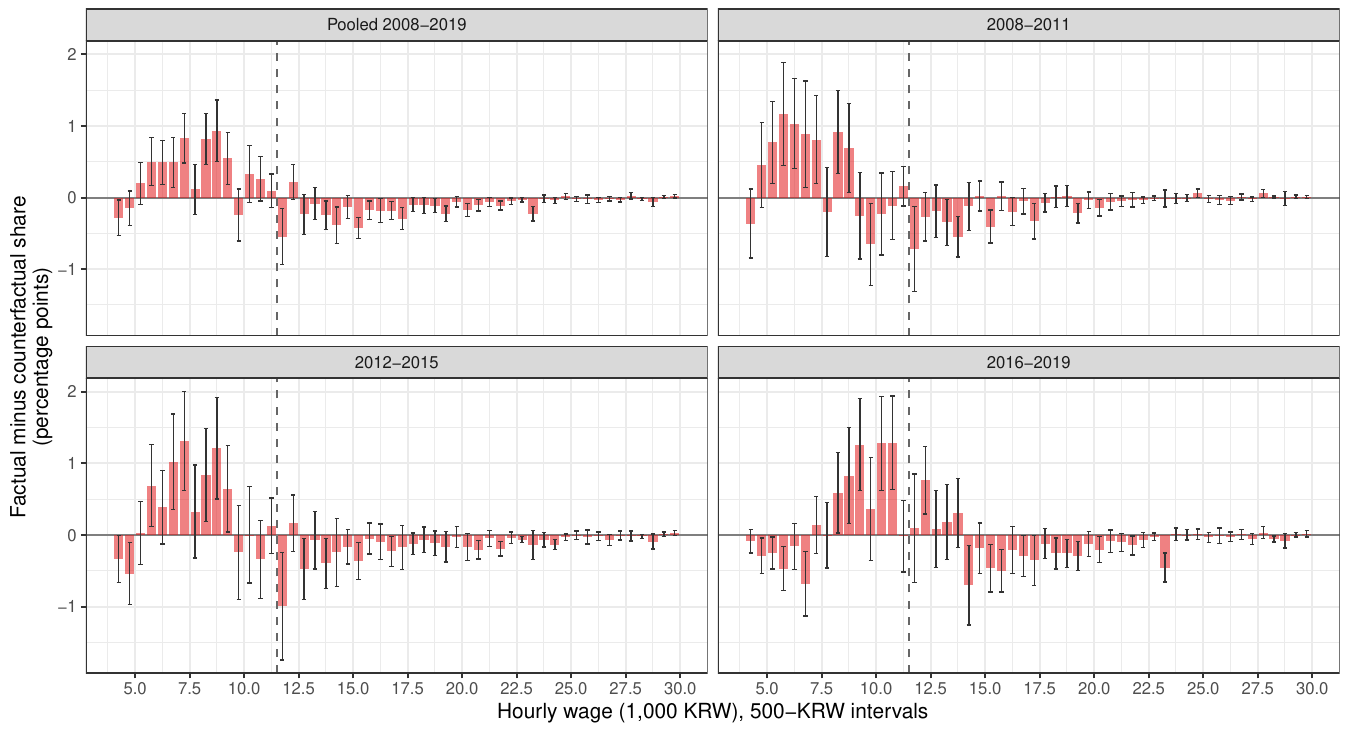}
\caption{Distribution-regression effects on women's hourly wage distribution, pooled and by subperiod, GOMS. Each bar shows the observed share of employed women in a 500-KRW wage interval minus the predicted share when the female indicator is set to zero and all other covariates are held fixed, in percentage points, plotted at the interval midpoint; the interval below 4,000 KRW and the open interval above 30,000 KRW are omitted. Whiskers are 95\% bootstrap confidence intervals as in Figure~\ref{fig:unconditional_dr}. The dashed line marks 11,507 KRW, the hourly wage of 2,000,000 KRW a month for a 40-hour week.}\label{fig:dr_bins}
\end{figure}

I use distribution regression \citep{chernozhukov2013inference} to compare women's observed wage distribution with the distribution that the model predicts for men with the same observed characteristics. At each hourly-wage threshold, I estimate a logit regression for whether a worker earns at or below that threshold, using employed men and women together and the covariates in Table~\ref{tab:main}. For each employed woman, I then predict this probability twice: once with the female indicator equal to one and once with it set to zero, keeping all other covariates unchanged. Averaging the first set of predictions reproduces women's observed distribution, which I call the factual distribution. Averaging the second gives the counterfactual distribution. Both distributions therefore use women's own observed characteristics, so their difference does not reflect differences in the composition of male and female workers.

Figure~\ref{fig:dr_bins} shows where these two distributions differ. Subtracting the cumulative probabilities at adjacent thresholds gives the share of women in each 500-KRW wage interval. Each bar reports the factual share minus the counterfactual share, in percentage points. A positive bar means that more women earn wages in that interval than the model predicts for men with the same characteristics, and a negative bar means fewer do. These interval shares are useful because many graduates have exactly the same hourly wage. GOMS records monthly pay and weekly hours, and round monthly salaries (81\% are multiples of 100,000 KRW) combined with common working hours (about a third of employed graduates work exactly 40 hours a week) produce mass points in the hourly-wage distribution. The largest, 11,507 KRW, the hourly rate of 2,000,000 KRW for a 40-hour week, holds 7\% of employed graduates, and the twelve largest together hold 29\%. When a quantile falls on a mass point, a change in the share of women earning that wage moves the quantile very little. Following \cite{biewen2026distribution}, I therefore examine the shares in wage intervals alongside the quantile comparisons in Figure~\ref{fig:unconditional_dr}.

The thresholds run from 4,000 to 30,000 KRW in steps of 500 KRW, a grid on which no mass point falls on a threshold, and the lowest interval runs from 2,000 KRW, half of the 2009 minimum wage that applies to the 2008 wave and the lowest wage the sample admits, to 4,000 KRW. I monotonize the counterfactual distribution by rearrangement and compute the quantile effects in Figure~\ref{fig:unconditional_dr} by linear interpolation within wage intervals, as in \cite{biewen2026distribution}. Confidence intervals come from a bootstrap over graduates (200 replications) that re-estimates every regression.

In the pooled sample, women are more concentrated in lower wage intervals and less likely to earn high wages than the counterfactual predicts (Figure~\ref{fig:dr_bins}). The share earning at most 11,500 KRW is 4.3 percentage points higher (standard error 0.3), while the share earning more than 20,000 KRW is 1.0 point lower (0.1). The pattern at the bottom of the distribution differs across periods. In 2008--2011, the share of women earning at most 7,500 KRW exceeds the counterfactual by 3.5 percentage points (0.6). In 2016--2019, however, women are less likely than the counterfactual to earn at most 7,000 KRW, by 2.1 points (0.3), and the excess builds between 9,000 and 14,000 KRW, raising the cumulative difference to 5.1 points (0.5) at 14,000 KRW. Their share earning more than 20,000 KRW is 1.4 points lower (0.2), against 0.4 points (0.2) in 2008--2011. Women in the later period are therefore concentrated in the middle of the distribution, with fewer at either end than the counterfactual predicts. Their shortfall at high wages has grown even as they have become less likely to be among the lowest-paid workers.

Distribution regression and RIF regression give a similar picture in the pooled sample and the two later subperiods: women have no wage disadvantage at the bottom of the distribution and the largest disadvantage at the top (Figure~\ref{fig:unconditional_dr}). The two methods differ by less than one log point at the 10th percentile in every sample, and they put the pooled gap at the 90th percentile at 7.5 and 7.9 log points. In 2016--2019, both estimate a female advantage at the bottom and a disadvantage at the top. They agree less closely in 2008--2011, when distribution regression gives a flatter profile and a 90th-percentile gap of 3.6 log points, compared with 7.3 from RIF regression. The estimates differ more noticeably near the two largest mass points above the median, 11,507 and 14,384 KRW, which hold women's 70th or 80th percentile depending on the sample; there the differences reach 3.3--6.2 log points. At these wage levels, the interval shares in Figure~\ref{fig:dr_bins} show distributional differences that the quantile estimates can obscure. The RIF estimates also depend on the bandwidth used to estimate the wage density, because the effect at a quantile is scaled by the inverse of the estimated density there. Halving or doubling the baseline bandwidth changes the estimates by at most 1.4 log points, with the largest changes between the 60th and 80th percentiles. Changes are smaller at the two ends: less than half a log point at the 10th percentile and at most 1.2 log points at the 90th. The bandwidth therefore matters most where the large mass points lie and least at the 10th percentile.

\section{Excluded Variables and the Quantile Selection Estimator}\label{sec:appendix_iv}

\textit{Excluded variables.} In GOMS the candidate excluded variables come from the family block of the questionnaire, which every graduate answers. Parental monthly income at admission enters as bracket midpoints in 10,000 KRW, deflated to 2015 prices with the consumer price index of the graduation year, and parental income at the survey as nominal bracket midpoints, zero if the parents have no income or are both deceased, with an indicator for deceased parents. Father's and mother's education each have seven levels. Household structure is co-residence with parents and the number of co-resident household members. The spouse's monthly income bracket is zero for unmarried graduates and for spouses without income, and the number of dependent children is zero for unmarried graduates and is not asked in the 2008 wave. Parental assets are collected only from the 2015 wave onward and are not used.

In KLIPS I construct the excluded variables from the person and household files of each wave. To construct spouse earnings, I link the household head and the spouse (relationship codes 10 and 20 within a household); spouse earnings equal the spouse's monthly wage for a wage worker, monthly earnings for a self-employed spouse, and zero for a spouse who is not working. Children by age group (0--2, 3--5, 6--11, 12--18) count household members by birth year from the roster. The counts are assigned to the head and the spouse, set to zero for unmarried members, and missing for the few married members who are neither. Asset income is the household's previous-year financial and real estate income, and transfer income its previous-year public and social transfers, both divided by twelve. Spouse earnings and asset income enter as the log of one plus their value. For the young KLIPS groups, household size is the number of household members, and co-residence with parents equals one when the person and at least one of his or her own parents currently live in the household, based on the relationship codes and the living-together item of all members. As in the GOMS question, members temporarily living elsewhere, for example for study or work in another region, do not count.

In Table~\ref{tab:iv_goms} the GOMS excluded variables enter without female interactions, whereas in the KLIPS tables each excluded variable is interacted with the female indicator, except in the household-structure rows. Adding the female interactions to the GOMS household variables moves the Kim-Lee estimates by at most $0.004$ and the Heckman estimates and the quantile selection medians by at most $0.001$. Table~\ref{tab:iv_firststage} reports the first-stage coefficients of the main blocks of excluded variables, and Table~\ref{tab:iv_wald} the joint Wald statistics of each GOMS set of excluded variables by sample.

\begin{table}[htbp]
\centering
\caption{First-Stage Coefficients of the Excluded Variables}\label{tab:iv_firststage}
{\small
\begin{tabular}{lrr}
\toprule
 & \multicolumn{1}{c}{Coefficient} & \multicolumn{1}{c}{SE} \\
\midrule
\multicolumn{3}{l}{\textit{GOMS, pooled: family background (probit, $N = 186{,}033$)}} \\
Parental income at admission (10,000 KRW) & $+$0.000 & (0.000) \\
Parental income at the survey (10,000 KRW) & $-$0.000 & (0.000) \\
Parents deceased & $+$0.068 & (0.075) \\
Father's education (1--7) & $-$0.026 & (0.003) \\
Mother's education (1--7) & $-$0.022 & (0.004) \\
\midrule
\multicolumn{3}{l}{\textit{GOMS, pooled: household structure (probit, $N = 201{,}745$)}} \\
Co-residence with parents & $+$0.003 & (0.013) \\
Household size & $-$0.148 & (0.005) \\
\midrule
\multicolumn{3}{l}{\textit{KLIPS, pool 24--55: children by age group (probit, $N = 98{,}632$)}} \\
Children aged 0--2 & $+$0.404 & (0.033) \\
Children aged 0--2 $\times$ female & $-$0.974 & (0.041) \\
Children aged 3--5 & $+$0.214 & (0.026) \\
Children aged 3--5 $\times$ female & $-$0.608 & (0.032) \\
Children aged 6--11 & $+$0.081 & (0.022) \\
Children aged 6--11 $\times$ female & $-$0.357 & (0.026) \\
Children aged 12--18 & $+$0.021 & (0.019) \\
Children aged 12--18 $\times$ female & $-$0.112 & (0.025) \\
\bottomrule
\end{tabular}
}

\vspace{0.5em}
\parbox{\textwidth}{\scriptsize\textit{Notes:} Probit coefficients of the excluded variables in the employment equation, which also contains the selection covariates (GOMS: female and the Table~\ref{tab:main} covariates other than the job characteristics; KLIPS: as in Section~\ref{sec:klips}). In KLIPS each excluded variable also enters interacted with the female indicator. Heteroskedasticity-robust standard errors for GOMS and standard errors clustered by person for KLIPS.}
\end{table}

\begin{table}[htbp]
\centering
\caption{First-Stage Wald Statistics of the Excluded Variables, GOMS}\label{tab:iv_wald}
{\small
\begin{tabular}{lrrrr}
\toprule
 & \multicolumn{1}{c}{Pooled} & \multicolumn{1}{c}{2008--2011} & \multicolumn{1}{c}{2012--2015} & \multicolumn{1}{c}{2016--2019} \\
\midrule
\multicolumn{5}{l}{\textit{Panel A: Sieve probit (Kim-Lee first stage)}} \\
Household structure (2 df) & 3,601 & 1,066 &   642 & 2,959 \\
Family background (5 df) & 265 & 101 & 124 &  97 \\
All candidate excluded variables (9 df) & 3,514 & 1,180 &   762 & 2,590 \\
\midrule
\multicolumn{5}{l}{\textit{Panel B: Probit (Heckman first stage)}} \\
Household structure (2 df) & 3,371 & 1,056 &   612 & 2,976 \\
Family background (5 df) & 260 &  98 & 127 &  94 \\
All candidate excluded variables (9 df) & 3,285 & 1,199 &   757 & 2,613 \\
\bottomrule
\end{tabular}
}

\vspace{0.5em}
\parbox{\textwidth}{\scriptsize\textit{Notes:} Heteroskedasticity-robust Wald statistics for the joint significance of the excluded variables in the first stages of Panels A and B of Table~\ref{tab:iv_goms}, with degrees of freedom in parentheses. Sets of excluded variables and samples as in Table~\ref{tab:iv_goms}.}
\end{table}

\begin{figure}[htbp]
\centering
\includegraphics[width=\textwidth]{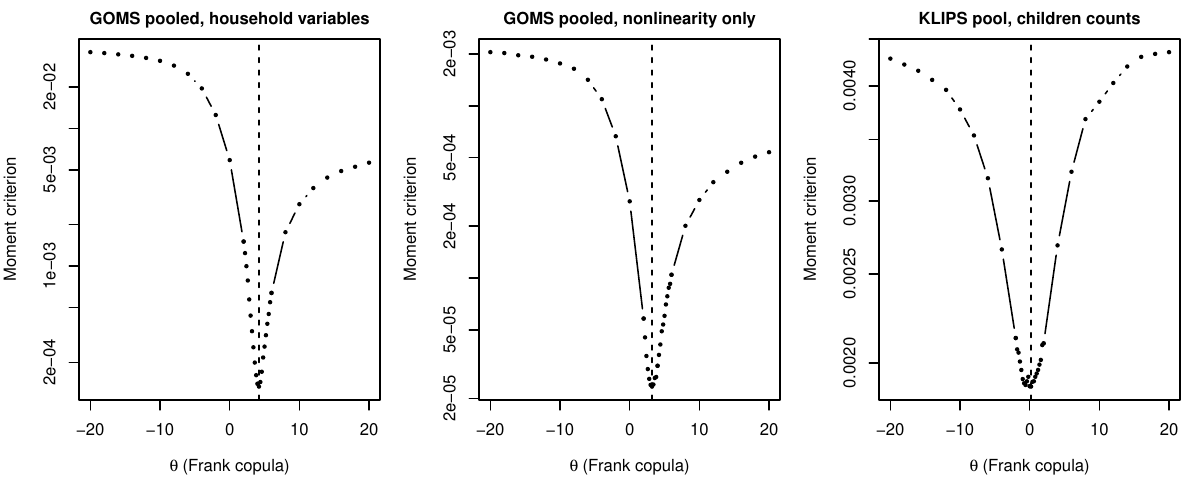}
\caption{Moment criterion of the quantile selection model as a function of the Frank copula parameter $\theta$: pooled GOMS with the excluded household variables (left), pooled GOMS with nonlinearity only (center), and the KLIPS prime-age pool with the children counts as excluded variables (right). The dashed line marks the estimate; the vertical axis is logarithmic.}\label{fig:ab_criterion}
\end{figure}

\textit{Quantile selection estimator.} The \cite{arellano2017quantile} model writes the latent wage as $Y_i^* = X_i'\beta(U_i)$ with $U_i$ uniform, employment as $D_i = \mathbbm{1}\{V_i \le p(Z_i)\}$ with $V_i$ uniform, and the joint distribution of $(U_i, V_i)$ given $Z_i$ as a copula $C(\cdot, \cdot; \theta)$. Then $P(Y_i \le X_i'\beta(\tau) \mid D_i = 1, Z_i) = C(\tau, p(Z_i); \theta)/p(Z_i) =: G_i(\tau)$, so $\beta(\tau)$ solves a quantile regression in which observation $i$ is evaluated at the level $G_i(\tau)$. The copula parameter solves the moment conditions $E[D_i \phi(Z_i)(\mathbbm{1}\{Y_i \le X_i'\beta(\tau_l)\} - G_i(\tau_l))] = 0$ over a grid of quantiles $\tau_l \in \{0.1, 0.3, 0.5, 0.7, 0.9\}$, with $\phi(Z_i)$ the excluded variables and $p(Z_i)$, or $p(Z_i)$ and $p(Z_i)^2$ when there is no excluded variable. I use the Frank copula, whose closed form makes the criterion cheap to evaluate, estimate $p(Z_i)$ by a probit with the excluded variables (or by the sieve probit of the Kim-Lee first stage in the nonlinearity-only variant), minimize the criterion over $\theta$ by a grid search with a refinement pass, and solve the rotated quantile regression by the Frisch-Newton interior point method with the observation-specific levels entering through the dual constraint. Job characteristics are observed only for the employed, so, as in the Kim-Lee estimates, they enter the quantile regressions but not the selection probability; this assumes that, given the other covariates and the excluded variables, the selection error and its copula with the wage rank do not depend on job characteristics. Figure~\ref{fig:ab_criterion} shows the criterion as a function of $\theta$ for the pooled GOMS sample with the excluded household variables and with nonlinearity only, and for the KLIPS prime-age pool with the children counts as excluded variables. The minima are interior. With the family-background variables as the only excluded variables, the criterion reaches its minimum at or near the boundary of the search range, so I do not report those estimates. Standard errors for the quantile selection estimates come from a bootstrap over graduates in GOMS and over persons in KLIPS (200 replications) that re-estimates the first stage and the copula parameter in each replication.

\section{Supplementary Selection Estimates}\label{sec:appendix_selection}

Table~\ref{tab:iv_goms_q} reports the GOMS median estimates behind the comparison in Section~\ref{sec:quantile}. The medians with the excluded household variables and with nonlinearity only agree within 0.8 log points in every period, and both differ from uncorrected median regression by less than two log points.

\begin{table}[t!]
\centering
\caption{Exclusion-Based Estimates of the Entry-Level Gender Gap, GOMS: Quantile Selection Model}\label{tab:iv_goms_q}
{\small\setlength{\tabcolsep}{4pt}\renewcommand{\arraystretch}{1.1}
\begin{tabular}{lrrrr}
\toprule
 & \multicolumn{1}{c}{Pooled} & \multicolumn{1}{c}{2008--2011} & \multicolumn{1}{c}{2012--2015} & \multicolumn{1}{c}{2016--2019} \\
\midrule
\multicolumn{5}{l}{\textit{Panel A: Female coefficient at the median}} \\
Household structure & $-$0.054 & $-$0.046 & $-$0.068 & $-$0.042 \\
 & (0.003) & (0.005) & (0.005) & (0.004) \\
All candidate excluded variables & $-$0.056 & $-$0.049 & $-$0.069 & $-$0.046 \\
Without the income variables & $-$0.054 & $-$0.045 & $-$0.066 & $-$0.042 \\
Household size only & $-$0.054 & $-$0.047 & $-$0.068 & $-$0.041 \\
Nonlinearity only (sieve $\hat{p}$) & $-$0.053 & $-$0.052 & $-$0.060 & $-$0.040 \\
 & (0.003) & (0.005) & (0.005) & (0.004) \\
Uncorrected median regression & $-$0.048 & $-$0.054 & $-$0.052 & $-$0.036 \\
\midrule
\multicolumn{5}{l}{\textit{Panel B: Copula dependence (Spearman)}} \\
Household structure & $+$0.58 & $+$0.64 & $+$0.69 & $+$0.45 \\
Nonlinearity only & $+$0.47 & $+$0.26 & $+$0.42 & $+$0.40 \\
\bottomrule
\end{tabular}
}

\vspace{0.5em}
\parbox{\textwidth}{\scriptsize\textit{Notes:} \cite{arellano2017quantile} estimator (Appendix~\ref{sec:appendix_iv}): female coefficient at the median of log hourly wages with the Table~\ref{tab:main} controls. Sets of excluded variables as in Table~\ref{tab:iv_goms}; ``without the income variables'' is co-residence, household size, and parents' education. ``Nonlinearity only'' uses the Kim-Lee sieve-probit selection probability and no excluded variable. Bootstrap standard errors over graduates in parentheses, where reported. Panel B: Spearman correlation implied by the estimated copula parameter.}
\end{table}

Tables~\ref{tab:klips_hetero}, \ref{tab:iv_klips}, and \ref{tab:iv_klips_q} report the KLIPS heterogeneous-correction and exclusion-based estimates summarized in Section~\ref{sec:klips}. The Wald test of a common correction and the Huber-Melly test concern different restrictions: the first tests the shape of the mean correction, the second the stronger conditional-independence restriction. The excluded variables have substantial first-stage relevance (Panel C of Table~\ref{tab:iv_klips}), but their validity still requires that children, spouse earnings, household resources, and household structure have no direct effect on offered wages conditional on the controls. In the prime-age pool the copula dependence of the quantile selection model is close to zero with the children counts as excluded variables and positive with children and transfer income, with all four groups, and with nonlinearity only. The corrected medians differ accordingly, from $-0.336$ with the children counts to $-0.288$ with all four groups (Table~\ref{tab:iv_klips_q} and Figure~\ref{fig:klips_quantile}), and I report all of them.

\begin{table}[t!]
\centering
\caption{Heterogeneous Selection in KLIPS}\label{tab:klips_hetero}
{\small\setlength{\tabcolsep}{4pt}
\begin{tabular}{lrrrr}
\toprule
 & \multicolumn{1}{c}{Pool 24--55} & \multicolumn{1}{c}{Young 24--35} & \multicolumn{1}{c}{College 24--55} & \multicolumn{1}{c}{College 24--35} \\
\midrule
\multicolumn{5}{l}{\textit{Panel A: Kim-Lee female coefficient}} \\
Single correction & $-$0.260 & $-$0.078 & $-$0.156 & $-$0.070 \\
 & (0.012) & (0.012) & (0.014) & (0.013) \\
Cell-specific corrections & $-$0.222 & $-$0.070 & $-$0.144 & $-$0.081 \\
 & (0.014) & (0.013) & (0.016) & (0.014) \\
Wald test of the interaction terms & 265 & 105 & 141 & 80 \\
 & [55 df] & [55 df] & [50 df] & [50 df] \\
\midrule
\multicolumn{5}{l}{\textit{Panel B: Huber-Melly test, all slopes}} \\
Kolmogorov-Smirnov & 7.66 & 6.75 & 8.29 & 7.61 \\
 & [$<$0.01] & [0.005] & [$<$0.01] & [$<$0.01] \\
Cram\'er-von Mises & 2,126 & 1,378 & 2,144 & 1,492 \\
 & [$<$0.01] & [$<$0.01] & [$<$0.01] & [$<$0.01] \\
\multicolumn{5}{l}{\textit{Panel C: Huber-Melly test, female coefficient}} \\
Kolmogorov-Smirnov & 5.44 & 1.98 & 3.08 & 1.21 \\
 & [$<$0.01] & [0.236] & [0.015] & [0.774] \\
Cram\'er-von Mises & 78 & 11 & 39 & 3 \\
 & [$<$0.01] & [0.216] & [$<$0.01] & [0.859] \\
\bottomrule
\end{tabular}
}

\vspace{0.5em}
\parbox{\textwidth}{\scriptsize\textit{Notes:} The checks of Table~\ref{tab:cellcf} (Panel A) and Table~\ref{tab:hm} (Panels B and C) for the four KLIPS groups, with the Table~\ref{tab:klips} controls. Panel A: the cell-specific corrections interact the spline in $\hat{p}$ with the 2-year, 4-year, married, and field-of-study indicators that vary within the group; standard errors in parentheses and the Wald test are first-stage corrected and clustered by person. Panels B and C: $p$-values in brackets from a bootstrap over persons (199 replications).}
\end{table}

\begin{table}[p]
\centering
\caption{Exclusion-Based Estimates of the Gender Gap over the Lifecycle, KLIPS}\label{tab:iv_klips}
{\scriptsize\setlength{\tabcolsep}{4pt}
\begin{tabular}{lrrrr}
\toprule
 & \multicolumn{1}{c}{Pool 24--55} & \multicolumn{1}{c}{Young 24--35} & \multicolumn{1}{c}{College 24--55} & \multicolumn{1}{c}{College 24--35} \\
\midrule
\multicolumn{5}{l}{\textit{Panel A: Kim-Lee with excluded variables}} \\
Children by age group & $-$0.244 & $-$0.090 & $-$0.152 & $-$0.076 \\
 & (0.011) & (0.011) & (0.013) & (0.013) \\
All four groups & $-$0.255 & $-$0.099 & $-$0.161 & $-$0.082 \\
 & (0.010) & (0.011) & (0.013) & (0.013) \\
Children and transfer income & $-$0.238 & $-$0.089 & $-$0.151 & $-$0.076 \\
 & (0.011) & (0.011) & (0.013) & (0.013) \\
Household structure (GOMS set) &  & $-$0.076 &  & $-$0.071 \\
 &  & (0.012) &  & (0.013) \\
\midrule
\multicolumn{5}{l}{\textit{Panel B: Heckman maximum likelihood with excluded variables}} \\
Children by age group & $-$0.317 & $-$0.119 & $-$0.179 & $-$0.098 \\
 & (0.011) & (0.010) & (0.015) & (0.011) \\
All four groups & $-$0.298 & $-$0.123 & $-$0.177 & $-$0.101 \\
 & (0.010) & (0.010) & (0.013) & (0.011) \\
Children and transfer income & $-$0.312 & $-$0.118 & $-$0.178 & $-$0.098 \\
 & (0.011) & (0.010) & (0.015) & (0.011) \\
Household structure (GOMS set) &  & $-$0.113 &  & $-$0.090 \\
 &  & (0.012) &  & (0.013) \\
\midrule
\multicolumn{5}{l}{\textit{Panel C: First-stage Wald statistics of the excluded variables}} \\
Children by age group (8 df), sieve probit & 568 & 497 & 382 & 359 \\
\quad probit & 747 & 751 & 624 & 608 \\
All four groups (14 df), sieve probit & 998 & 630 & 577 & 451 \\
\quad probit & 1,338 & 1,010 &   984 &   824 \\
Children and transfer income (10 df), sieve probit & 600 & 513 & 392 & 365 \\
\quad probit & 775 & 770 & 632 & 615 \\
Household structure, GOMS set (2 df), sieve probit &  & 185 &  & 160 \\
\quad probit &  & 244 &  & 186 \\
\bottomrule
\end{tabular}
}

\vspace{0.5em}
\parbox{\textwidth}{\scriptsize\textit{Notes:} Female coefficient on log hourly wages with the Table~\ref{tab:klips} controls. Excluded variables (Appendix~\ref{sec:appendix_iv}) enter the selection equation only, interacted with the female indicator except in the household-structure rows (co-residence with parents and household size, as in Table~\ref{tab:iv_goms}). Among college graduates aged 24 to 35, all 13 men with children aged 12 to 18 are wage-employed, so the men's term for that age group is omitted and the Panel C degrees of freedom are one lower in that column. ``All four groups'' are children by age group, spouse earnings, asset income, and transfer income. Each row uses the observations with non-missing excluded variables. Standard errors and the Panel C Wald statistics are clustered by person; Kim-Lee standard errors are first-stage-corrected.}
\end{table}

\FloatBarrier
\begin{table}[!ht]
\centering
\caption{Exclusion-Based Estimates of the Gender Gap over the Lifecycle, KLIPS: Quantile Selection Model}\label{tab:iv_klips_q}
{\footnotesize\setlength{\tabcolsep}{4pt}
\begin{tabular}{lrrrr}
\toprule
 & \multicolumn{1}{c}{Pool 24--55} & \multicolumn{1}{c}{Young 24--35} & \multicolumn{1}{c}{College 24--55} & \multicolumn{1}{c}{College 24--35} \\
\midrule
\multicolumn{5}{l}{\textit{Panel A: Female coefficient at the median}} \\
Children by age group & $-$0.336 & $-$0.108 & $-$0.137 & $-$0.100 \\
 & (0.023) & (0.014) & (0.026) & (0.012) \\
All four groups & $-$0.288 & $-$0.120 & $-$0.136 & $-$0.103 \\
Children and transfer income & $-$0.304 & $-$0.105 & $-$0.137 & $-$0.100 \\
Household structure (GOMS set) &  & $-$0.075 &  & $-$0.065 \\
Nonlinearity only (sieve $\hat{p}$) & $-$0.302 & $-$0.074 & $-$0.107 & $-$0.076 \\
 & (0.018) & (0.017) & (0.017) & (0.013) \\
Uncorrected median regression & $-$0.341 & $-$0.148 & $-$0.209 & $-$0.126 \\
\midrule
\multicolumn{5}{l}{\textit{Panel B: Copula dependence (Spearman)}} \\
Children by age group & $+$0.03 & $+$0.52 & $+$0.42 & $+$0.47 \\
All four groups & $+$0.34 & $+$0.42 & $+$0.45 & $+$0.42 \\
Children and transfer income & $+$0.26 & $+$0.54 & $+$0.42 & $+$0.47 \\
Household structure (GOMS set) &  & $+$0.63 &  & $+$0.67 \\
Nonlinearity only (sieve $\hat{p}$) & $+$0.26 & $+$0.61 & $+$0.49 & $+$0.52 \\
\midrule
$N$ with non-missing children counts & 98,632 & 30,937 & 55,235 & 24,218 \\
\bottomrule
\end{tabular}
}

\vspace{0.5em}
\parbox{\textwidth}{\scriptsize\textit{Notes:} \cite{arellano2017quantile} estimator as in Table~\ref{tab:iv_goms_q}, with the Table~\ref{tab:klips} controls and the excluded variables of Table~\ref{tab:iv_klips}. The nonlinearity-only and household-structure rows use all observations of the group, the other rows the observations with non-missing excluded variables, and the uncorrected median regression the observations with non-missing children counts. Bootstrap standard errors over persons in parentheses, where reported. Panel B: Spearman correlation implied by the estimated copula parameter.}
\end{table}

\end{document}